\documentclass[jmp,twocolumn,preprintnumbers,aps,groupedaddress,amsmath,amssymb]{revtex4}

\usepackage{graphicx}
\usepackage{xspace}
\usepackage{dcolumn}
\usepackage{bm}
\usepackage{amscd}
\usepackage{amsthm}
\usepackage{german}
\usepackage{color}

\input xy
\xyoption{all}

\newcommand{\unity}{\ensuremath{{\rm 1 \negthickspace l}{}}}

\renewcommand{\Re}{\operatorname{Re}}
\renewcommand{\Im}{\operatorname{Im}}
\renewcommand{\vec}{\operatorname{vec}}
\newcommand{\tr}{\operatorname{tr}}
\newcommand{\Hess}{\mathrm{Hess}}
\newcommand{\Hessop}{\mathrm{\bf Hess}}

\newcommand{\grad}{\operatorname{grad}}
\newcommand{\D}{\operatorname{D}}
\newcommand{\reach}{\operatorname{Reach}}
\newcommand{\step}{\operatorname{Step}}
\newcommand{\ad}[1]{\operatorname{ad}_{#1}}
\newcommand{\Ad}[1]{\operatorname{Ad}_{#1}}

\newcommand{\C}{\mathbb{C}}
\newcommand{\K}{\mathbb{K}}
\newcommand{\N}{\mathbb{N}}
\newcommand{\Z}{\mathbb{Z}}
\newcommand{\bG}{\mathbf{G}}
\newcommand{\bH}{\mathbf{H}}
\newcommand{\bK}{\mathbf{K}}

\newcommand{\SU}{\mathrm{SU}}
\newcommand{\U}{\mathrm{U}}

\newcommand{\rD}{\ensuremath{{\rm D}{}}}

\newcommand{\fg}{\mathfrak{g}}
\newcommand{\fk}{\mathfrak{k}}
\newcommand{\fh}{\mathfrak{h}}
\newcommand{\fp}{\mathfrak{p}}

\newcommand{\LAG}{{Lagrange}\xspace}

\newcommand{\adr}{\operatorname{ad}}
\newcommand{\Adr}{\operatorname{Ad}}

\newcommand{\Cd}{\ensuremath{C^{\dagger}}{}}
\newcommand{\Dd}{\ensuremath{D^{\dagger}}{}}
\newcommand{\Ed}{\ensuremath{E^{\dagger}}{}}
\newcommand{\Ud}{\ensuremath{U^{\dagger}}{}}

\newcommand{\GEE}{\ensuremath{\comm{UEU^{\dagger}}{E^\dagger}}{}}

\newcommand{\SUloc}{SU_{\rm loc}}
\newcommand{\suloc}{\mathfrak{su}_{\rm loc}}
\newcommand{\su}{\mathfrak{su}}

\newcommand{\gl}{\mathfrak{gl}}
\newcommand{\diag}{{\rm diag\,}}
\newcommand{\e}{{\rm e}}
\newcommand{\ri}{{\rm i}}
\newcommand{\rT}{{\rm T}}

\newcommand{\g}{\mathfrak{g}}
\newcommand{\h}{\mathfrak{h}}
\newcommand{\p}{\mathfrak{p}}

\newcommand{\R}{\mathbb{R}}

\newcommand{\maxover}[1]{\ensuremath{\underset{#1}{\rm max} \,\,}}

\newcommand{\comm}[2]{\ensuremath{[#1,#2]}}

\newcommand{\norm}[1]{\ensuremath{\Vert #1 \Vert{}}}

\newcommand{\fnormsq}[1]{\ensuremath{\Vert #1 \Vert{}}_2^2}

\newcommand{\ket}[1]{\ensuremath{| #1 \rangle}{}}
\newcommand{\braket}[2]{\ensuremath{\langle #1 | #2 \rangle}{}}

\newcommand{\roundbraket}[2]{\ensuremath{( #1 | #2 )}{}}
\newcommand{\ketbra}[2]{\ensuremath{| #1 \rangle \langle #2 | }{}}

\newcommand{\argmax}{\operatorname{argmax}{}}

\newcommand{\argmaxover}[1]{\ensuremath{\underset{#1}\argmax}}

\theoremstyle{plain}

\newtheorem{corollary}{Corollary}[section]
\newtheorem{lemma}{Lemma}[section]
\newtheorem{proposition}{Proposition}[section]
\newtheorem{theorem}{Theorem}[section]
\newtheorem{claimA}{Claim \!A}

\theoremstyle{definition}
\newtheorem{definition}{Definition}[section]
\newtheorem{remark}{Remark}[section]
\newtheorem{fact}{Fact}
\newtheorem{app:fact}{Fact}
\newtheorem{remarkA}{Remark \!A}

\theoremstyle{remark}

\begin{document}


\title{Gradient Flows for Optimisation and Quantum Control:\\
 Foundations and Applications}

\author{T.~Schulte-Herbr{\"u}ggen}\email{tosh@ch.tum.de}
\affiliation{Department of Chemistry, Technical University Munich,
Lichtenbergstrasse 4, D-85747 Garching, Germany}

\author{S.J.~Glaser}
\affiliation{Department of Chemistry, Technical University Munich,
Lichtenbergstrasse 4, D-85747 Garching, Germany}

\author{G.~Dirr}\email{dirr@mathematik.uni-wuerzburg.de}
\affiliation{Institute of Mathematics, University of W{\"u}rzburg,\\
Am Hubland, D-97074 W{\"u}rzburg, Germany}

\author{U.~Helmke}
\affiliation{Institute of Mathematics, University of W{\"u}rzburg,\\
Am Hubland, D-97074 W{\"u}rzburg, Germany}

\date{\today}

\begin{abstract}
For addressing optimisation tasks on finite dimensional quantum
systems, we give a comprehensive account on the foundations of gradient flows
on Riemannian manifolds including new applications to quantum control:
we extend former results on unitary groups to closed subgroups with tensor-product structure,
where the finest product partitioning consists of purely local unitary operations.
Moreover, the framework is kept sufficiently general for setting
up gradient flows on (sub-)manifolds, Lie (sub-)groups, 
and (reductive) homogeneous spaces. Relevant convergence conditions are
discussed, in particular for gradient flows on compact and analytic
manifolds. This part of the paper is meant to serve as foundation
for some recent and new achievements, and as setting for further research.

Exploiting the differential geometry of
quantum dynamics under different scenarios helps to provide highly
useful algorithms: (a) On an abstract level, gradient flows may
establish the exact upper bounds of pertinent quality functions,
i.e. upper bounds reachable within the underlying manifold of the
state space dynamics;
(b) in a second stage referring to a concrete experimental setting, 
gradient flows on the space of piecewise constant control
amplitudes in $\R^m$ may be set up to yield (approximations to)
optimal control for quantum devices under realistic conditions. 

Illustrative examples and new applications are given,
such as figures of merit on the subgroup of local unitary action
on $n$ qubits relating to distance measures of pure-state entanglement.
We establish the correspondence to best rank"~1 approximations of
higher order tensors and show applications from quantum information,
where our gradient flows on the subgroup of 
local unitary operations provide a numerically stable 
alternative to tensor-{\sc svd} techniques.\\[4mm]


\end{abstract}
\maketitle

\bigskip
{\footnotesize
\tableofcontents
}


\section{Introduction}
\label{Sec:intro}

Controlling quantum systems offers a great potential
for performing computational tasks or for simulating the behaviour of
other quantum systems (which are difficult to handle experimentally) 
or classical systems \cite{Fey82, Fey96}, when the complexity of a problem
reduces upon going from a classical to a quantum setting \cite{Kit02}.
Important examples are known
in quantum computation, quantum search and quantum simulation. 
Most prominently, there is the exponential speed-up by Shor's quantum
algorithm of prime factorisation \cite{Shor94, Shor97}, which
on a general level relates to the class of quantum algorithms
\cite{Jozsa88, Mosca88} solving hidden subgroup problems in an efficient
way \cite{EHK04}. In Grover's quantum search algorithm \cite{Grov96, Grov97}
one still finds a quadratic acceleration compared to classical approaches
\cite{Pap95}. Recently, the simulation of quantum phase-transitions
\cite{Sachdev99} has shifted into focus \cite{JC03,PC04}.

Among the generic tools needed for 
advances in quantum technology, for a survey see, e.g., \cite{DowMil03}, 
quantum control plays a major role. Its key concern is not only to
find (optimal) control strategies for quantum dynamical systems
such that a certain performance index is maximised (typical examples
being  quantum gate fidelities, efficiencies of state transfer or
coherence transfer, as well as distances related to Euclidean 
entanglement measures) but, moreover,
to develop constructive ways for implementing controls under realistic
experimental settings.
Usually, such figures of merit depend on terminal conditions and 
running cost, like time or energy, cf.~Section \ref{Sec:overview}.
In quantum control, however, important classes of performance
indices are completely determined by some quality function
and its value at the system's final state. 

Since realistic quantum systems are mostly beyond analytical
tractability, numerical methods are often indispensable. 
A good strategy is to proceed in two steps:
(a) firstly, by exploring the possible gains on an abstract and
computationally cheap level, i.e. by maximising the quality function
either over the entire state space or over the set of possible
states---the so-called reachable set;
(b) secondly, by going into optimising the experimental controls
(\/`pulse shapes\/') in a concrete setting. However, (b) is often
computationally expensive and highly sensitive, as it actually
consists of solving an infinite dimensional constrained
variational problem.

By merely depending on the geometry of the underlying state-space
manifold the first instance (a) allows for analysing in advance and
on an abstract level
the limits of what one can achieve in step (b). 
We therefore refer to (a) as the {\em abstract optimisation task}.
The second instance, in contrast, hinges on introducing the specific
time scales of an experimental setting for finding steerings of the
quantum dynamical system 
such that the optima determined in (a) are actually assumed.
This is why we term (b) the dynamic {\em optimal control task}. 
Certainly, one can approach the entire problem only in terms of (b) and
sometimes one is even forced to do so, e.g.~if nothing is known about
the geometric structure of the reachable set.
Yet, the above two-fold strategy may serve to yield a benchmark in (a)
for judging the reliability of the numerical results of (b). 
In both regards, gradient-flow methods will prove to be particularly
useful. 

In a pioneering paper \cite{Bro88+91}, Brockett introduced
the idea of exploiting gradient flows on the orthogonal group
for diagonalising real symmetric matrices and for sorting lists.
In a series of subsequent papers he extended the concept to intrinsic
gradient methods  for (constrained) optimisation \cite{Bro89,Bro93}.
Soon these techniques were generalised to Riemannian manifolds,
their mathematical and numerical details were worked out
\cite{Diss-Smith,Smith94,HM94} and thus they turned out to be
applicable to a broad array of optimisation tasks including
eigenvalue and singular-value problems, principal component analysis,
matrix least-squares matching problems,
balanced matrix factorisations, and combinatorial optimisation---for
an overview see, e.g., \cite{Bloch94,HM94}. 
Implementing gradient flows
for optimisation on smooth manifolds, such as unitary orbits,
inherently  ensures the discretised flow remains within 
the manifold by virtue of Riemannian exponential map.
Alternatively, formulating the optimisation problem on 
some embedding Euclidean space comes at the expense of additional constraints
(e.g. enforcing unitarity) to be taken care of by Lagrange-type or
penalty-type techniques.
In this sense, gradient flows on manifolds are {\em intrinsic} optimisation
methods \cite{ChouDriessel90}, whereas {\em extrinsic} optimisations
on the embedding spaces require non-linear projective techniques in order to stay
on the (constraint) manifold.

Using the differential geometry of matrix
manifolds has thus become a field of active research. For new
developments (however without exploiting the Lie structure to the
full extent) see, e.g., \cite{AMS08}.  ---
Even beyond manifolds, gradient flows have recently been described for 
metric spaces with applications of probability theory \cite{AGS0508}.

Meanwhile, gradient flows and their discrete numerical integration
schemes have also proven powerful tools of optimisation in quantum
dynamics. This holds in both of the two types of tasks:
(a) for exploring the maxima of pertinent quality functions on
the reachable set of a quantum system, e.g.~on the unitary group
and its orbits \cite{Science98,TOSH-Diss,PRA_inv} and
(b) for arriving at concrete experimental steerings
(i.e.~\/`pulse shapes\/') actually achieving the quality limits
established in (a) under given experimental conditions for closed
\cite{grape,PRA05,PRA07} or open \cite{PRL_decoh,PRL_decoh2,Rabitz07}
systems.
Recently we thus gave upper limits for time complexities of
implementing unitary modules \cite{PRA05}, or for exploring the
potential of quantum compilation with complex instruction sets
\cite{CISC07}.

Moreover, in view of using gradient techniques for unifying variational 
approaches to ground-state calculations \cite{Eisert07}, it will be useful 
to exploit a common framework of gradient flows on Riemannian manifolds as
well as projective techniques on their tangent spaces. To this end,
we also show how gradient flows
can readily be restricted, e.g., from Lie groups $\bG$ to any closed subgroup $\bH$,
in particular any closed subgroup of $SU(N)$. 
Since quantum dynamics often takes place in a subspace of the entire
Hilbert space so that long-range entangling correlations can be neglected
on the basis of area laws (see, e.g.~\cite{Plenio05,Plenio06,Wolf07b}), 
truncating the Hilbert space to the pertinent subspaces is tantamount to
representing dynamics of large systems. For instance,
unitary networks use consecutive partitionings into different
subgroups that can be applied to efficiently compute ground states of
large-scale quantum systems with a cost increasing polynomially with
system size while retaining sufficiently good approximations.
Current approaches of truncating the full-scale Hilbert space into
the according subspaces include matrix-product states (MPS)
\cite{Fannes92a, Fannes92b}
of density-matrix renormalisation groups (DMRG) \cite{LNP528,Schollwoeck05},
quantum cellular automata with Margolus partitionings \cite{Werner04},
projected entangled pair states (PEPS) \cite{Verstraete04} 
weighted graph states (WGS) \cite{Anders06},
multi-scale entanglement renormalisation approaches (MERA) \cite{Vidal07},
string-bond states (SBS) \cite{Schuch07} as well as
combinations of different techniques \cite{Eisert07,Eisert08}. --- 
It is noteworthy, however, that in many-particle physics 
gradient flows for diagonalising Hamiltonians were re-introduced 
independently of Brockett's work \cite{Bro88+91} by Wegner \cite{Weg94} 
and were further elaborated on again independently of the monography by
Helmke and Moore \cite{HM94} or the one by Bloch \cite{Bloch94} in the tract by Kehrein \cite{Kehr06}. 
Suffice this to illustrate the need for making the mathematical methods
available to the physics community in a comprehensive way.

Another field of applications of restricting flows to closed
subgroups of $SU(N)$ is entanglement of multi-partite quantum systems
\cite{PV07,HHHH07}: we present a connection from gradient flows on
the subgroup of local unitary operations to best rank--$1$
approximations of higher order tensors as well as a relation
to tensor"~{\sc svd}s. They are of importance, e.g., in view of
optimisation of entanglement witnesses \cite{Guehne04}. 
Gradient flows on partitionings of the full unitary group are
anticipated to be of use also for classifying multi-partite systems
according to their mutual separability, an example being three-tangles
of GHZ-type and W-type states \cite{LOSU06,OSU07,EOSU07}.

Moreover, with the framework of treating gradient flows on
Riemannian manifolds being very general, we will also show how
they can be carried over to homogeneous spaces
that do no longer form Lie groups themselves. Standard examples
are coset spaces $\bG/\bH$, i.e.~the quotient of a Lie group $\bG$
by a closed (yet not necessarily normal) subgroup $\bH$.
In particular,  naturally reductive homogeneous space
are in the focus of interest.
The well-known double-bracket flows will be demonstrated to
form a special case precisely of this kind.

Though gradient flows on the set of control amplitudes
can be seen as another instance of flows on Riemannian manifolds,
our paper does not primarily focus on optimal control. The goal
is rather to give a comprehensive account of the foundations of gradient 
flows---and thus the justification for some recent developments---as well as 
to present new flows for intrinsic or extrinsic constraints and
new schemes of flows on reductive homogeneous spaces.
Terms are kept general enough to trigger future developments, since we
elucidate the necessary
requirements for implementing gradient-based optimisation methods
in different geometric settings: Riemannian manifolds and submanifolds,
Lie groups and homogeneous spaces. 

A separate paper on open quantum systems \cite{DHKS08}
sets up a formal approach within the framework of {\em Lie semigroups}
accounting for Markovian quantum evolutions (or Markovian channels).
There we also show the current limits of abstract optimisation over
reachable sets specifically arising in open systems.
The differential geometry of the set 
of all completely positive, trace-preserving invertible maps is analysed
in the framework of {\em Lie semigroups}. In particular, the set of
all Kossakowski-Lindblad generators is retrieved
as its tangent cone (Lie wedge). Moreover, it shows how the
Lie-semigroup structure corresponds to the Markov properties recently
studied in terms of divisibility \cite{Wolf08a}.
It illustrates why {\em abstract optimisation tasks for open systems}
are much more intricate than in the case of closed system, while 
{\em dynamic optimal control tasks for open systems} can be handled 
completely analogously.
It specifies algebraic conditions for time-optimal controls to be the method
of choice in open systems.
Finally it draws perspectives to new algorithmic approaches on
semigroup orbits combining (abstract) knowledge of
the respective Lie wedges with elements of numerical optimal control.

\subsection*{Outline}

To begin with, we consider flows on (Riemannian) manifolds and recall some
basic terminology on dynamical systems and Riemannian geometry.
Then the aim is to provide the differential geometric tools for setting up
gradient flows in different scenarios ranging from optimisation over the
entire unitary group to subgroups (e.g.~of local actions) or homogeneous spaces. 
Finally we give a number of applications including worked examples.

More precisely,
the paper is organised as follows: Section~\ref{Sec:overview} draws
a general sketch of dynamical systems and flows on manifolds 
including issues of reachability and controllability. 
It provides the manifold setting for gradient-flow-based algorithms 
like steepest ascent, conjugate gradients, Jacobi-type, and
Newton methods.

A detailed analysis is then given in Section~\ref{Sec:theory}, where
(1) we resume the general
preconditions for gradient flows on smooth manifolds.
In particular, we recall the role of a Riemannian metric
that allows for identifying the cotangent bundle $T^*M$
with the tangent bundle $TM$.
Large parts of the foundations can be found in
~\cite{HM94,Udriste94,AMS08}, but here we additionally provide
a comprehensive overview of the interplay between Riemannian geometry,
Lie groups, and (reductive) homogeneous spaces. 
(2) We give examples of gradient flows on compact
Lie groups as well as their closed subgroups.
(3) In view of further developments, we address gradient flows
on several types of reductive homogeneous spaces: Cartan-like,
naturally reductive ones and merely reductive ones. In particular,
double-bracket flows turn out as gradient flows on naturally reductive
homogeneous spaces.
(4) Examples interdispersed in the main text
illustrate the relevance in a plethora of different settings.

Section~\ref{Sec:applications} is
dedicated to specific applications in quantum control
and quantum information.
(1) We show how gradient flows on the subgroup of
local unitaries $SU_{\rm loc}(2^n)$ in $n$ qubits
do not only provide a valuable tool in witness optimisation,
but relate to generalised singular-value decompositions
({\sc svd}), namely the tensor-{\sc svd}.
Here, our gradient flows yield an alternative to common algorithms
for best rank"~1 approximations of higher-order tensors,
e.g.~higher-order power methods ({\sc hopm}) or higher-order
orthogonal iteration ({\sc hooi}).
(2) Flows on $SU_{\rm loc}(2^n)$
also serve as a convenient tool to decide whether 
Hamiltonian interactions can be time-reversed solely by local
unitary manipulations thus complementing the algebraic
assessment given in \cite{PRA_inv}.
(3) Optimisation tasks with (i) intrinsic and (ii) extrinsic
constraints are addressed by tailored gradient flows on
the respective subgroups (i) or with Lagrange-type penalties (ii).
By including practical applications and worked examples we illustrate 
the ample range of problems to which gradient flows on manifolds 
provide valuable solutions. ---
To this end, we start out by an extended
overview on techniques on (Riemannian) manifolds
with particular emphasis on gradient techniques.

\section{Overview}
\label{Sec:overview}


\subsection*{Flows and Dynamical Systems}
In this paper, we treat various optimisation tasks for 
quantum dynamical systems in a common framework, namely by gradient flows
on smooth manifolds.
Let $M$ denote a smooth manifold, e.g. the unitary orbit of all quantum
states relating to an initial state $X_0$. By a continuous-time dynamical
system or a {\em flow} one defines a smooth map 
\begin{equation}
\Phi: \R{}\times M\to M 
\end{equation}
such that for all states $X\in M$ and times $t,\tau \in \R{}$ one has
\begin{equation}
\begin{split}
\Phi(0,X) &= X\\
\Phi(\tau,\Phi(t,X)) &= \Phi(t+\tau,X)\quad.
\end{split}
\end{equation}
Since these equations hold for any $X\in M$, one gets the operator
identity
\begin{equation}\label{eqn:flow-semi}
\Phi_\tau\circ\Phi_t = \Phi_{t+\tau}
\end{equation}
for all $t,\tau\in \R$, thus showing the flow acts as a one-parameter
{\em group}, and for positive times $t,\tau\geq 0$ as a one-parameter
{\em semigroup of diffeomorphisms} on $M$.


\subsection*{Gradient Flows for Optimisation}

Now, the general idea for optimising a scalar quality function
on a smooth manifold $M$ (which might either arise naturally or from
including smooth equality constraints, {\em vide infra})
by dynamical systems is as follows:
Let $f: M \to \R{}$ be a smooth quality function on $M$.
The differential of $f: M \to \R{}$ is a mapping $D f: M \to T^*M$ of
the manifold to its cotangent bundle $T^*M$, while the gradient vector
field is a mapping $\grad f: M \to TM$ to its tangent bundle $TM$.
So the \emph{gradient} of $f$ at $X \in M$, denoted
$\grad f\,(X)$, is the vector in $T_XM$ uniquely determined by 
\begin{equation}
\D f\,(X)\cdot\xi = \braket{\grad f\,(X)}{\xi}_X
\quad \mbox{for all $\xi \in T_XM$}.
\end{equation}
Here, the scalar product $\braket{\cdot}{\cdot}_X$ plays a
central role: it allows for identifying $T_X^*M$ with $T_XM$.
The pair $\big(M,\braket{\cdot}{\cdot}\big)$ is called a
\emph{Riemannian manifold} with  \emph{Riemannian metric}
$\braket{\cdot}{\cdot}$.
In view of gradient flows, the convenience of
Riemannian manifolds lies in the fact that by duality
in particular
the differential $\D f\,(X)$ of $f$ at $X$ can be identified
with a tangent vector of $T_XM$.

Then, the flow $\Phi: \R{}\times M\to M$ determined by the
ordinary differential equation
\begin{equation}
\label{eqn:grad_system}
\dot X = \grad f(X) 
\end{equation}
is termed {\em gradient flow}. Formally, it is obtained by
integrating Eqn. (\ref{eqn:grad_system}), i.e. 
\begin{equation}
\Phi(t,X) = \Phi(t,\Phi(0,X)) = X(t)\;,
\end{equation}
where $X(t)$ denotes the unique solution of Eqn. (\ref{eqn:grad_system})
with initial value X(0) = X. Observe this ensures that $f$ does
increase along trajectories $\Phi$ of the system by virtue of following
the gradient direction of $f$. 


\subsection*{Discretised Gradient Flows}

\begin{figure}[Ht!]
\includegraphics[scale=0.3]{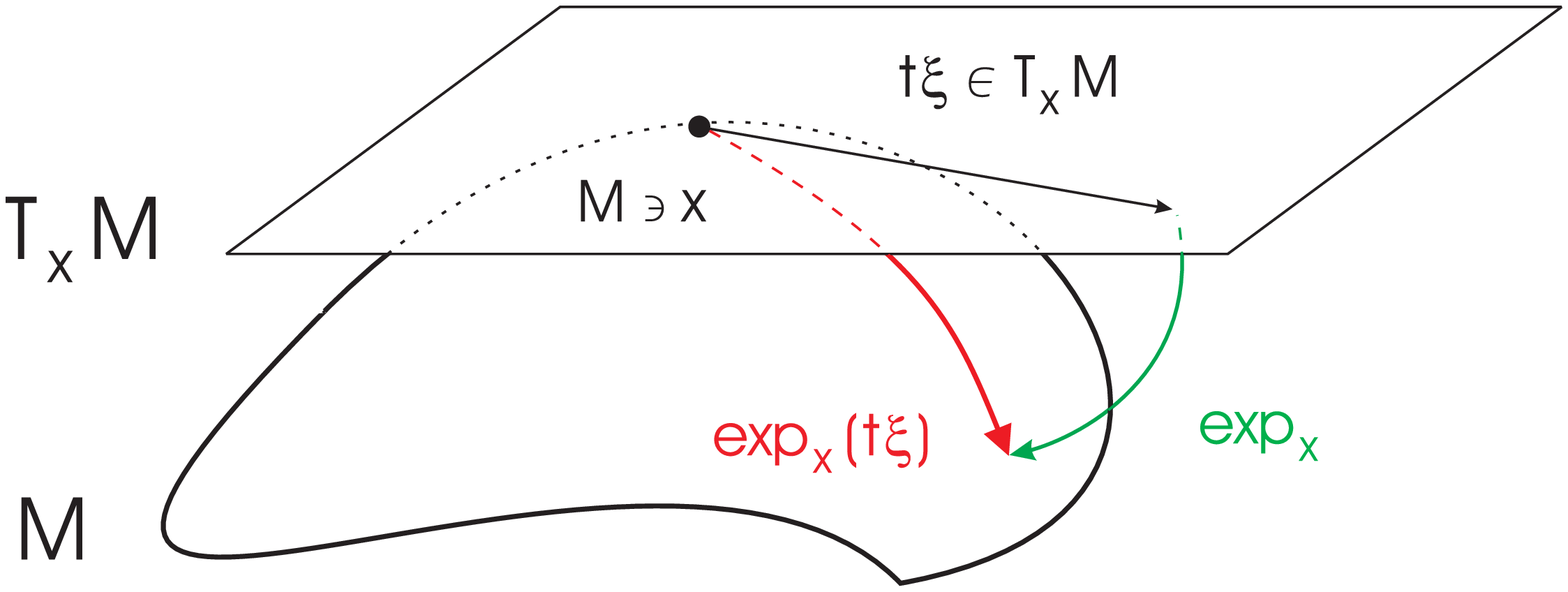}
\caption{\label{fig:mfds} (Colour online)
The Riemannian exponential $\exp_X$ is a smooth map taking the
tangent vector $t\xi \in T_XM$ at $X\in M$ to $\exp_X(t\xi) \in M$.
By varying $t \in \R$, it yields the unique geodesic with initial value
$X \in M$ and \/`initial velocity\/' $\xi \in T_XM$.}
\end{figure}

Gradient flows may be envisaged as natural continuous versions of
the steepest ascent method for optimising a real-valued function
$f : \R^m \to \R$ by moving along its gradient $\grad f \in \R^m$, i.e.
\begin{enumerate}
\item[]{\em  Steepest Ascent Method}
\begin{equation}
\label{eqn:eulerI}
X_{k+1} = X_k + \alpha_k \grad f(X_k),
\end{equation}
where $\alpha_k \geq 0$ is an appropriate step size. 
\end{enumerate}
Here, the right hand side of Eqn. (\ref{eqn:eulerI}) does make sense,
as the manifold $M=\R^m$ coincides with its tangent space $T_XM = \R^m$
containing $\grad f(X)$. Clearly, a generalisation is required as
soon as $M$ and $T_XM$ are no longer identifiable.  This gap is filled
by the {\em Riemannian exponential map} defined by
\begin{equation}\label{eqn:RiemExp}
\exp_X:T_XM \to M\,, \quad \xi \mapsto \exp_X(\xi)
\end{equation}
so that $t \mapsto \exp_X(t \xi)$ describes the unique geodesic with
initial value $X \in M$ and \/`initial velocity\/' $\xi \in T_XM$ as
illustrated in Fig.~\ref{fig:mfds}. 

If the manifold $M$ carries the structure of a matrix Lie group
$\mathbf G$, we may identify the tangent space element $\xi \in T_X
\mathbf{G}$ with $\Omega X$, where $\Omega$ is itself
an element of the Lie algebra $\mathfrak g$, i.e. the tangent space at
the unity element $\mathfrak g = T_\unity\mathbf G$.
Moreover, if the Lie-group structure matches with the Riemannian framework
in the sense that the metric is bi-invariant (as will be explained in more
detail later), then the Riemannian exponential of $\xi=\Omega X$
can readily be calculated explicitly. This is done in three steps by
(i) right translation with the inverse group element $X^{-1}$,
(ii) taking the conventional exponential map of the Lie algebra element
$\Omega$,
(iii) right translation with the group element $X$ as summarised in the
following diagram
\begin{equation}\label{eqn:RiemExp2}
\begin{CD}
{\xi = \Omega X \in T_X \mathbf{G}}
@>{\quad\exp_X\quad}>>
{e^{\Omega} X \in \mathbf{G} }\\
@V{R_{X^{-1}}}VV              @AA{R_X}A   \\
{\Omega\in\mathfrak g} @>\qquad \exp\qquad>> {e^{\Omega}\in\mathbf G \;.}
\end{CD}
\end{equation}

\begin{figure}[Ht!]
\includegraphics[scale=0.35]{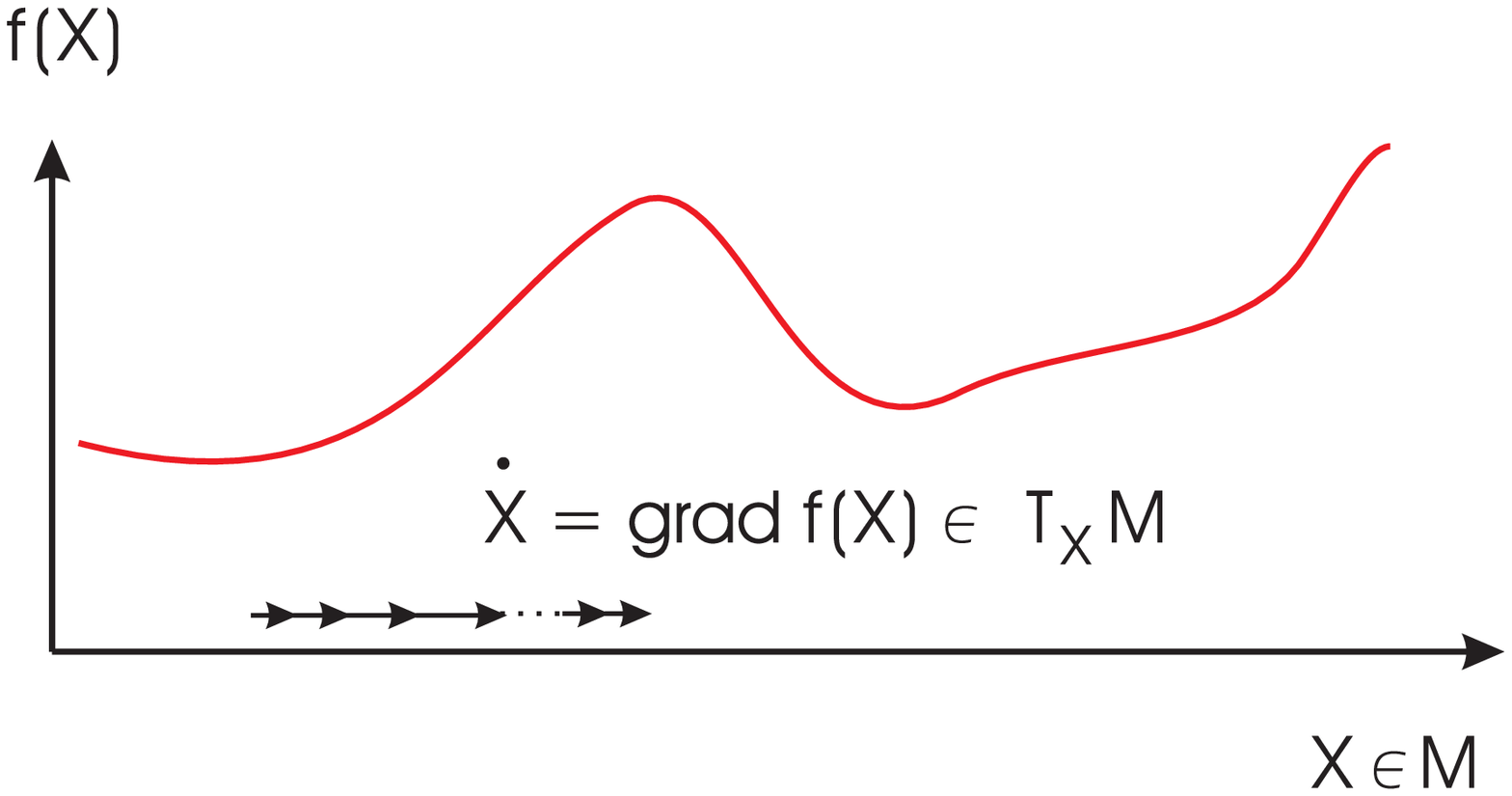}\\[5mm]
\Large{\hspace{44mm}{\Huge $\uparrow$}$\;f\hfill$}\\[9mm]
\includegraphics[scale=0.4]{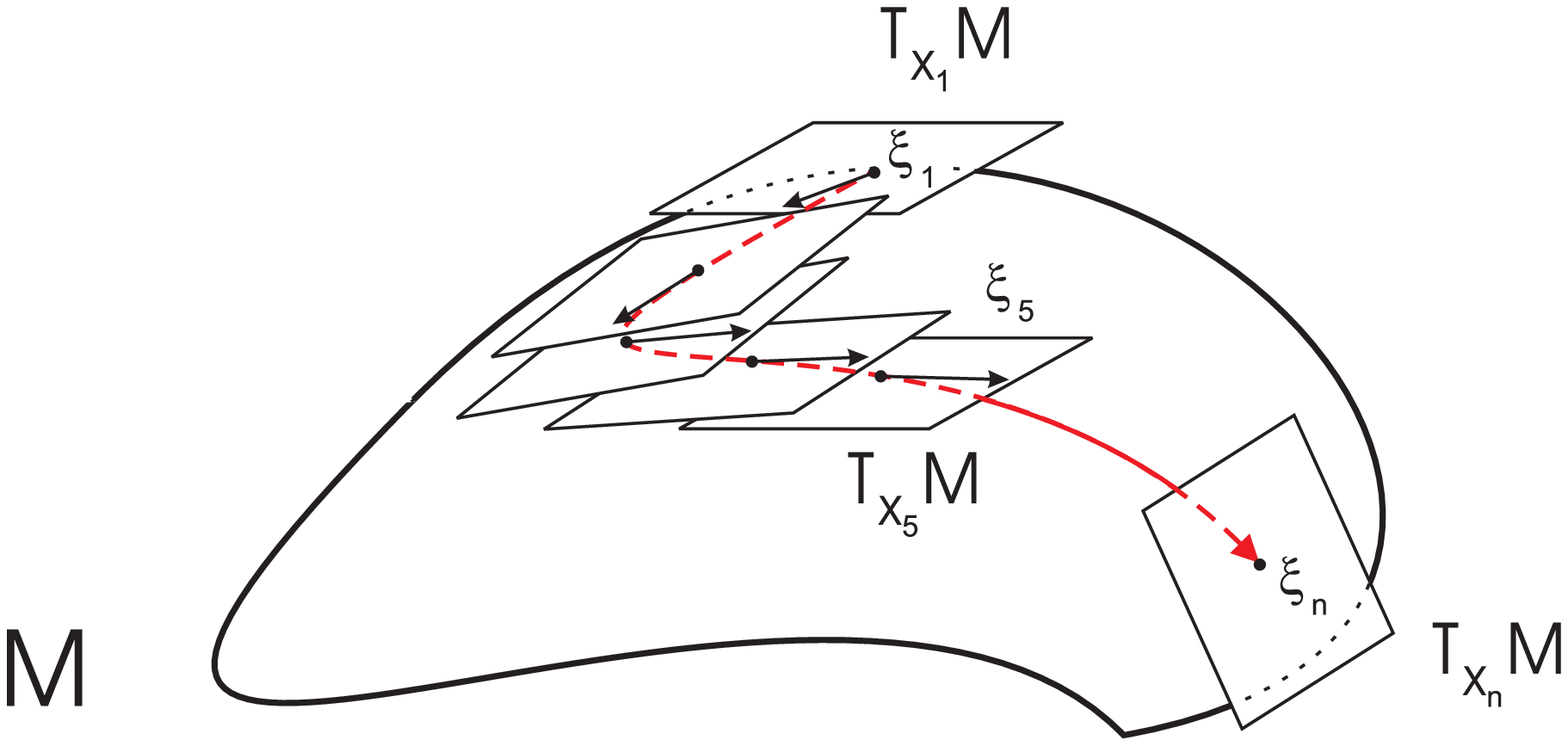}\\[2mm]
\caption{\label{fig:mfds_flow} (Colour online)
Abstract optimisation task:
the quality function $f: M \to \R, X \mapsto f(X)$ (top trace)
is driven into a (local) maximum by following the gradient
flow $\dot X = \grad f(X)$ on the manifold $M$ (lower trace).}
\end{figure}

Next, the gradient system (\ref{eqn:grad_system})
will be integrated (to sufficient approximation)
by a discrete scheme that can be seen as an \emph{intrinsic Euler
step method}. 
This can be performed by way of the Riemannian exponential map,
which is to say straight line segments used in
the standard method are replaced by geodesics on $M$.
This leads to the following integration scheme which is well-defined
on any Riemannian manifold $M$.
\begin{enumerate}
\item[1.] {\em Riemannian Gradient Method} 
\begin{equation}
\label{eq:gradI}
\begin{split}
 X_{k+1} :&= {\exp}_{X_k} \big(\alpha_k \grad f(X_k)\big)
\end{split}
\end{equation}
where $\alpha_k \geq 0$ denotes a step size appropriately selected
to guarantee convergence, cf.~Section \ref{Sec:theory}. 
\end{enumerate}
For matrix Lie groups $\mathbf G$ with bi-invariant metric,
Eqn.~(\ref{eq:gradI}) simplifies to 
\begin{enumerate}
\item[1'.] {\em Gradient Method on a Lie Group}
\begin{equation}
\label{eq:gradII}
\begin{split}
 X_{k+1} 
         :&= {\exp} \big(\alpha_k \grad f(X_k)\;X_k^{-1}\big) X_k\;, 
\end{split}
\end{equation}
where $\exp: \mathfrak g \to \mathbf G$ denotes the conventional
exponential map. 
\end{enumerate}
In either case, the iterative procedure can be pictured as follows:
at each point $X_k \in M$ one evaluates
$\grad f(X_k)$ in the tangent space $T_{X_k}M$. Then one moves via the
Riemannian exponential map in direction $\grad f(X_k)$ to the next
point $X_{k+1}$ on the manifold so that the quality function $f$ improves,
$f(X_{k+1}) \geq f(X_k)$, as shown in Fig.~\ref{fig:mfds_flow}.

The steepest ascent approach just outlined is most basic
for addressing abstract optimisation tasks intrinsically. 
Other intrinsic iterative schemes exploiting the underlying
Riemannian geometry like conjugate gradients, Jacobi-type methods
or Newton's method can be obtained similarly. For an introduction to
these more advanced topics beyond the subsequent sketch see, e.g.,  
Refs.~\cite{Gabay, Diss-Smith, Diss-Kleinsteuber}.

\begin{enumerate}
\item[2.] {\em Conjugate Gradient Method}
\begin{equation}
\begin{split}
X_k^{l+1}
& := \argmaxover{t \geq 0} f \big(\exp_{X_k^l} (t\,\Omega_k^l)\big)\\
X_{k+1}^{0}
& := X_{k}^{n},
\end{split}
\end{equation}
\begin{equation*}
 \Omega_k^l :=
\begin{cases}
 \grad f(X_k^l)  \hfill \mbox{for $l=0$}\\[2mm]
 \grad f(X_k^l)+ \alpha_k^l{\Pi}_{X_{k^{l-1}},X_k^l}(\Omega_{k-1}^l)\\[2mm]
 \quad\hfill \mbox{for $l=1, \dots, n-1$},
\end{cases}
 \end{equation*}
where $\alpha_k^l$ is a real parameter and ${\Pi}_{X,Y}(\Omega)$
denotes the parallel transport of $\Omega$ along the geodesic
from $X$ to $Y$.
\item[3.] {\em Jacobi-Type Method}
\begin{equation}
\begin{split}
X_k^{l+1}
& := \argmaxover{t\in\R{}} f
\Big({\exp}_{X_k^l} \big(t\,\Omega_l(X_k^l)\big)\Big)\\
X_{k+1}^{0}
& := X_{k}^{s},
\end{split}
\end{equation}
where $\Omega_0, \Omega_1, \dots, \Omega_{s-1}$ are vector fields
such that
$\Omega_0(X), \Omega_1(X), \dots, \Omega_{s-1}(X)$ span $T_{X}M$
for all $X \in M$. The integer $s$ is called \emph{sweep length}.
\item[4.] {\em Newton's Method}
\begin{equation}
 X_{k+1} := {\exp}_{X_k} \Big(\,-\big(\Hess f(X_k)\big)^{-1}\grad f(X_k)\Big),
\end{equation}
where $\Hess f(X)$ denotes the Hessian of $f$ at $X$.
\end{enumerate}


\subsection*{Gradient-Based Methods for Optimal Control}

Up to now we have addressed optimisation tasks over state spaces forming
abstract manifolds $M$---hence the term \emph{abstract optimisation task}
({\sc aot}). In this subsection, we briefly describe how gradient methods
like the one in Ref.~\cite{grape} arise in the context of 
{\em optimal control tasks} ({\sc oct}).
These algorithms are of practical relevance, as in many 
applications, the entire state space $M$ of a physical system
$(\Sigma)$ evolving under some internal dynamics and 
external controls is not accessible.

Now the general optimal control task amounts to finding the
time course of controls 
to achieve a maximum of a given quality functional $l$,
which may depend on the time evolution of the controls and the
state---so-called \/`running costs\/'---as well as on
the terminal state.  Here, we consider only quality functionals
$l$ which are determined by the terminal state of $(\Sigma)$, i.e.
$l$ is given by some smooth quality function $f$ on the state space
$M$. Note, however, that this is in general not a restriction as
running costs can often be reduced to an end point condition,
see below.

Thus, the optimal control task reduces to finding controls that
drive the initial state $X_0$ to a maximum of 
\begin{equation}
f|_{\reach (X_0)},
\end{equation}
i.e. to a maximum of $f$ restricted to what is known as the
\emph{reachable set} $\reach (X_0)$ of $X_0$.
It is the set of all possible states the system can be driven
into within positive time.
Moreover, the set of all states which can be reached just in time $T$
will be denoted by $\reach(X_0,T)$. Thus 
\begin{equation}
\label{reachI}
\reach (X_0) := \bigcup_{T \geq 0} \reach(X_0,T).
\end{equation}
A more detailed discussion on this topic will be given in the next
subsection. 

Choosing an (equidistant) sufficiently small partitioning
$0=t_0, \dots, t_{n_p}=T$  of the interval $[0,T]$
and assumimg that the set of all states which can be reached
in time $T \geq 0$ by {\em piecewise constant} controls is dense
in $\reach(X_0,T)$, the optimal control task translates into
a maximisation over the set
\begin{equation}
M_{cp}:=\step(t_0, \dots, t_{n_p};\R^{n_c})
\end{equation}
of all piecewise constant controls on $[0,T]$ with \emph{fixed}
switching points $t_1, \dots, t_{n_p-1}$. It is important to note
that $M_{cp}$ is a vector space isomorphic to $\R^{n_c \cdot n_p}$, 
where $n_c$ and $n_p$ are the respective numbers of controls and
subintervals. 

\begin{figure}[Ht!]
\includegraphics[scale=0.35]{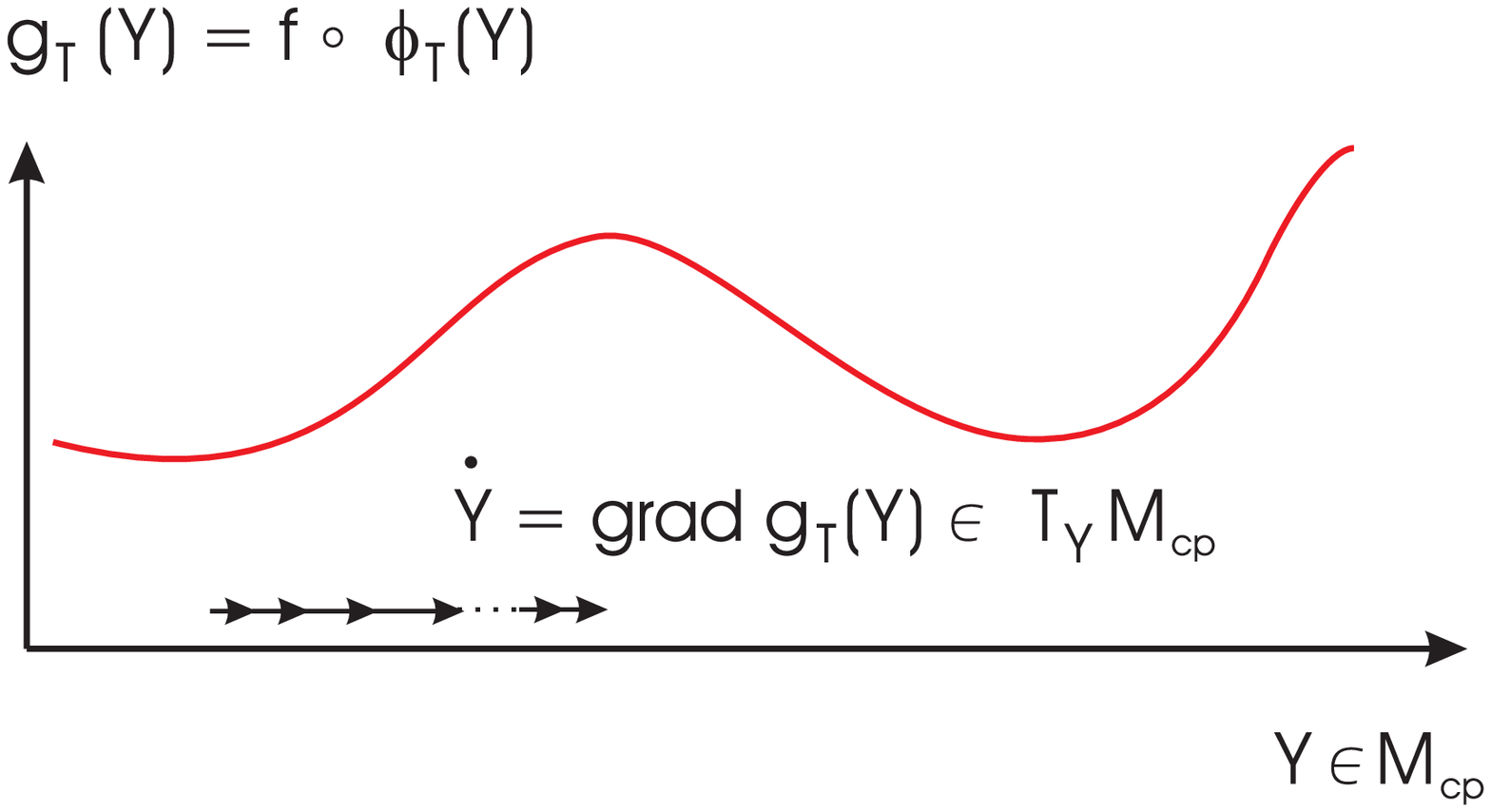}\\[5mm]
\Large{\hspace{44mm}{\Huge $\uparrow$}$\;f\hfill$}\\[9mm]
\includegraphics[scale=0.4]{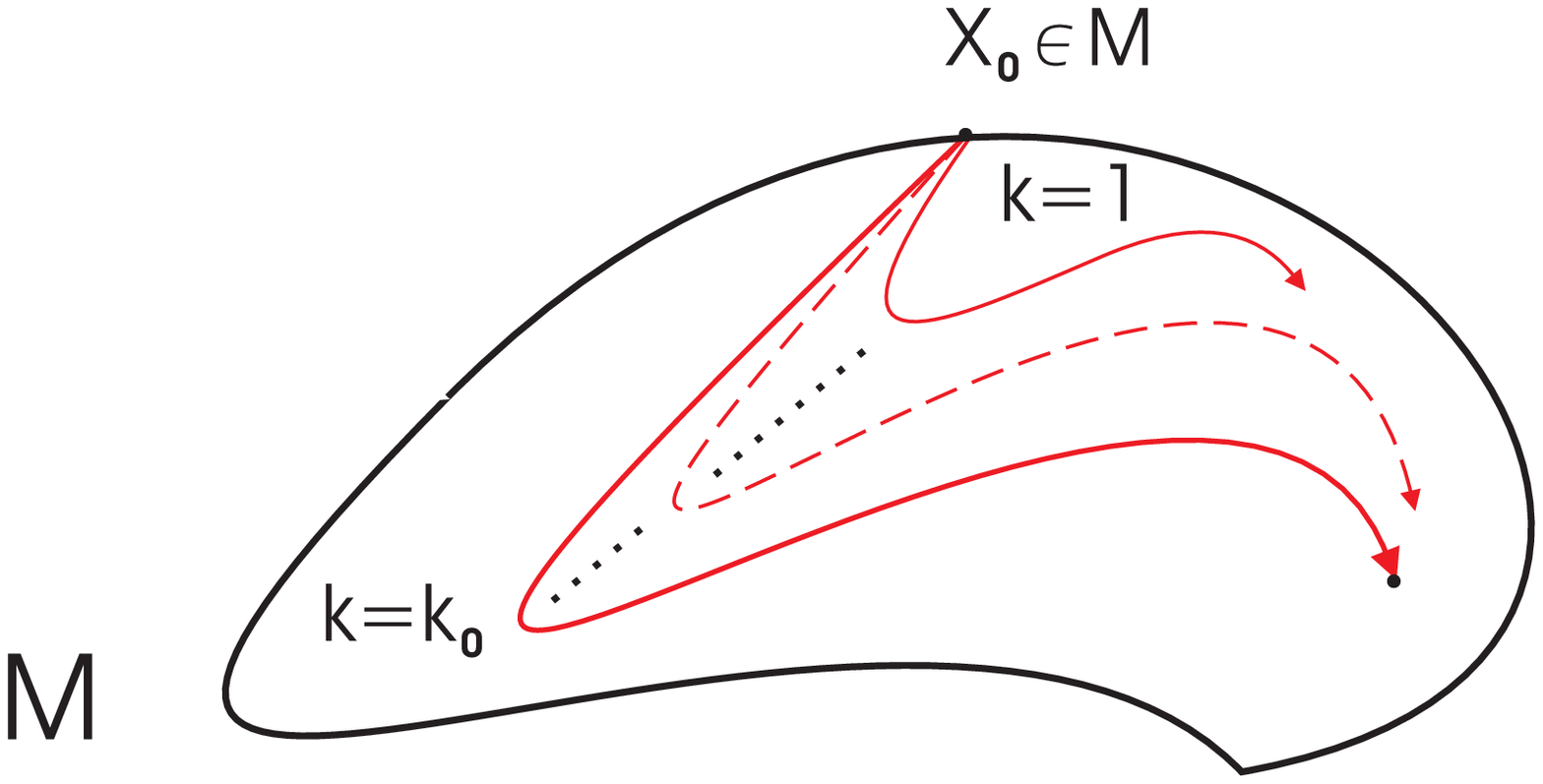}\\[6mm]
\Large{\hspace{44mm}{\Huge $\uparrow$}$\;\phi_T\hfill$}\\[9mm]
\Large{\hspace{11mm}${\sf M_{cp}} \cong \R^{n_c\cdot n_p}{}\hfill$}\\[4mm]
\includegraphics[scale=0.24]{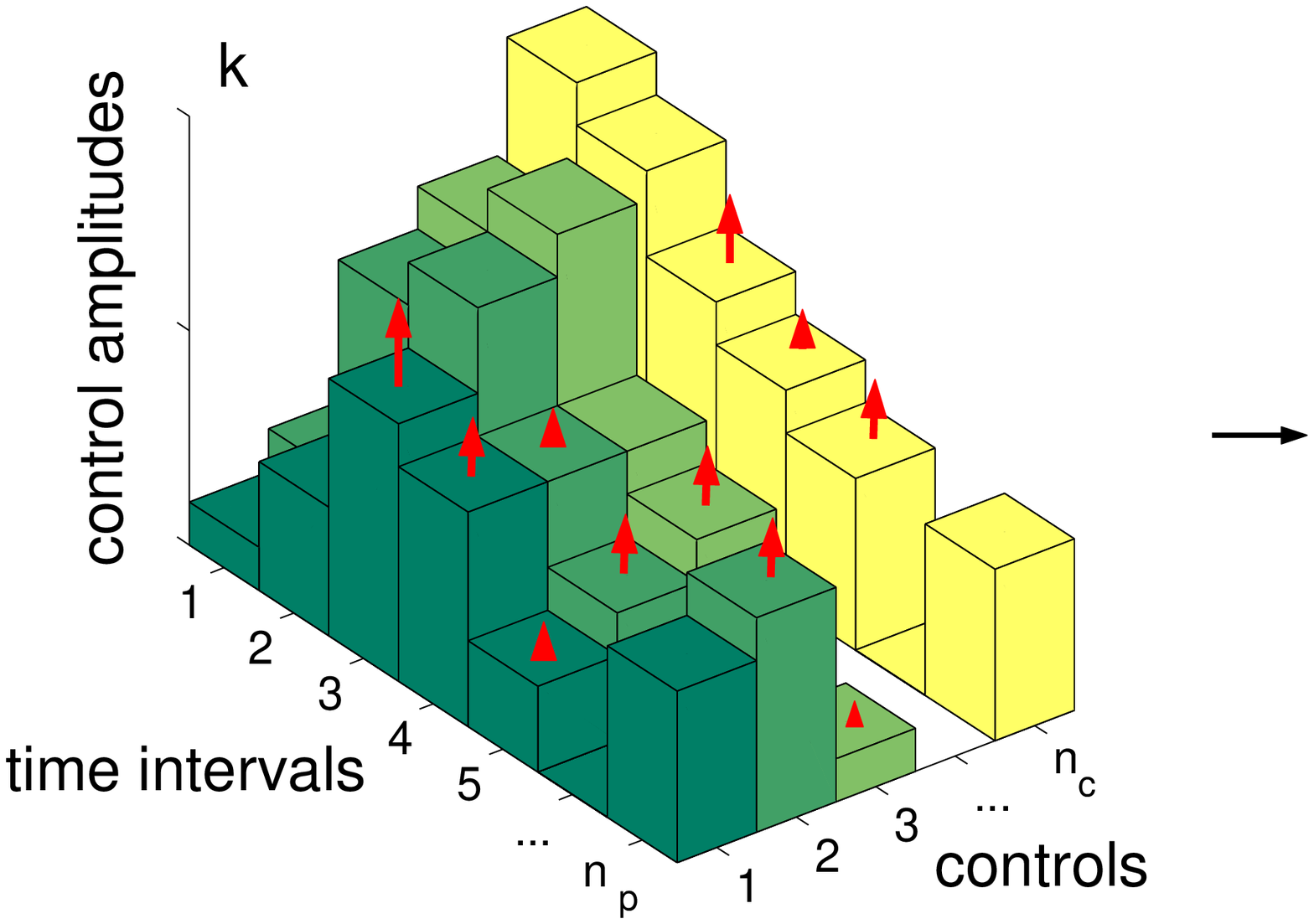} 
\includegraphics[scale=0.24]{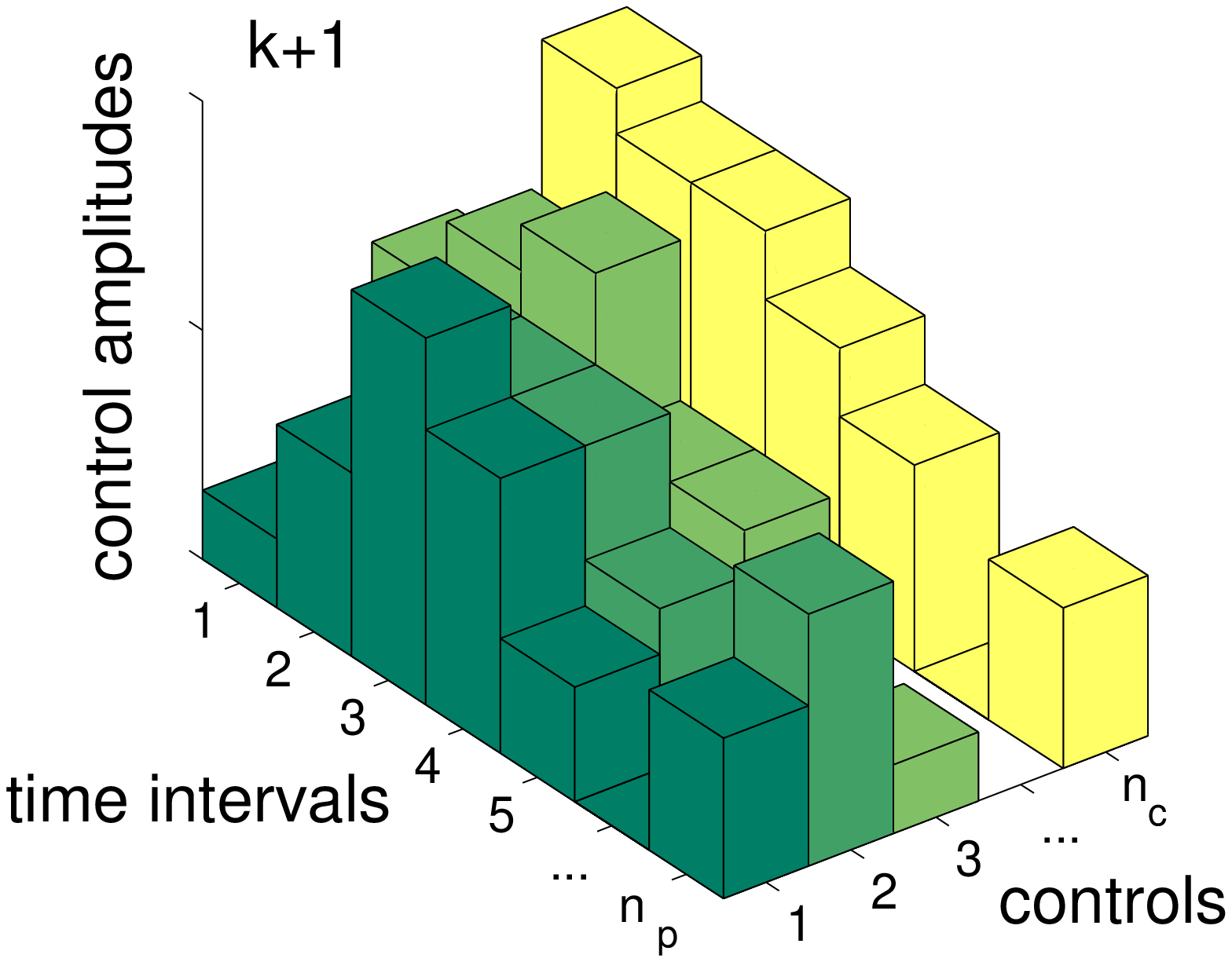}
\caption{\label{fig:mfds_control} (Colour online)
Optimal control task: the quality function $f: M \to \R, X \mapsto f(X)$
is driven into a (local) maximum on the reachable set
$\reach(X_0)\subseteq M$ by following an implicit
procedure  (intermediate panel). It is brought about by a
gradient flow $\dot Y = \grad g_T(Y)$
on the level of experimental control amplitudes $Y \in M_{cp}$
(lower traces) so that a standard steepest ascent method applies.}
\end{figure}
In order to express the effect of a piecewise constant control on the 
final state of $(\Sigma)$, we define a mapping 
\begin{equation}\label{eqn:phi_mapI}
\phi_T: M_{cp} \to M, \quad Y \mapsto X(T,Y),
\end{equation}
where $X(t, Y)\equiv X(t, Y_1, \dots, Y_{n_p})$ denotes the
unique solution of the control system $(\Sigma)$ with initial value
$X(0, Y) = X_0$ and piecewise constant control $Y_j \in \R^{n_c}$
on $[t_{j-1},t_j]$. 
Hence, the impact of the controls on the quality function
is described by the composition 
\begin{equation}
\label{eqn:indqua}
g^{\phantom|}_T := f \circ \phi_T: M_{cp} \to \R\quad.
\end{equation}
Often, the map $\phi_T: M_{cp} \to M$ is \/`almost\/' {\em surjective}
in the sense that the interior of its image $\phi_T(M_{cp})$
relative to $\reach(X_0,T)$ is open and dense. Therefore,
$\phi_T$ can be interpreted as a highly non-regular
over--parametrisation of the set $\reach(X_0,T)$. Now, any
standard non-linear optimisation tool, e.g. an Euclidian gradient
method allows to compute the (local) maximum of $g_T$ and thus
to numerically evaluate the overall quality function $g$ defined by 
\begin{equation*}
g(T) := \max_{T' \leq T} \Big(\max_{Y}\, g^{\phantom|}_{T'}(Y)\Big),
\end{equation*}
where $Y$ varies over all piecewise constant controls in
$\step(t_0, \dots, t_{n_p'},\R^{n_c})$ with $t_{0}=0$ and
$t_{n_p'}=T'$. Hence, for solving the {\sc oct}, at last one
has to increment $T \geq 0$ until $g(T)$ remains constant. 
Moreover, \emph{time-optimal} solutions of the {\sc oct} are obtained
while searching for the minimal time $T^* > 0$ such that $g(T^*)$
is equal to the maximum of $g$. 

In order to be more explicit, assume that the system $(\Sigma)$
is given by a time-continuous control system, i.e. by an (ordinary)
differential equation
\begin{equation}
\label{sigmaI}
(\Sigma)
\quad\quad
\dot{X} = F(X,u),
\quad
u \in \R^m
\end{equation}
depending on some (unrestricted) control parameters $u \in \R^m$.
Thus the general {\sc oct} amounts to maximising a functional
\begin{equation}
\label{costfunctional}
l(X,u,T) :=
f(X(T)) + \int_{0}^{T}h\big(X(t),u(t)\big){\rm d} t
\end{equation}
under the constraint Eqn. \eqref{sigmaI}. 
Here, the integral term in Eqn. \eqref{costfunctional}
represents the so-called  \/`running costs\/'. We therefore suppose
that $h$ vanishes. This, however, is not really a restriction
as mentioned above. By augmenting an auxiliary state
variable and state equation
to $(\Sigma)$ one can reduce  \/`running costs\/' to a terminal
state condition, cf. \cite{dAll08,Jurdjevic97}.
Moreover, note that in the setting of time-continuous control systems
the final scanning for the optimal value of $g$ can be accomplished
by introducing another auxiliary optimisation parameter $u_0 > 0$
in $(\Sigma)$ such that
\begin{equation}
\label{scaledsigma}
(\Sigma_0)
\quad\quad
\dot{X} = u_0 F(X,u),
\quad
(u_0,u) \in \R^{m+1}
\end{equation}
while keeping the time interval $[0,T]$ fixed, e.g. $T=1$.
Further numerical methods from optimal control can be found in
e.g. \cite{BryHo,Krotov,Rabitz87,Theo,MaTur03,dAll08}.
Some of these more involved techniques
require \emph{Pontryagin's maximum principle}, which
can be viewed as Lagrange-multiplier method for constrained
variational problems. 
These techniques often come at the cost of additional 
boundary-value problems.

To sum up, the quality function $f|_{\reach (X_0)}$ is driven
into a (local) maximum by an implicit procedure in
the sense that it is not explicitly defined on the reachable set
$\reach(X_0)\subseteq M$. Rather it is the result of a gradient flow
\begin{equation}
\dot Y = \grad g_T(Y).
\end{equation}
on the level of control amplitudes $Y \in M_{cp}\cong \R^{n_p\cdot n_c}$
so that one finally gets the discretised version
\begin{equation}
\label{eqn:Euler_controls}
Y_{k+1} = Y_k + \alpha_k \grad g_T(Y).
\end{equation}
Thus the iterative scheme (\ref{eqn:Euler_controls}) reads as a
standard steepest ascent method.
These ideas are illustrated in Fig.~\ref{fig:mfds_control}; they are
exploited in numerical optimal-control methods like the {\sc grape}
algorithm \cite{grape}.

Finally, note that the {\sc oct} relates to an {\sc aot} if the
reachable set $\reach(X_0)$ itself has the geometric structure of a
submanifold $N$ of the state-space manifold $M$.
However, in general this is not the case and would have to be ensured
{\em a priori}. Therefore, if one seeks to reduce the {\sc oct} to an
{\sc aot}, geometric control and reachability aspects come into play
as discussed in the next subsection.
For instance, in the Hamiltonian unitary evolution
of a finite dimensional closed quantum system, the closure of the
reachable set takes always the form of a \emph{group orbit} of some
initial state \cite{DiHeGAMM08},
e.g. $\mathcal O(\rho_0):=\{U\rho_0 U^\dagger\;|\; U\;\text{unitary}\}$.
Its Riemannian geometry is well understood. In open dissipative
systems, however, the dynamics is governed, e.g., by a Markovian
quantum Master equation and thus by a \emph{semigroup} of completely
positive operators.
As will be illustrated in Ref.~\cite{DHKS08}, it is much harder to give an explicit
characterisation of their reachable sets. 
Apart from utterly simple scenarios, an abstract approach is 
often unviable in dissipative systems, and thus implicit methods by
optimal control techniques may become indispensible.


\subsection*{Reachability and Controllability}
\label{sec:reach}

In the following, some general remarks on reachable sets and
controllability will clarify the previous, slightly sloppy
introduction of these term.
For simplicity, let $(\Sigma)$ denote a smooth
\emph{control system} on the manifold $M$, i.e. a familiy of
(ordinary) differential equations
\begin{equation}
\label{sigma}
(\Sigma)
\quad\quad
\dot{X} = F(X,u),
\quad
u \in \mathcal U \subset \R^m
\end{equation}
with control parameters $u \in \mathcal U$ and smooth vector fields
$F_u := F(\cdot,u)$ on $M$.
While the vector fields $F_u$ are assumed to be time-independent,
the controls are allowed to vary in time. For convenience,
the resulting control function $t \mapsto u(t) \in \mathcal U$ is denoted
again by $u$. Moreover, the set $\mathcal{U}$ of all
\emph{admissible controls functions} is supposed to contain at
least all piecewise constant ones.

For $u \in \mathcal{U}$, we refer to $X(t,X_0,u)$ as the unique
solution of \eqref{sigma} with initial value $X_0$. Therewith, the
\emph{reachable set} of $X_0$ is defined by
\begin{equation}
\label{reachII}
\reach (X_0) := \bigcup_{0 \leq T} \reach(X_0,T).
\end{equation}
Here $\reach(X_0,T)$ denotes the set of all states which can be
reached in time $T$, i.e.
\begin{equation}
\reach(X_0,T)
:= \big\{X(T,X_0,u) \in M \;|\; u \in \mathcal{U}\big\},
\end{equation}
cf. \eqref{reachI}. 
The system $(\Sigma)$ is said to be \emph{controllable}, if
$\reach (X_0) = M$ for all $X_0 \in M$, i.e. if for each pair
$(X_0, Y_0)$ there exists an admissible control $u$ and a
time $T \geq 0$ such that  $X(T,X_0,u) = Y_0$.

In general, it is hard to decide whether a given system $(\Sigma)$ is
controllable or not. However, for dynamics expressed on some Lie group
$\mathbf G$, the situation is much easier. 
Let $(\Sigma_{\mathbf G})$ be a \emph{bilinear} or, equivalently,
a \emph{right invariant, control-affine} system on a matrix
Lie group $\mathbf G$ with Lie algebra $\mathfrak g$, i.e.
\begin{equation}
(\Sigma_{\mathbf G})
\quad\quad
\dot{X} = (A_0 + \sum_{j=1}^{m}u_jA_j)X,
\quad
u \in \R^m
\end{equation}
with drift $A_0 \in \g$ and control directions $A_j \in \g$. For
\emph{compact} Lie groups $\mathbf G$, a simple algebraic test for
controllability is known: If the \emph{system Lie algebra} 
\begin{equation}
\label{systemLiealgebra}
\mathfrak{s} := \langle A_0, \dots, A_m \rangle_{\rm Lie}
\end{equation}
generated by $A_0, \dots, A_m$ via nested commutators coincides with
$\g$, then the corresponding system $(\Sigma_{\mathbf G})$ is controllable,
cf. \cite{SJ72,JS72}. In particular, there exists a finite time $T' > 0$,
such that the entire group $\mathbf G$ can be reached from any initial
point $X_0 \in {\mathbf G}$ within this time,
i.e.
\begin{equation}
\label{reachability0}
\mathbf G = \bigcup_{0 \leq T \leq T'} \reach(X_0,T)
=: \reach(X_0,\leq T')
\end{equation}
for all $X_0 \in {\mathbf G}$.

In contrast, for \emph{non-compact} groups $\mathbf G$, which are 
indispensible for the description of dissipation in open quantum
systems, the situation gets more involved.
Here, $\mathfrak{s} = \g$ implies only \emph{accessibility} of
$(\Sigma_{\mathbf G})$, i.e. that all reachable sets $\reach(X_0)$
have non-empty interior. This follows from a more general result on
smooth non-linear control systems---the so-called \emph{Lie algebra
rank condition} ({\sc larc})
\begin{equation}
\label{larc}
\big\{F(X) \;|\; F \in \langle F_u \;|\; u \in \mathcal U\rangle_{\rm Lie}\big\}
= T_XM,
\end{equation}
where $\langle F_u \;|\; u \in \mathcal U\rangle_{\rm Lie}$ denotes the Lie
subalgebra of vectors fields generated by $F_u$, $u \in \mathcal U$ via Lie
bracket operation, cf.~\cite{Jurdjevic97,CK00}. Note that for right
invariant vector fields on $\bG$, the Lie bracket essentially
coincides with the commutator such that \eqref{larc} boils down
to $\mathfrak{s} = \g$. Moreover, by exploiting the identity
\begin{equation}
\label{reachabilityII}
\reach(\unity, T_1) \cdot \reach(\unity, T_2)
= \reach(\unity, T_1+T_2),
\end{equation}
one can show that $\reach(\unity)$ is always a {\em Lie subsemigroup}
of $\mathbf G$ \cite{DHKS08}. Here, a {\em subsemigroup} is a subset
$\mathbf S \subset \mathbf G$ which contains the unity and
is closed under multiplication, 
i.e. $\unity\in \mathbf S$ and $\mathbf S\cdot \mathbf S \subseteq \mathbf S$.
However, the geometry of subsemigroups is rather subtle compared to
Lie subgroups and therefore at present not amenable to intrinsic
optimisation methods, as will be shown in more detail in a separate
paper dwelling on open systems \cite{DHKS08}.


\subsection*{Settings of Interest}

In terms of reachability, there are different scenarios that
structure the subsequent line of thought:
we start out with {\em fully controllable} or 
{\em operator controllable} quantum systems
\cite{JS72,SJ72,Bro72,Bro73,BW79,TOSH-Diss,AA03}
represented as spin- or pseudo-spin systems. Then, neglecting 
decoherence, to any initial state represented by its density
operator $A$, the entire unitary orbit
$\mathcal O(A):= \{UAU^{-1}\, |\, U \; {\rm unitary}\}$
can be reached \cite{AA03}.
In systems of $n$ qubits ({\em e.g.} spin-$\frac{1}{2}$ particles), this
is the case under
the following mild conditions \cite{TOSH-Diss,Science98,GA02}: 
(1) the qubits have to be inequivalent, i.e., distinguishable and
selectively addressable, and
(2) they have to be pairwise coupled (e.g., by Ising or
Heisenberg-type interactions),
where the coupling topology may take the form of any connected graph.
In other instances not the entire (unitary) group, but just a subgroup
$\mathbf K$ can be reached.
This is the case, if the coupling topology is not
a connected graph or 
the $n$ qubits cannot be addressed by separate controls. 

Otherwise, the system itself can be fully controllable, but the
the focus of interest may be reduced: 
e.g., the subgroup 
$\mathbf K =
SU_{\rm loc}(2^n) :=SU(2)\otimes SU(2)\otimes\cdots\otimes SU(2) $
of (possibly fast) local actions on each qubit is of interest to
study local reachability, or whether an effective multi-qubit
interaction Hamiltonian is locally reversible in the sense of Hahn's
spin echo \cite{PRA_inv}. 
Or, one may ask what is the Euclidean distance of some pure
state to the nearest point on the local unitary orbit of a pure product
state. This may be useful when optimising entanglement
witnesses \cite{Guehne04}.
Likewise, one may address other than the finest partitioning of the
entire unitary group, e.g.
$\bK = SU(2^{n_1})\otimes \cdots\otimes SU(2^{n_j})
\otimes\cdots\otimes SU(2^{n_r})\subset SU(2^n)$,
where $\sum_{j=1}^{r} n_j = n$.

Another type of reduction arises not by restriction to a subgroup
$\bH$, but by the fact that the quality function of interest
$f$ is {\em equivariant}, i.e.~constant on cosets $\bH G$. 
Consider, for instance, a fully controllable system
where $f$ is {equivariant} with respect to the closed subgroup
$\mathbf H \subset \mathbf G$.
Then, it may be favourable to transfer the optimisation problem
from the original Lie group $\mathbf G$ to the homogeneous space
$\mathbf G/\mathbf H$.




\section{Theory: Gradient Flows}
\label{Sec:theory}

\noindent
Gradient systems are a standard tool of Riemannian optimisation
for maximizing smooth quality functions on a manifold $M$.
Thus the manifold structure of $M$ arises either naturally
by the problem itself or by smooth
equality-constraints imposed on a previously unconstrained
problem.
Note that in general inequality-constraints would entail
manifolds with a boundary---and thus are a much more subtle issue 
not to be developed any further here.

The case $M = \R^m$---sometimes refered to as the unconstrained 
case---is well-known and can be found in many texts on ordinary
differential equations or nonlinear programming,
cf.~\cite{HirschSmale74, Irwin, Fletcher87, Bazaraa93}.
However, gradient systems on abstract Riemannian manifolds provide
a rather new approach to constrained optimisation problems.
Although the resulting numerical algorithms are in general only
linearly convergent, their global behaviour is often much better
then the global behaviour of locally quadratic methods.

Textbooks combining the different areas of Riemannian geometry, gradient
systems and constrained optimisation are quite rare. The best choices
to our knowledge are \cite{HM94, Udriste94}. For further reading
we also suggest the papers \cite{Diss-Smith, Gabay, Bro93}.
Nevertheless, most of the material which is necessary to understand
the intrinsic optimisation approach applied in Section
\ref{Sec:applications} is scattered in many different
references. For the reader's convenience, we therefore
review the basic ideas on these topics. First, we discuss the general
setting on Riemannian manifolds, then we proceed with Lie groups and
finally summarize some more advanced results on homogeneous spaces.
For standard definitions and terminology from Riemannian geometry
we refer to any modern text on this subject such as
\cite{Boothby75,Gallot04,Spivak99}.


\subsection{Gradient Flows on Riemannian Manifolds}
\label{sec:mfds}



In the following, let $M$ denote a finite dimensional
smooth manifold $M$ with tangent and cotangent bundles $\rT M$ and
$\rT^*M$, respectively. Moreover, let $M$ be equipped with a
\emph{Riemannian metric} $\braket{\cdot}{\cdot}$, i.e. with a
scalar product $\braket{\cdot}{\cdot}_X$ on each tangent space
$\rT_XM$ varying smoothly with $X \in M$. More precisely,
$\braket{\cdot}{\cdot}$ has to be a smooth, positive definite
section in the bundle of all symmetric bilinear forms over $M$.
Such sections always exist for finite dimensional smooth manifolds,
cf. \cite{AMR88}. 
The pair $\big(M,\braket{\cdot}{\cdot}\big)$ is called a
\emph{Riemannian manifold}. 

Let $ f : M \to \R$ be a smooth quality function on $M$ with differential 
$D f : M \to T^*M$. Then the \emph{gradient} of $f$ at $X \in M$, denoted
by $\grad f\,(X)$, is the vector in $\rT_XM$ uniquely determined
by the equation
\begin{equation}
\label{grad}
\D f\,(X)\cdot\xi = \braket{\grad f\,(X)}{\xi}_X
\end{equation}
for all $\xi \in T_XM$.
Equation (\ref{grad}) naturally defines a vector field on $M$ via 
\begin{equation}
\label{gradfield}
\grad f \;:\; M \to \rT M,\quad X \mapsto \grad f\,(X)
\end{equation}
called the \emph{gradient vector field} of $f$. 
The corresponding ordinary differential equation
\begin{equation}
\label{gradflow}
\dot{X} = \grad\, f\,(X),
\end{equation}
and its flow are referred to as the \emph{gradient system} and the
\emph{gradient flow} of $f$, respectively.

Obviously, the \emph{critical points} of $ f\,:\,M\to\R$ coincide with the
equilibria of the gradient flow. Moreover, the quality function $f$ is
monotonically increasing along solutions $X(t)$ of (\ref{gradflow}), i.e.
the real-valued function $t \mapsto f(X(t))$ is monotonically increasing
in $t$, as
\begin{eqnarray*}
\frac{{\rm d}}{{\rm d}t}f\big(X(t)\big)
& = &
\braket{\grad f(X(t))}{\dot X(t)}_{X(t)}\\
& = & ||\grad f(X(t))||^2_{X(t)} \geq 0.
\end{eqnarray*}
Here $||\cdot||_X$ denotes the norm on $\rT_XM$ induced by
$\braket{\cdot}{\cdot}_X$, i.e. $||\xi||_X := \sqrt{\braket{\xi}{\xi}_X}$
for all $\xi \in T_XM$.


\subsubsection*{Convergence of Gradient Flows}

Recall that the asymptotic behaviour (for $t \to +\infty$) of a
solution of (\ref{gradflow}) is characterised by its
$\omega$-\emph{limit set} 
\begin{equation*}
\label{omegalimit}
\omega(X_0) := \bigcap_{0<t<t^+(X_0)}
\overline{\big\{X(\tau,X_0)\;\big|\; t \leq \tau < t^+(X_0)\big\}},
\end{equation*}
where $\overline{\{\cdots\}}$ denotes the closure of the set
$\{\cdots\}$ and $X(t,X_0)$ the unique solution of (\ref{gradflow})
with initial value $X(0) = X_0$ and positive escape time $t^+(X_0) > 0$.  




The following result gives a sufficient condition for solutions
of Eqn. \eqref{gradflow} to converge to the set of critical points
of $f$.

\begin{proposition}
\label{propgradomegalimit}
If $f$ has compact a superlevel set, i.e. if the sets
$\{X \in M \;|\; f(X) \geq C\}$ are compact for all $C \in \R$,
then any solution of Eqn.~(\ref{gradflow}) exists for $t \geq 0$
and its $\omega$-limit set is a non-empty compact and connected
subset of the set of critical points of $f$.
\end{proposition}

{\bf Proof.}\hspace{.5em}
Since solutions of Eqn.~(\ref{gradflow}) are monotonically
increasing in $t$, the compact sets $\{X \in M \;|\; f(X) \geq C\}$
are positively invariant, i.e. invariant for $t \geq 0$ under the
gradient flow of Eqn.~(\ref{gradflow}).
Thus, the assertion follows from standard results on $\omega$-limit
sets and Lyapunov theory, cf. \cite{HM94, Irwin}.
\hfill$\square$

\medskip


Although, Proposition \ref{propgradomegalimit}
guarantees that $\omega(X_0)$ is contained in the set of critical points
of $f$, this does not imply convergence to a critical point.
Indeed, there are smooth gradient systems which exhibit solutions converging
only to the set of critical points, cf. \cite{PalisdeMelo82}.
The next two results provide sufficient conditions for convergence to a
single critical point under different settings. In particular, Threorem
\ref{thmIIpointconvergence} yields a powerful tool for analysing
real analytic gradient systems.

\begin{corollary}
\label{corpointwiseconvergence}
If $f$ has compact superlevel sets and if all critical points
are isolated, then any solution of (\ref{gradflow}) converges
to a critical point of $f$ for $t \to +\infty$.
\end{corollary}

{\bf Proof.}\hspace{.5em}
This is an immediate consequence of Proposition
\ref{propgradomegalimit}.
\hfill$\square$

\begin{theorem}[{\L}ojasiewicz]
\label{thmIIpointconvergence}
If $\big(M, \braket{\cdot}{\cdot}\big)$ and $f$ are real analytic, then
all non-empty $\omega$-limit sets $\omega(X_0)$ of Eqn.~(\ref{gradflow})
are singletons, i.e. $\omega(X_0) \neq \emptyset$ implies that $X(t,X_0)$
converges to a single critical point $X^*$ of $f$ for $t \to +\infty$.
\end{theorem}

{\bf Proof.}\hspace{.5em}
The main argument is based on {\L}ojasiewicz's inequality which
says that in a neighbourhood of $X^*$ an estimate of the type 
\begin{equation*}
\label{Loja-inequality}
|f(X)|^p \leq C \|\grad f(X)\|
\end{equation*}
for some $p<1$ and $C >0$ holds. A complete proof can be found in
\cite{Loja84, Kurdyka}.
\hfill$\square$


\subsubsection*{Restriction to Submanifolds}

Now, consider the restriction of $f$ to a smooth submanifold
$N \subset M$. Obviously, the Riemannian metric $\braket{\cdot}{\cdot}$
on $M$ restricts to a Riemannian structure on $N$. Thus
$\big(N, \braket{\cdot}{\cdot}|_{\rT N}\big)$
constitutes a Riemannian manifold in a canonical way.
Moreover, the equality $\D f|_N\,(X) = \D f\,(X)|_{\rT_XN}$
immediately implies that the gradient of the restriction $f|_N$ at
$X \in N$ is given by the orthogonal projection of $\grad f(X)$ onto
$\rT_XN$, i.e.
\begin{equation}
\label{restrictedgrad}
\grad f|_N(X) = P_X\big(\grad f(X)\big),
\end{equation}
where $P_X$ denotes the orthogonal projector onto $\rT_XN$.
Hence the gradient system of $f|_N$ on an arbitrary submanifold $N$ is
well-defined and reads
\begin{equation}
\label{restrictedgradflow}
\dot{X} = 
P_X\big(\grad f(X)\big).
\end{equation}



\subsubsection*{Analysing Critical Points by the Hessian}

Subsequently, we address the problem, how to define and compute
the \emph{Hessian} of $f$, as its knowledge is essential for a deeper
insight of (\ref{gradflow}). For instance, the stability of critical
points is determined by its eigenvalues or the computation
of explicit discretisation schemes, preserving the convergence behaviour
of (\ref{gradflow}), can be based on it, cf. \cite{Bro88+91, HM94}.

At critical points $X^* \in M$ of $f$, the \emph{Hessian} is given by
the symmetric bilinear form
$\Hess f (X^*): \rT_{X^*} M \times \rT_{X^*}M \to \R$, 
\begin{eqnarray}
\label{hessianI}
\lefteqn{\Hess f(X^*)(\xi,\eta) \;:=\;}
\\\nonumber
&  & 
(\D \varphi(X^*)\xi)^{\top}
\Hess {(f\circ\varphi^{-1})}\big(\varphi(X^*)\big)
\D \varphi(X^*)\eta,
\end{eqnarray}
where $\varphi$ is any chart around $X^*$ and
$\Hess {(f\circ\varphi^{-1})}$ denotes the ordinary
Hesse matrix of $f\circ\varphi^{-1}$. It is straightforward to show
that (\ref{hessianI}) is independent of $\varphi$. Equivalently,
$\Hess f (X^*)$ is uniquely determined by
\begin{equation}
\label{hessianII}
\Hess f(X^*)(\xi,\xi) :=
\frac{\mathrm{d}^2 (f \circ \alpha)}{\mathrm{d}t^2}\Big|_{t=0}\;,
\end{equation}
where $\alpha$ is any smooth curve with $X^* = \alpha(0)$ and
$\dot{\alpha}(0)=\xi$. While the remaining values of $\Hess f (X^*)$
can be obtained by a standard polarisation argument, i.e. via the
formula
\begin{eqnarray}
\label{polar}
& 2 \Hess f(X^*)(\xi,\eta) = \Hess f(X^*)(\xi+\eta,\xi+\eta)&
\nonumber\\
&  - \Hess f(X^*)(\xi,\xi) - \Hess f(X^*)(\eta,\eta)&.
\end{eqnarray}

\noindent
However, the previous definition does not apply to regular points of
$f$. In general, one has to establish the concept of \emph{geodesics},
cf. Remark \ref{remgeodesics}
More precisely, the Hessian of $f$ at an arbitrary point $x \in M$
is defined by 
\begin{equation}
\label{hessianIII}
\Hess f(X)(\xi,\xi) :=
\frac{\mathrm{d}^2 (f \circ \gamma)}{\mathrm{d}t^2}\Big|_{t=0},
\end{equation}
where $\gamma$ is the unique geodesic with $X = \gamma(0)$ and
$\dot{\gamma}(0)=\xi$, cf. \footnote{ In general, 
Eqn. \eqref{hessianIII} is not used as a definition, but
obtained as a consequence.  Other definitions, e.g.~via
a linear connection, have the advantage that they guarantee
bi-linearity straight away, cf. Remark \ref{remgeodesics}.}.
Again, the remaining values can be computed
by (\ref{polar}). As usual, we associate to $\Hess f(X)$ a unique
selfadjoint linear operator $\Hessop f(X): \rT_{X} M \to \rT_{X} M$
such that 
\begin{equation}
\label{hessianIV}
\braket{\xi}{\Hessop f(X)\eta}_X = 
\Hess f(X)(\xi,\eta)
\end{equation}
holds for all $\xi,\eta \in \rT_{X} M$. It is called the \emph{Hessian
operator} of $f$ at $X \in M$.

\begin{remark}
\label{remgeodesics}
In modern textbooks, geodesics are defined via \emph{linear connections}
on $M$, cf. \cite{KobNom96, Spivak99}. 
For Riemannian manifolds $M$, however, it is possible to introduce
(Riemannian) geodesics as curves of minimal arc length. Both concepts
coincide (locally) if we choose the so-called \emph{Riemannian} or
\emph{Levi-Civita connection} on $M$.
Unfortunately, the computation of geodesics is in general a highly
non-trivial problem. However, on compact Lie groups their calculation 
is much easier as we will see at the end of Section \ref{sec:liegroups}.
\end{remark}

The above concepts yield the following generalisation of a familiar
result from elementary calculus.

\begin{theorem}
\label{hessianvsmaximum}
Let $M$ be a Riemannian manifold and let $X_*$ be a critical point
of the quality function $f : M \to \R$. If $\Hess f(X^*)$ or,
equivalently, $\Hessop f(X)$ are negative definite, then $X^*$ is
a strict local maximum of $f$.
\end{theorem}

{\bf Proof.}\hspace{.5em}
Use local coordinates, then the result follows straightforwardly
from Eqn. \eqref{hessianI}.
\hfill$\square$

\medskip

In general, (asymptotic) stability of an equilibrium $X^* \in M$
of (\ref{gradflow}) may dependent on the Riemannian metric
$\braket{\cdot}{\cdot}$. 
However, the property of being a strict local maximum or an
isolated critical point of a smooth function $f$ is obviously
not up to the choice of any Riemannian metric.
Therefore, the following result shows that in fact certain
(asymptotically) stable equilibria $X^* \in M$ of (\ref{gradflow})
are independent of the Riemannian metric.

\begin{theorem}
\label{thmIpointconvergence}
\begin{enumerate}
\item[(a)]
If $X^* \in M$ is a strict local maximum of $f$, then $X^*$ is a
stable equilibrium of (\ref{gradflow}).
In particular, for any neighbourhood $U$ of $X^*$ there exists
a neighbourhood $V$ of $X^*$ such that the $\omega$-limit sets
$\omega(x_0)$ are non-empty and contained in $U$ for all $x_0 \in V$.
\item[(b)]
If $X^* \in M$ is a strict local maximum and an isolated critical
point of $f$, then  $X^*$ is an asymptotically stable equilibrium
of (\ref{gradflow}).
In particular, there is a neighbourhood $V$ of $X^*$ such that
$\omega(x_0) = \{X^*\}$ for all $X_0 \in V$, i.e. all solutions 
$X(t,X_0)$ with initial value $X_0 \in V$ converge to $X^*$ for
$t \to +\infty$.
\end{enumerate}
\end{theorem}

{\bf Proof.}\hspace{.5em}
Both assertions follow immediately from classical stability theory
by taking $f$ as Lyapunov function, cf. \cite{HM94, Irwin}
\hfill$\square$

\medskip

\noindent
Note that the convergence analysis near arbitrary equilibria,
i.e. near arbitrary critical points of $f$ is quite subtle and
may depend on the Riemannian metric, cf.~\cite{Takens71}.


\subsubsection*{Discretised Gradient Flows}

Finally, we approach the problem of finding discretisations of
(\ref{gradflow}) which lead to convergent \emph{gradient ascent
methods}. The ideas presented below can be traced back to R. Brockett,
cf. \cite{Bro88+91}. Let 
\begin{equation}
\label{riemannexp}
\exp_X:\rT_XM \to M 
\end{equation} 
be the \emph{Riemannian exponential map} at $X \in M$, i.e.
$t \mapsto \exp_X(t \xi)$ denotes the unique geodesic with initial
value $X \in M$ and initial velocity $\xi \in \rT_XM$. Moreover, we
assume that $M$ is (geodesically) \emph{complete}, i.e. any
geodesic is defined for all $t \in \R$. Hence, \eqref{riemannexp}
is well-defined for the entire tangent bundle $\rT M$.



The simplest discretisation approach---a scheme that can be seen as an
\emph{intrinsic Euler step method}---leads to

\begin{enumerate}
\item[] {\em Riemannian Gradient Method}
\begin{equation}
\label{eq:grad}
\begin{split}
 X_{k+1} :&= {\exp}_{X_k} \big(\alpha_k \grad f(X_k)\big)
\end{split}
\end{equation}
where $\alpha_k$ denotes an appropriate step size. 
\end{enumerate}
In order to guarantee convergence of \eqref{eq:grad} to the set of
critical points, it is sufficient to apply the \emph{Armijo rule}
\cite{Bazaraa93}.
An alternative to Armijo's rule provides the step size selection
suggested by Brockett in \cite{Bro93}, see also \cite{HM94}.
Convergence to a single critical point is a more subtle issue.
If $\big(M, \braket{\cdot}{\cdot}\big)$ and $f$ are analytic,
and the step sizes are chosen according to a version of the first
\emph{Wolfe-Powell} condition for Riemannian manifolds,
then pointwise convergence holds. A detailed proof can be found in
\cite{Diss-Lageman}.




\subsection{Gradient Flows on Lie Groups}
\label{sec:liegroups}

In the following, we apply the previous results to Lie groups and
Lie subgroups. However, to fully exploit Lie-theoretic tools, the
Riemannian structure and the group structure have to match,
i.e. the metric  $\braket{\cdot}{\cdot}$ has to be invariant under
the group action.
For basic concepts and results on Lie Groups and their Riemannian
geometry we refer to \cite{Helgason78,Gallot04}. In particular, we
recommend the book of Arvanitoyeorgos \cite{Arv03} for a rather
condensed overview.

Let $\bG$ denote a finite dimensional \emph{Lie group},
i.e. a group which carries a smooth manifold structure such that the
group operations are smooth mappings. For notational convenience we
will assume that $\bG$ can be represented as a (closed)
\emph{matrix Lie group}, i.e. as an (embedded) Lie subgroup of
some \emph{general linear group} $GL(N,\K)$ of invertible
$N \!\times\! N$-matrices over $\K = \R$ or $\C$.

\begin{remark}
\label{remmatrixLiegroup}
According to a well-known result by Cartan, a subgroup
$\bG \subset GL(N,\K)$ is an (embedded) Lie subgroup, i.e. a
smooth submanifold of $GL(N,\K)$, if and only if it is closed
in $GL(N,\K)$, cf. \cite{Knapp02}.
Note, however, that there is a subtle difference between embedded
and immersed Lie subgroups.
Moreover, not every abstract Lie group admits a faithful representation
as a matrix Lie group. Nevertheless, the class of 
matrix Lie groups is rich enough for all of our subsequent applications. 
For more details on these topics we refer to \cite{Hall03}.
\end{remark}


\subsubsection*{Invariant Metrics}

A Lie group $\bG$ can be endowed in a canonical way with a
Riemannian metric $\braket{\cdot}{\cdot}$. Let $\g := \rT_{\unity}\bG$
be the Lie algebra of $\bG$, i.e. the tangent space of $\bG$ at the unity
$\unity$. From the fact that the right multiplication $r_H: \bG \to \bG$
and left multiplication $l_H: \bG \to \bG$ are diffeomorphisms of $\bG$
for all $H \in \bG$, it follows
\begin{equation}
\label{tangenspacerelation}
\rT_H\bG = \g H = H \g
\end{equation}
for all $H \in \bG$. Now, let $\roundbraket{\cdot}{\cdot}$ be any
scalar product on $\g$. Then
\begin{equation}
\label{invariantmetric}
\braket{g}{h}_G := \roundbraket{g G^{-1}}{h G^{-1}}
\end{equation}
for all $G \in \bG$ and $g,h \in \rT_G\bG$ yields a \emph{right invariant}
metric on $\bG$, where right invariance stands for  
\begin{equation}
\label{invariance}
\braket{g}{h}_G = \braket{g H}{h H}_{GH}
\end{equation}
for all $G,H \in \bG$ and $g,h \in \rT_G\bG$. Thus, right multiplication
$r_H$ represents an \emph{isometry} of $G$.  In the same way, one could
obtain \emph{left invariant} metrics on $\bG$. 

\begin{remark}
\label{multiplicationmap}
In an abstract setting, one has to replace (\ref{tangenspacerelation})
by
\begin{equation}
\label{tangenspacerelationI}
\rT_H\bG = \D r_H(\unity)\g = \D l_H(\unity)\g
\end{equation}
for all $H \in \bG$, where $\D r_H$ and $\D l_H$ denote the tangent maps
of $r_H$ and $l_H$, respectively.
For a matrix Lie group, however, the respective tangent maps
are given by $\D r_H(G)\xi = \xi H$ and $\D l_H(G)\xi = H \xi $
for all $G \in \bG$ and $\xi \in \rT_G\bG$. Hence
(\ref{tangenspacerelationI})  reduces to  (\ref{tangenspacerelation}).
\end{remark}

The construction of \emph{bi-invariant}, i.e. right and left
invariant metrics is much more subtle and in general even impossible.
To summarise the basic results on this topic we need some further
terminology. The \emph{adjoint} maps $\Ad{}: \bG \to GL(\g)$ and 
$\ad{} : \g \to \mathfrak{gl}(\g)$ are defined by
\begin{equation*}
\Ad{G}h := GhG^{-1}
\end{equation*}
and
\begin{equation*}
\ad{g}h := [g,h] := gh-hg 
\end{equation*}
for all $G \in \bG$ and all $g,h \in \g$, where $GL(\g)$ and
$\mathfrak{gl}(\g)$ denote the set of all automorphisms and,
respectively, endomorphisms of $\g$. Note both notations $\ad{g}h$
and $[g,h]$ are used interchangeably in the literature.
A bilinear form $\roundbraket{\cdot}{\cdot}$ on $\g$ is called
\begin{enumerate}
\item[(a)] 
$\Ad{\bG}$-\emph{invariant} if the identity
\begin{equation}
\label{Ad-invariance}
\roundbraket{g}{h} = \roundbraket{\Ad{G}g}{\Ad{G}h}
\end{equation}
is satisfied for all $g,h \in \g$ and $G \in \bG$. 
\item[(b)]
$\ad{\mathfrak{g}}$-\emph{invariant} if the identity
\begin{equation}
\label{ad-invariance}
\quad\quad\quad
\roundbraket{\ad{g}h}{k} = -\roundbraket{h}{\ad{g}k}
\end{equation}
is satisfied for all $g, h, k \in \g$.
\end{enumerate}

\begin{proposition}
\label{propalgebravsgroup}
The following statements are equivalent:
\begin{enumerate}
\item[(a)]
There exists a bi-invariant Riemannian metric $\braket{\cdot}{\cdot}$
on $\bG$.
\item[(b)]
There exists an $\Ad{\bG}$-invariant scalar product
$\roundbraket{\cdot}{\cdot}$ on $\g$. 
\end{enumerate}
Moreover, each of the statements (a) and (b) imply
\begin{enumerate}
\item[(c)]
There exists an $\ad{\mathfrak{g}}$-invariant scalar product
$\roundbraket{\cdot}{\cdot}$ on $\g$. 
\end{enumerate}
If $\bG$ is also connected, then (c) is equivalent to
(a) and (b), respectively,
\end{proposition}

{\bf Proof.}\hspace{.5em}
The equivalence (a) $\Longleftrightarrow$ (b) follows easily
by exploiting Eqn.~\eqref{invariance} at $G = \unity$. Moreover,
applying (b) to a one-parameter subgroup $t \mapsto \exp(tg)$
and taking the derivative with respect to $t=0$ yields (c).
The implication (c) $\Longrightarrow$ (b) is obtained in the
same way, i.e. by taking the derivative of
\begin{equation*}
t \mapsto \big(\Ad{\e^{tg}} h|\Ad{\e^{tg}} k\big).
\end{equation*}
with respect to $t$, cf. \cite{Arv03}. Note, however, that this
implies $\Ad{\bG}$-invariance only on the connected component
of the unity. Therefore, connectedness is necessary for the
implication (c) $\Longrightarrow$ (b) as counter-examples show.
\hfill$\square$

\medskip

\noindent
Now, the main result on the existence of bi-invariant metrics reads as
follows.





\begin{theorem}
\label{thmbi-invariantmetricI}
A connected Lie group $\bG$ admits a bi-invariant Riemannian metric
if and only if $\bG$ is the direct product of a compact Lie group $\bG_0$
and an abelian one, which is isomorphic to some $(\R^m,+)$, i.e.
$\bG \cong \bG_0 \times \R^m$
\end{theorem}

{\bf Proof.}\hspace{.5em}
Cf. \cite{Gallot04, Milnor76}. 
\hfill$\square$

\medskip

Finally, we focus on a special class of Lie groups. A connected Lie group
$\bG$ is called \emph{semisimple} if the Killing form, i.e. the bilinear
form $(g,h) \mapsto \kappa(g,h) := \tr(\ad{g} \ad{h})$
is non-degenerate on $\g$. Most prominent representatives of this class
are $SL(N,\R)$, $SL(N,\C)$, $SO(N,\R)$ and $SU(N)$.
More on semisimple Lie groups and their algebras can be found in
\cite{Knapp02, Helgason78}.

\begin{theorem}
\label{killingform}
\begin{enumerate}
\item[(a)]
If $\bG$ is semisimple then the Killing form $\kappa$ defines an
$\ad{\mathfrak{g}}$-invariant  bilinear form on $\g$. 
\item[(b)]
If $\bG$ is semisimple and compact then $-\kappa$ defines
an $\ad{\mathfrak{g}}$-invariant scalar product on $\g$. Thus
$-\kappa$ induces a bi-invariant Riemannian metric on $\bG$.
\end{enumerate}    
\end{theorem}

{\bf Proof.}\hspace{.5em}
Cf. \cite{Arv03, Knapp02}.
\hfill$\square$


\subsubsection*{Gradient Flows with Respect to an Invariant Metric}

Next, we study gradient flows on $\bG$ or on a closed subgroup
$\bH \subset \bG$ with respect to an invariant metric
$\braket{\cdot}{\cdot}$.
Therefore, let $f : \bG \to \R$ be a smooth quality function and let
$\varphi:\bG \to \bG$ be any diffeomorphism. Using the identity
\begin{equation}
\label{diffeomorphism}
\grad\,(f\circ\varphi)(G) =
\big(\D\varphi(G)\big)^* \grad\,f\big(\varphi(G)\big)
\end{equation}
for all $G \in \bG$, where $(\cdot)^*$ denotes the adjoint operator,
we obtain by the right invariance of the metric
\begin{equation}
\label{Liegrad}
\grad\, f\,(G) = \grad (f\circ r_G)\,(\unity) G
\end{equation}
for all $G \in \bG$. Hence 
\begin{equation}
\label{Liegradflow}
\dot{G}  = \grad\, f(G).
\end{equation}
can be rewritten as
\begin{equation}
\label{LiegradflowII}
\dot{G}  = \grad\, (f\circ r_G)(\unity) G.
\end{equation}
Thus, the gradient flow of $f$ is completely determined by the
mapping $G \mapsto \grad (f\circ r_G)(\unity) \in \g$.
To study its asymptotic behaviour of Eqn.~(\ref{Liegradflow})
we can apply the results of the previous section. For instance,
for compact Lie grops we have.

\begin{corollary}
\label{corIpointwiseconvergence}
Let $\bG$ be a compact Lie group with a right invariant
Riemannian metric $\braket{\cdot}{\cdot}$ and let $f: \bG \to \R$
be a real analytic quality function. Then any solution of
Eqn.~(\ref{Liegradflow}) converges to a critical point of $f$ for
$t \to +\infty$.
\end{corollary}

{\bf Proof.}\hspace{.5em}
This follows immediately from Proposition \ref{propgradomegalimit}
and Theorem \ref{thmIIpointconvergence}, as the pair
$\big(\bG, \braket{\cdot}{\cdot}\big)$ constitutes a real analytic
Riemannian manifold whenever the metric $\braket{\cdot}{\cdot}$ is
invariant, cf. \cite{DuiKolk00}.
\hfill$\square$

\medskip

Now, let $\bH$ be a closed subgroup of $\bG$. By Remark
\ref{remmatrixLiegroup}, we know that $\bH$ is actually
an (embedded) submanifold of $\bG$. 
%
%
%
%
Therefore, the gradient flow of $f|_{\bH}$ with respect to
$\braket{\cdot}{\cdot}|_{H}$ is well-defined and can be given
explicitly via the orthogonal projectors $P_{H}$, cf.
(\ref{restrictedgradflow}). However, for an invariant metric
the computation of $P_{H}$ simplifies considerably, as all
calculations can be carried out on the Lie algebra $\g$
of $\bG$.  

\begin{lemma}
\label{projection}
Let $\bG$ be a Lie group with a right invariant Riemannian metric
$\braket{\cdot}{\cdot}$ and let $\bH$ be a closed subgroup of $\bG$.
Furthermore, let $\g$ and $\h$ their corresponding Lie algebras and
denote by $P_{\h}$ the orthogonal projection of $\g$ onto $\h$. Then
the orthogonal projection $P_H$ in (\ref{restrictedgrad}) is given
by 
\begin{equation}
\label{projector}
P_H(gH) := P_{\h}(g)H
\end{equation}
for all $gH \in \rT_H{\bG}$.
\end{lemma}

{\bf Proof.}\hspace{.5em}
This is a straightforward consequence of the identity
$\rT_H{\bH} = \h H$ and the right invariance of $\braket{\cdot}{\cdot}$.
\hfill$\square$

\medskip

\noindent 
According to (\ref{restrictedgradflow}), (\ref{Liegrad}) and
(\ref{projector}), the gradient flow of $f|_H$ finally reads 
\begin{equation}
\label{restictedLiegradflow}
\dot{H} 
 = P_{\h}\big(\grad (f \circ r_H)(\unity)\big)H.
\end{equation}


\subsubsection*{Geodesics with Respect to an Invariant Metric}

The remainder of this subsection is devoted to the issue: How to
compute geodesics and the Hessian of a smooth quality function with
respect to an invariant metric. The main results for the forthcoming
applications are summarised in Theorem \ref{thmLiegeodesics}(b) and
Proposition \ref{hessianvsrestriction}. For readers with basic
differential geometric background we provide some details of the 
proof which however can be skipped, so as not to lose the thread. 
First, we need some further notation. Let
\begin{equation}
\mathcal{X}^{r}_g: G \mapsto gG
\quad\mbox{and}\quad
\mathcal{X}^{l}_g: G \mapsto Gg
\end{equation}
be the \emph{right} and \emph{left invariant vector fields} on $\bG$
which are uniquely determined by $\mathcal{X}^{r}_g(\unity) = g$ and
$\mathcal{X}^{l}_g(\unity) =g$, respectively.
Moreover, let $L_\mathcal{X}(\cdot)$ denote the Lie derivative with
respect to the vector field $\mathcal{X}$, i.e. for a smooth function
$f: \bG \to \R$ one has
\begin{equation*}
L_\mathcal{X}(f)(G) := \D f(G) \cdot \mathcal{X}(G).
\end{equation*}
On vector fields $\mathcal{Y}$, the action of $L_\mathcal{X}(\cdot)$
is given by
\begin{equation*}
\begin{split}
L_\mathcal{X}(\mathcal{Y})(G) & := \\
- \lim_{t \to \infty} &
\frac{\big(\D \Phi_\mathcal{X}(t,G)\big)^{-1} \cdot
\mathcal{Y}\big(\Phi_\mathcal{X}(t,G)\big) - \mathcal{Y}(G)}{t},
\end{split}
\end{equation*}
where $\Phi_\mathcal{X}(t,\cdot)$ denotes the corresponding flow
of $\mathcal{X}$. 

Next, we recall two basic facts from differential geometry which
play a key role for the proof of Theorem \ref{thmLiegeodesics}.
The first one shows that the set of right/left invariant
vector fields is invariant under Lie derivation,
cf.~\cite{Arv03, Helgason78}. The second one relates a Riemannian
metric of a manifold $M$ with a particular linear connection on $M$.
For more details see e.g.~\cite{Gallot04}.

\begin{fact}
\label{fac:Lie-deri}
The Lie derivative of a right/left invariant vector field
is again right/left invariant and satisfies
\begin{equation}
\label{eq:Lie-deri}
L_{\mathcal{X}^{r}_g} \mathcal{X}^{r}_h  = - \mathcal{X}^{r}_{[g,h]} 
\quad\mbox{and}\quad
L_{\mathcal{X}^{l}_g} \mathcal{X}^{l}_h  = \mathcal{X}^{l}_{[g,h]}. 
\end{equation}
\end{fact}


\begin{fact}
On any Riemannian manifold $M$ there exists a unique
\emph{Riemannian connection} $\nabla$ determined by
the properties
\begin{equation}
\label{fac:connection}
L_{\mathcal{X}} \mathcal{Y} = 
\nabla_{\mathcal{X}} \mathcal{Y} - \nabla_{\mathcal{Y}} \mathcal{X} 
\end{equation}
and
\begin{equation}
\label{eq1:connection}
\nabla_{\mathcal{X}}\braket{\mathcal{Y}}{\mathcal{Z}} =
\braket{\nabla_{\mathcal{X}}\mathcal{Y}}{\mathcal{Z}}
+ \braket{\mathcal{Y}}{\nabla_{\mathcal{X}}\mathcal{Z}}.
\end{equation}
\end{fact}

Now, combining both facts yields the main result about
geodesics on Lie groups.

\begin{theorem}
\label{thmLiegeodesics}
Let $\bG$ be a Lie group with a bi-invariant metric $\braket{\cdot}{\cdot}$
and let $\nabla$ denote the unique Riemannian connection on $\bG$ induced
by $\braket{\cdot}{\cdot}$.
\begin{enumerate}
\item[(a)]
For right/left invariant vector fields
the Riemannian connection $\nabla$ is given by
\begin{equation}
\label{eq:Riemannianconnection}
\nabla_{\mathcal{X}^{r}_g} \mathcal{X}^{r}_h =
-\frac{1}{2}\mathcal{X}^{r}_{[g,h]}
\quad\mbox{and}\quad
\nabla_{\mathcal{X}^{l}_g} \mathcal{X}^{l}_h = 
\frac{1}{2}\mathcal{X}^{l}_{[g,h]}.
\end{equation}
\item[(b)]
The geodesics through any $G \in \bG$ are of the form $t \mapsto G\exp(t g)$
or $t \mapsto \exp(t g)G$ with $g \in \g$. In particular, the geodesics
through the unity $\unity$ are precisely the one-parameter subgroups of
$\bG$.  
\end{enumerate} 
\end{theorem}

{\bf Proof.}\hspace{.5em}
(a) Applying Koszul's identity, cf. \cite{Arv03, Gallot04},
\begin{eqnarray*}
\label{Koszul}
2 \braket{\nabla_{\mathcal{X}} \mathcal{Y}}{\mathcal{Z}}
& = &
L_\mathcal{X}\braket{\mathcal{Y}}{\mathcal{Z}} +
L_\mathcal{Y}\braket{\mathcal{Z}}{\mathcal{X}} -
L_\mathcal{Z}\braket{\mathcal{X}}{\mathcal{Y}}\\
&&
-\braket{\mathcal{X}}{L_\mathcal{Y} \mathcal{Z}} +
\braket{\mathcal{Y}}{L_\mathcal{Z} \mathcal{X}} +
\braket{\mathcal{Z}}{L_\mathcal{X} \mathcal{Y}},
\end{eqnarray*}
to $\mathcal{X}^{r}_g, \mathcal{X}^{r}_h$ and $\mathcal{X}^{r}_k$
we obtain 
\begin{eqnarray*}
\lefteqn{
2 \braket{\nabla_{\mathcal{X}^{r}_g} \mathcal{X}^{r}_h}{\mathcal{X}^{r}_k}
=}\\
&&
+ \braket{\mathcal{X}^{r}_g}{\mathcal{X}^{r}_{[h,k]}}
- \braket{\mathcal{X}^{r}_h}{\mathcal{X}^{r}_{[k,g]}}
- \braket{\mathcal{X}^{r}_k}{\mathcal{X}^{r}_{[g,h]}}.
\end{eqnarray*}
Now Proposition \ref{propalgebravsgroup} and Fact \ref{fac:Lie-deri}
implies
\begin{equation*}
2 \braket{\nabla_{\mathcal{X}^{r}_g} \mathcal{X}^{r}_h}{\mathcal{X}^{r}_k} =
- \braket{\mathcal{X}^{r}_k}{\mathcal{X}^{r}_{[g,h]}}
\end{equation*}
and hence
\begin{equation*}
2 \nabla_{\mathcal{X}^{r}_g} \mathcal{X}^{r}_h =
-\frac{1}{2}\mathcal{X}^{r}_{[g,h]}.
\end{equation*}
Obviously, for left invariant vector fields the same arguments apply.

\medskip
\noindent
(b) Let $\gamma(t) := \exp(t g)G$. Part (a) implies that
the covariant derivative
$\nabla_{\dot{\gamma}(t)} \dot{\gamma}(t)
= \nabla_{\mathcal{X}^r_g} \mathcal{X}^r_g (\gamma(t))$
of $\gamma$ vanishes and thus $\gamma$ represents the unique
geodesics through $G$ with \/`initial velocity\/' $\xi = gG$. The
same holds for $\widetilde \gamma(t) := G\exp(t g)$,
cf.~\cite{Arv03} or \cite{Helgason78}. 
\hfill$\square$

\medskip

\noindent
Observe that the bi-invariance of the metric and the invariance
of the vector fields are essential for the above result. 
For example Eqn. (\ref{eq:Riemannianconnection}) fails, if
the Riemannian metric is just right invariant. More details on
this topic can be found in \cite{Helgason78, CheegerEbin75}

Finally, by Theorem \ref{thmLiegeodesics}, the Hessian of the
restriction $f|_{\bH}$ can easily be obtained by restricting
the Hessian of $f$ to $\rT\bH$. More precisely, we have.  

\begin{proposition}
\label{hessianvsrestriction}
Let $f : \bG \to \R$ be a smooth quality function on a Lie group with 
bi-invariant metric $\braket{\cdot}{\cdot}$ and let $\bH$ be a closed
subgroup. Then the Hessian of $f|_{\bH}$ at $H$ is given by
\begin{equation}
\label{restrictedhessian}
\Hess f|_\bH (H) = \Hess f(H)\Big|_{\rT_H\bH \times \rT_H\bH} 
\end{equation}
\end{proposition}

\noindent
Note that in general Eqn.~(\ref{restrictedhessian}) is sheer nonsense
unless $\bH$ is a Lie subgroup. Counter-examples can be obtained easily
for $\bG = \R^m$.


\subsection{Gradient Flows on Homogeneous Spaces}
\label{sec:flows_homsp}

The subsequent section on homogeneous spaces is motivated by the
following observation, cf.~Subsection \ref{sec:ex}.
As before, let $f : \bG \to \R$ be a smooth quality function. In many
applications $f$ can be decomposed into a function $F$
defined on a smooth manifold $M$ and a (right) \emph{group action}
$\alpha: (X,G) \mapsto X \cdot G$ on $M$ such that
\begin{equation}
\label{groupaction}
f(G) := F(X \cdot G)
\end{equation}
for some fixed $X \in M$. Then we can think of $f$ as defined on the
orbit of $X$. More precisely, let $\widehat{f} = F|_{\mathcal{O}(X)}$,
where $\mathcal{O}(X):=\{X \cdot G \;|\; G \in \bG\}$ denotes the
\emph{orbit} of $X$. Thus,
\begin{equation}
\label{inducedfunction}
\widehat{f}(Y) = f(G)
\end{equation}
for $Y = X \cdot G$ with $G \in \bG$. Such quality functions $f$ are
called \emph{induced} by $F$, cf. Subsection \ref{sec:ex}.
By construction, we have
\begin{equation}
\label{equalityofmaxima}
\max_{G \in \bG}f(G) = \max_{Y \in \mathcal{O}(X)}\widehat{f}(Y).
\end{equation}
Moreover, let  $\bH_X := \{G \in \bG \;|\; X \cdot G = X\}$ denote
the \emph{stabiliser} or, equivalently,
\emph{isotropy subgroup} of $X$. Then $\widehat{f}$ can also be viewed
as a function on the 
\emph{right coset space}
\begin{equation}
\bG/\bH_X := \{\bH_XG \;|\; G \in \bG\},
\end{equation}
cf. \footnote{ Note that the
coset-terminoloy in the group
literature is not consistent, i.e. right cosets are sometimes called left
cosets and vice versa. Here, we stick to the term right coset, if the group
element in on the right side, i.e. $[G] = \bH G$.},
which is equivalent to say that $f$ is \emph{equivariant} with respect
to $\bH_X$, i.e. 
\begin{equation}
\label{equivariance}
f(G) = f(HG)
\end{equation}
for all $H \in \bH_X$. Therefore, coset space show up quite naturally
in optimising equivariant quality functions. Note that passing from
$\bG$ to $\bG/\bH_x$ can be rather useful in order to avoid
certain degeneracies such as continua of critical points.

\subsubsection*{Coset Spaces}

We first collect the fundamental facts on the differential structure of
$\bG/\bH$, where $\bH$ is any closed subgroup of $\bG$. Detailed
expositions can be found in \cite{Helgason78, Warner83, ONeill83, BroeDie85, Arv03}.

\begin{theorem}
\label{thmcosetspace}
Let $\bG$ be a Lie group with Lie algebra $\mathfrak{g}$ and let
$\bH \subset \bG$ be a closed subgroup with Lie algebra $\mathfrak{h}$.
Moreover, let $\mathfrak{p}$ be any complementary subspace to $\mathfrak{h}$,
i.e. $\mathfrak{g} = \mathfrak{h} \oplus \mathfrak{p}$. Then the following holds:
\begin{enumerate}
\item[(a)]
The quotient topology turns the set of right cosets
$\bG/\bH := \{[G]:= \bH G \;|\; G \in \bG\}$ into a locally
compact Hausdorff space. 
\item[(b)]
There exists a unique manifold structure on $\bG/\bH$ such that
the canonical projection $\Pi: \bG \to \bG/\bH$, $G \mapsto [G]$
is a submersion. In particular, the tangent space of $\bG/\bH$ at
$[\unity]$ is isomorphic to $\mathfrak{p}$ via the canonical
identification $p \mapsto \frac{\rm d}{{\rm d}t}[\exp tp]\big|_{t=0}$
and thus $\mathrm{dim} \bG/\bH = \mathrm{dim}\bG - \mathrm{dim}\bH$.
\end{enumerate} 
The following statements refer to the unique manifold structure on
$\bG/\bH$ given in part (b).
\begin{enumerate}
\item[(c)]
The Lie group $\bG$ acts smoothly from the right on $\bG/\bH$ via
\begin{equation}
\label{eq:cosetaction}
\big([G'],G\big) \mapsto [G'G]
\end{equation}
such that
\begin{equation} 
\label{eq:righttranslation}
\widehat{r}_{G} : \bG/\bH \to \bG/\bH,
\quad
[G'] \mapsto [G'G]
\end{equation}
are diffeomorphisms for all $G \in \bG$. Moreover, 
\begin{equation} 
\label{eq:lefttranslation}
\Pi\circ l_{G}: \bG \to \bG/\bH,
\quad
G' \mapsto [GG']
\end{equation}
are submersions for all $G' \in \bG$.
Thus, the tangent space $\rT_{[G]}\bG/\bH$ is given by
\begin{equation}
\label{eq:tangentspace-GH}
\begin{split}
\rT_{[G]}\bG/\bH &= \D \widehat{r}_G([\unity])\,\rT_{[\unity]}\bG/\bH\\
&= \D (\Pi\circ l_G)(\unity)\g\\
&= \D (\Pi\circ l_G)(\unity)(\Ad{G^{-1}}\p).
\end{split}
\end{equation}
\item[(d)]
Moreover, if $\bH$ is a normal subgroup, i.e. $G\bH G^{-1} = \bH$
for all $G \in \bG$, then the multiplication $[G]\cdot[G'] := [GG']$
is well-defined and yields a Lie group structure on $\bG/\bH$.
\end{enumerate} 
\end{theorem}

{\bf Proof.}\hspace{.5em}
Cf. \cite{Helgason78, DuiKolk00, BroeDie85}.
\hfill$\square$
 
\medskip
\noindent
The Lie group $\bG/\bH$ given by Theorem \ref{thmcosetspace} (d) is
called the \emph{quotient Lie group} of $\bG$ by $\bH$. Moreover,
the result provides the possibility to extend the well-known
\emph{First Isomorphism Law} to the category of Lie groups.

\begin{theorem}
\label{thmisomorphismlaw}
Let $\Phi: \bG \to \bG'$ be a smooth surjective Lie group homomorphism.
Then there exists a well-defined Lie group isomorphism
$\widehat{\Phi}: \bG/\bH \to \bG'$ with $\bH := \mathrm{ker}\,\Phi$
such that the diagram
\begin{equation}
\label{isomorphismlawI}
\xymatrix{
\mathbf{G} \ar[d]_\Pi \ar[r]^{\Phi} & \mathbf{G}' \\
{\mathbf{G}/{\mathbf H}}\ar[ur]_{\widehat{\Phi}} &}
\end{equation}
commutes. Moreover, let $\g$, $\g'$ and $\h$ denote the corresponding
Lie algebras and let $\p$ be any complementary space to $\h$. Then
$\D\Phi(\unity)$ is a surjective Lie algebra homomorphism with
$\mathrm{ker}\,\D\Phi(\unity) = \h$ and commutative diagram
\begin{equation}
\label{isomorphismlawII}
\xymatrix{
\g \ar[d]_{\D\Pi(\unity)} \ar[r]^{\D\Phi(\unity)} & \g' \\
{\p \cong \g/\h \;.}\ar[ur]_{\D\widehat{\Phi}(\unity)} &}
\end{equation}
\end{theorem}

{\bf Proof.}\hspace{.5em}
Note that $\bH = {\rm ker}\,\Phi$ is a closed normal subgroup
of $\mathbf{G}$. Thus, by the First Isomorphism Law
$\widehat{\Phi}([G]) := \Phi(G)$ for $[G] \in \bG/\bH$ is a
well-defined group isomorphism. Moreover, $\widehat{\Phi}$
is smooth, since $\Pi$ is a smooth submersion by Theorem
\ref{thmcosetspace}. The assertion that $\D\Phi(\unity)$ is a
surjective Lie algebra homomorphism, follows easily from 
the properties of the exponential map. Finally, a straightforward
application of the chain rule yields Eqn. \eqref{isomorphismlawII}.
\hfill$\square$


\subsubsection*{Orbit Theorems and Homogeneous Spaces}

Next, we analyse the relation between group actions and
coset spaces. A smooth right \emph{Lie group action} is a smooth
map $\alpha: M \times \bG \to M$, $(X,G) \mapsto X \cdot G$
which satisfies
$$
(X \cdot G) \cdot H = X \cdot (GH)
\quad\mbox{and}\quad 
X \cdot \unity = X
$$
for all $X \in M$ and $G,H \in \bG$. The \emph{orbit} of $X \in M$
under the group action $\alpha$ is defined by 
$\mathcal{O}(X) := \{X \cdot G \;|\; G \in \bG\}$.
The action is called \emph{transitive} if $M = \mathcal{O}(X)$
for some and hence for all $X \in M$. Equivalently, one can say
that for all $X,Y \in M$ there exists an element $G \in \bG$ with
$Y = X \cdot G$. Moreover, for $X \in M$ let
$\bH_X := \{G \in \bG \;|\; X \cdot G = X\}$ denote the \emph{stabiliser}
of $X$ and $\alpha_X: \bG \to M$ the map $G \mapsto X \cdot G$. 
Then the \emph{canonical map} $\widehat{\alpha}_X: \bG/\bH_X \to M$
is defined by $[G] \mapsto X \cdot G$.

\begin{theorem}[Orbit Theorem]
\label{thmorbit}
Let $\bG$ be a Lie group with Lie algebra $\g$ and let
$\alpha: M \times \bG \to M$ be a smooth right action of $\bG$
on a smooth manifold $M$. Moreover, let $X$ be any
point in $M$. Then the following statements are satisfied:
\begin{enumerate}
\item[(a)]
The stabiliser subgroup $\bH_X$ is a closed subgroup of
$\bG$.
\item[(b)]
Let $\h_X$ be the Lie algebra of $\bH_X$. Then
\begin{equation}
\ker \D\alpha_X(\unity) = \h_X.
\end{equation}
In particular, the canonical map $\widehat{\alpha}_X: \bG/\bH_X \to M$
is an injective immersion.
\item[(c)]
The canonical map $\widehat{\alpha}_X$
is an embedding, i.e. $\mathcal{O}(X)$ is a submanifold of $M$
diffeomorphic to $\bG/\bH_X$, if and only if $\widehat{\alpha}_X$ is proper,
cf. \footnote{A map $\varphi$ is called \emph{proper} if the pre-image
$\varphi^{-1}(K)$ of any compact set $K$ is also compact.}.
In this case, the tangent space of  $\mathcal{O}(X)$ at $Y = X \cdot G$
is given by
\begin{equation}
\label{orbittangentspace}
\begin{split}
\rT_Y \mathcal{O}(X)
& = \D\alpha_X(G)\,\rT_{G}\bG\\
& = \D\alpha_Y(\unity)\,\g\\
& = \D\alpha_Y(\unity)\,\Ad{G^{-1}}\p_X,
\end{split}
\end{equation}
where $\p_X$ is any complementary subspace of $\h_X$, i.e.
$\g = \h_X \oplus \p_X$.
\end{enumerate} 
\end{theorem}

\medskip

{\bf Proof.}\hspace{.5em} (a) The continuity of $\alpha_X$
implies that $\bH_X = \alpha_X^{-1}(X)$ is closed.

\medskip

(b) In order to see that $\widehat{\alpha}_X$ is an injective immersion,
consider the identity
$\alpha_X \circ r_G = \alpha\big(\alpha_X(\cdot),G\big)$ and thus
\begin{equation*}
\D \alpha_X(G) \cdot gG = \D_1 \alpha (X,G) \circ \D \alpha_X(\unity) \,g.
\end{equation*}
Therefore, $\D \alpha_X(\unity) \,g = 0$ implies
\begin{equation*}
\frac{\rm d}{{\rm d}t}\, \alpha_X \big(\exp(tg)\big) = 0
\end{equation*}
for all $t \in \R$ and hence $\ker \D\alpha_X(\unity) \subset \h_X$.
As the inclusion $\h_X \subset \ker \D\alpha_X(\unity)$ is obvious,
we obtain $\ker \D\alpha_X(\unity) = \h_X$. Moreover, let $\p_X$ be
any complementary subspace of $\h_X$. Then, identifying $\p_X$ with
$\rT_{[\unity]}\bG/\bH_X$ 
yields
$D\widehat{\alpha}_X\big([\unity]\big) = \D\alpha_X(\unity) \big|_\p$,
cf. Theorem \ref{thmcosetspace}. Thus,
$D\widehat{\alpha}_X\big([\unity]\big)$ is injective and
the same holds for any other
$[G] \in \bG/\bH_X$ by right multiplication $\widehat{r}_{G}$.

\medskip

(c) The first part follows from a standard embedding criterion on immersed
manifolds, cf. \cite{AMR88}. The first equality of Eqn.
\eqref{orbittangentspace} is a straightforward consequence of
the identity $\alpha_X = \widehat{\alpha}_X \circ \Pi_X$, where
$\Pi_X: \bG \to \bG/\bH_X$ denotes the canonical projection. The second
one is obtained by
$\alpha_Y = \alpha_X \circ l_G = \alpha_X \circ r_G \circ \Ad{G}$,
while the third one follows from $\bH_Y = \Ad{G^{-1}} \bH_x$.
For further details see also
\cite{HM94}. \hfill$\square$


\medskip

\begin{corollary}
\label{cor:orbit}
Let $\alpha: M \times \bG \to M$ be as in Theorem \ref{thmorbit}
and let $X \in M$ be any point.
\begin{enumerate}
\item[(a)]
If $\bG$ is compact then $\bG/\bH_X$ is diffeomorphic to $\mathcal{O}(X)$.
\item[(b)]
If $\alpha$ is transitive then $\bG/\bH_X$ is diffeomorphic to $M$.
\end{enumerate}
\end{corollary}

{\bf Proof.}\hspace{.5em}
(a) This follows immediately from Theorem \ref{thmorbit}(c) and
the compactness of $\bG$.

\medskip
\noindent
(b) Observe that transitivity of $\alpha$ implies surjectivity
of $\D\alpha_X(G)$ and $\D\widehat{\alpha}_X\big([G]\big)$. Thus, Theorem
\ref{thmorbit}(b) yields the desired result,
cf.~\cite{BroeDie85}. \hfill$\square$

\medskip

\noindent
This gives rise to the following definition.
A manifold $M$ is called a \emph{homogeneous} $\bG$-\emph{space}
or for short a \emph{homogeneous space}, if there exists a transitive
smooth Lie group action of $\bG$ on $M$.
In particular, any coset space $\bG/\bH$ can be regarded as a
homogeneous space via the canonical action $([G'],G) \mapsto [G'G]$
for $[G'] \in \bG/\bH$ and $G \in \bG$.   
Further results on homogeneous spaces, orbit spaces and
principal $\bG$-bundles can be found in
\cite{KobNom96, DuiKolk00, BroeDie85}



\begin{remark}\label{rem:orbitmfd}
Note that by Theorem \ref{thmorbit} the orbit $\mathcal{O}(X)$
carries always a manifold structure the topology of which is equal
or finer than the topology induced by $M$.
\end{remark}


\subsubsection*{Reductive Homogeneous Spaces}

Let $M$ be homogeneous space with transitive Lie group action
$\alpha: M \times \bG \to M$ and let $\bH:= \bH_X$ be the stabiliser
subgroup of a fixed element $X \in M$. Next, we are interested in
carrying over the Riemannian structure of $\bG$ to $M$ or, equivalently,
to $\bG/\bH$.  First, we need some further terminology. As most of the
following terms are conveniently expressed via algebraic properties
of the pair $(\bG,\bH)$, we focus on the case $M = \bG/\bH$.
Nevertheless, one could restate all result in terms of an abstract
group action $\alpha$ on $M$.

A homogeneous space $\bG/\bH$ is \emph{reductive}, if the Lie algebra
$\h$ of $\bH$ has a complementary subspace $\mathfrak{p}$ in $\g$ such
that $\mathfrak{p}$ is $\Ad{\bH}$-invariant, i.e.
$H\mathfrak{p}H^{-1} \subset \mathfrak{p}$  for all $H \in \bH$. A Riemannian
metric $\braket{\cdot}{\cdot}$ on $\bG/\bH$ is called $\bG$-\emph{invariant}
if the mappings $\widehat{r}_G$ are isometries, i.e. if the identity
\begin{equation}
\label{quotientinvariance}
\braket{\xi}{\eta}_{[G']} =
\big\langle
{\D\widehat{r}_G([G'])\xi}\big|{\D\widehat{r}_G([G'])\eta}
\big\rangle_{[G'G]}
\end{equation}
is satisfied for all $\xi,\eta \in \rT_{[G']}\bG/\bH$ and $G,G' \in \bG$.
Moreover, a bilinear form $(\cdot|\cdot)$ on $\mathfrak{p}$ is called
\begin{enumerate}
\item[(a)]
$\Ad{\bH}$-\emph{invariant} if the identity
\begin{equation}
\label{Adquotientinvariance}
(p|p') =
(\Ad{H}p|\Ad{H}p')
\end{equation}
is satisfied for all $p,p' \in \mathfrak{p}$ and $H \in \bH$.
\item[(b)]
$\ad{\mathfrak{h}}$-\emph{invariant} if the identity
\begin{equation}
\label{adquotientinvariance}
(\ad{h}p|p') = - (p|\ad{h}p')
\end{equation}
is satisfied for all $p,p' \in \mathfrak{p}$ and $h \in \mathfrak{h}$.
\end{enumerate} 

\noindent
Note that $\bG/\bH$ is reductive, if $\bG$ has a bi-invariant metric,
as one can choose $\mathfrak{p} := \mathfrak{h}^{\perp}$. 
Next, we give a generalisation of Proposition \ref{propalgebravsgroup}
and Theorem \ref{thmbi-invariantmetricI} to homogeneous spaces.

\begin{proposition}
\label{propinvariantquotientmetric}
Let $\bG/\bH$ be a homogeneous space with reductive decomposition
$\g = \h \oplus \p$. The following statements are equivalent:
\begin{enumerate}
\item[(a)]
There exists a $\bG$-invariant metric $\braket{\cdot}{\cdot}$ on
$\bG/\bH$.
\item[(b)]
There exists an $\Ad{\bH}$-invariant scalar product
$(\cdot|\cdot)$ on $\mathfrak{p}$. 
\end{enumerate} 
In addition, if $\bH$ is connected then (a) and (b) are equivalent to
\begin{enumerate} 
\item[(c)]
There exists a $\ad{\mathfrak{h}}$-invariant scalar product
$(\cdot|\cdot)$ on $\mathfrak{p}$. 
\end{enumerate} 
\end{proposition}

{\bf Proof.}\hspace{.5em}
Cf.~\cite{Arv03} and Proposition \ref{propalgebravsgroup}.
\hfill$\square$

\medskip

\begin{theorem}
\label{thminvariantquotientmetric}
Let $\bG/\bH$ be a homogeneous space with reductive decomposition
$\g = \h \oplus \p$. Then $\bG/\bH$ admits a $\bG$-invariant
metric if and only if the closure of
$\Ad{\bH}|_{\mathfrak{p}} :=
\{\Ad{H}:\mathfrak{p}\to\mathfrak{p} \;|\; H \in \bH\}$
is compact in $GL(\mathfrak{p})$.
\end{theorem}

{\bf Proof.}\hspace{.5em}
Cf. \cite{Gallot04}. \hfill$\square$

\medskip

\begin{remark}\label{rem:reduchomspace}
\begin{enumerate}
\item[(a)]
As a special case, Theorem \ref{thminvariantquotientmetric} implies
the existence of bi-invariant metrics on compact Lie groups,
cf.~Theorem \ref{thmbi-invariantmetricI} and \cite{Gallot04}.
\item[(b)]
Replacing $\mathfrak{p}$ by the quotient space $\mathfrak{g}/\mathfrak{h}$,
allows to state Theorem \ref{thminvariantquotientmetric} without
referring to any reductive decomposition
$\mathfrak{g} = \mathfrak{h} \oplus \mathfrak{p}$ of $\mathfrak{g}$,
cf. \cite{Gallot04}.
Moreover, it can be shown that any homogeneous space $\bG/\bH$
which admits a $\bG$-invariant metric is reductive,
cf. \cite{KowalskiSzenthe}. 
\end{enumerate} 
\end{remark}

Theorem \ref{thminvariantquotientmetric} can easily be rephrased
for an arbitrary homogeneous $\bG$-space $M$ with transitive group
action $\alpha: M \times \bG \to M$, by choosing $\bH := \bH_X$ with
$X \in M$. Note however, for orbits $M:=\mathcal{O}(X)$ embedded
in some larger Riemannian manifold $N$, the invariant
metric given by Theorem \ref{thminvariantquotientmetric} does in
general not coincide with the induced metric. This gives rise to
the following definition.

A manifold $M$ is called a \emph{Riemannian homogeneous} $\bG$-{space}
or for short \emph{Riemannian homogeneous space}, if $M$ is a homogeneous
$\bG$-space with \emph{$\alpha$-invariant metric}, which
is to say that the mappings $\alpha_G: M \to M$,
$\alpha_G(X):= X \cdot G$ are isometries of $M$ for all
$G \in \bG$, i.e. 
\begin{equation}
\label{alphainvariance}
\braket{\xi}{\eta}_{X} =
\big\langle
{\D\alpha_G(X)\xi}\big|{\D\alpha_G(X)\eta}
\big\rangle_{X\cdot G}
\end{equation}
for all $\xi,\eta \in \rT_{X}M$ and $G \in \bG$.

\begin{proposition}
\label{prp:Riehomo}
\begin{enumerate}
\item[(a)]
Any homogeneous space of the form $\bG/\bH$ with a
$\bG$-invariant metric is a Riemannian homogeneous space.
\item[(b)]
Any Riemannian homogeneous space is isometric to a homogeneous
space of the form $\bG/\bH$ with a $\bG$-invariant metric.
\end{enumerate}
\end{proposition}

{\bf Proof.}\hspace{.5em}
This follows straightforwardly from the previous definitions and
Corollary \ref{cor:orbit}(b).
\hfill$\square$

\subsubsection*{Naturally Reductive Homogeneous Spaces and Geodesics}

Characterising the Riemannian connection of a homogeneous
space and its geodesics are in general advanced issues which
we do not want to address here, cf. \cite{Arv03} and the
references therein e.g.~\cite{Besse86}. However, there are
two cases -- see (a) and (b) below -- which are easy to handle.
A homogeneous space $\bG/\bH$ is called 
\begin{enumerate}
\item[(a)]
\emph{naturally reductive} if it is reductive with complementary
space $\mathfrak{p}$ and $\Ad{\bH}$-invariant scalar product
$(\cdot|\cdot)$ on $\mathfrak{p}$ such that the identity
\begin{equation}
\label{naturallyreductive}
(P\ad{g}h|k) = -(h|P\ad{g}k)
\end{equation}
is satisfied for all $g,h,k \in \mathfrak{p}$, where $P$
denotes the projection onto $\mathfrak{p}$ along $\mathfrak{h}$.
\item[(b)]
\emph{Cartan-like} if it is reductive with complementary space
$\mathfrak{p}$ and $\Ad{\bH}$-invariant scalar product
$(\cdot|\cdot)$ on $\mathfrak{p}$ such that the commutator
relations
\begin{equation}
\label{Cartan-like}
[\mathfrak{h},\mathfrak{h}] \subset \mathfrak{h},
\quad
[\mathfrak{h},\mathfrak{p}] \subset \mathfrak{p}
\quad \mbox{and} \quad
[\mathfrak{p},\mathfrak{p}] \subset \mathfrak{h}.
\end{equation}
are satisfied.
\end{enumerate}

\begin{remark}
\label{rem:naturallyreductive}
If, in definition (a), the complementary space $\p$ can be chosen
as the orthogonal complement of $\h$ with respect to some 
$\Ad{\bG}$-invariant scalar product $(\cdot|\cdot)$ on $\g$,
then condition \eqref{naturallyreductive} reduces to
\begin{equation}
\label{naturallyreductiveII}
(\ad{g}h|k) = -(h|\ad{g}k)
\end{equation}
for all $g,h,k \in \mathfrak{p}$.
\end{remark}

\begin{lemma}
\label{lemCartanlike}
Any Cartan-like Riemannian homogeneous space $\bG/\bH$ is naturally
reductive.
\end{lemma}

{\bf Proof.}\hspace{.5em}
By the commutator relation $[\mathfrak{p},\mathfrak{p}] \subset \mathfrak{h}$,
we have $P\ad{g}h = 0$ for all $g,h \in \mathfrak{p}$. Thus
Eqn.~(\ref{naturallyreductive}) is obviously satisfied.
\hfill$\square$

\begin{theorem}[Coset Version]
\label{thmgeodesicsquotientspaceA}
Let $\bG/\bH$ be naturally reductive. Then $\bG/\bH$ is Riemannian 
homogeneous space such that all geodesics through $[G] \in \bG/\bH$ are
of the form
\begin{equation}
\label{quotientgeodesicsA}
t \mapsto \big[G \exp(t\Ad{G^{-1}}p)\big]
= \big[\exp(t p)  G\big]
\end{equation}
with $p \in \mathfrak{p}$.
\end{theorem}

{\bf Proof.}\hspace{.5em}
Obviously, $\bG/\bH$ is Riemannian homogeneous space by 
Proposition \ref{propinvariantquotientmetric}. For a proof for
Eqn.\eqref{quotientgeodesicsA} we refer to \cite{Arv03, CheegerEbin75}. 
\hfill$\square$

\medskip



The above result can be restated for an arbitrary naturally reductive
Riemannian homogeneous $\bG$-space. 

\begin{theorem}[Orbit Version]
\label{thmgeodesicsquotientspaceB}
Let $M$ be a homogeneous $\bG$-space with transitive group
action $\alpha: M \times \bG \to M$. Assume that $\bG/\bH_X$
is naturally reductive with decomposition
$\mathfrak{g} = \mathfrak{h}_X \oplus \mathfrak{p}_X$.
Then $M$ is a Riemannian homogeneous $\bG$-space such that
all geodesics through $Y = X \cdot G \in M$ are of the form
\begin{equation}
\label{quotientgeodesicsB}
t \mapsto Y \cdot \exp(t\Ad{G^{-1}}p)
\end{equation}
with $p \in \mathfrak{p}_X$. 
\end{theorem}

{\bf Proof.}\hspace{.5em}
The result is a straightforward consequence of Theorem
\ref{thmgeodesicsquotientspaceA}.
\hfill$\square$

\medskip

%
%

\begin{figure}[Ht!]
\includegraphics[scale=0.32]{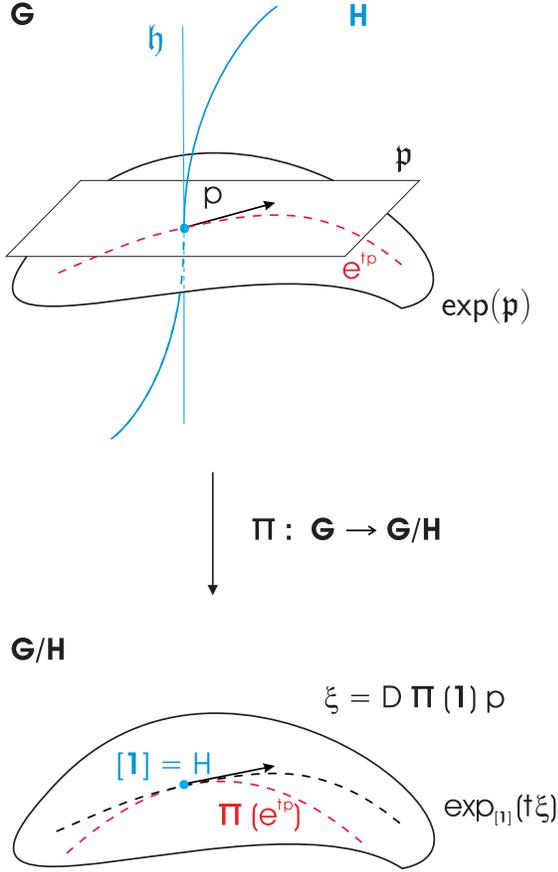}
\caption{\label{fig:geodesics} (Colour online)
Geodesics in reductive homogeneous spaces $\bG/\bH$. The tangent
vector $p\in \mathfrak p$ projects to the tangent vector $\xi$ at the
coset $[\unity] = \bH$. Note that only in {\em naturally} reductive
homogeneous spaces the geodesic in $\bG$ generated by $p$ projects
onto $\bG/\bH$ such that it coincides with the geodesic
of the projected tangent vector in the sense
$\Pi(e^{tp})= \exp_{[\unity]}(t\xi)$. In reductive homogeneous spaces
that are {\em not} naturally reductive, the projection yields 
only a first-order approximation at $[\unity] = \bH$ as shown in
the lower part, where $\Pi(e^{tp}) \neq \exp_{[\unity]}(t\xi)$.
}
\end{figure}

Thus, on {\em naturally reductive} spaces, the Riemannian exponential map
is particularly simple to compute.
By taking the basic framework of Ref.~\cite{ONeill83} further
to discuss geodesics, Figure \ref{fig:geodesics} illustrates
that only in naturally
reductive homogeneous spaces the geodesics project from the group
$\bG$ to give geodesics on $\bG/\bH$. Therefore, in this sense,
in naturally reductive homogeneous spaces
projection and exponentiation of tangent vectors commute.
However, on reductive homogeneous
spaces that are {\em not} naturally reductive, the problem is considerably
more involved.
A necessary and sufficient condition for $t \mapsto [G \exp(tg)]$
being a geodesic in $\bG/\bH$ can be found in \cite{Arv03,Kostant55}.

On the other hand, for numerical purposes it is often enough and even
advisible to approximate the Riemannian exponential map by another
computationally more efficient local parametrisation. Here, the map
\begin{equation}
\label{localpara}
\p \ni p \mapsto \Pi \circ l_G \circ \exp (\Ad{G^{-1}}p)
\end{equation}
might be a natural candidate, even if it fails to give the exact
Riemannian exponential map.
These issues are subject to current research, and
recent details can be found in \cite{jochen,AMS08}.
Figure \ref{fig:geodesics} also shows how in reductive
homogeneous spaces that are no longer naturally reductive, the
projected geodesic still provides a first-order
approximation to the geodesic generated by the projection of
the tangent vector.

%

\subsubsection*{Adjoint Orbits}

A prime example for naturally reductive homogeneous spaces is provided
by the adjoint action of a compact Lie group---a scenario which is of
major interest in the forthcoming applications. Therefore, we summarise
the previous results for the particular case of adjoint orbits.
Note that the adjoint action given by
$(X,G) \mapsto \Ad{G}X := GXG^{-1}$ is a \emph{left} action.
However, all previous statements and formulas remain valid 
{\em mutatis mutandis}, e.g., right cosets have to be replaced by left cosets,
etc.
 
\begin{corollary}
\label{adjointorbit}
Let $\bG$ be a Lie group with Lie algebra $\g$ and let $\bK \subset \bG$
be a compact subgroup with Lie algebra $\mathfrak{k}$ and bi-invariant
metric $\braket{\cdot}{\cdot}$. Moreover, let $\alpha: \g \times \bK \to \g$,
$(X,K) \mapsto \Ad{K}X := KXK^{-1}$ be the adjoint action of $\bK$ on $\g$
and denote by $\alpha_X: \bK \to \g$ the map $K \mapsto \Ad{K}X$. Then the
following assertions hold
\begin{enumerate}
\item[(a)]
The stabiliser group $\mathbf H:=\mathbf H_X$ of $X$
is a closed subgroup of $\bK$.
\item[(b)]
The coset space $\bK/\bH$ is diffeomorphic to the adjoint orbit
$\mathcal O(X):= \{\Ad{K}X \;|\; K \in \bK\}$ of $X$. In particular,
the map $\widehat{\alpha}_X: \bK/\bH \to \mathcal O(X)$,
$[K] \mapsto \Ad{K}X$ is a well-defined diffeomorphism
satisfying the commutative diagram
\begin{equation}
\label{isomorphismlawI-h}
\xymatrix{
\mathbf{K} \ar[d]_\Pi \ar[r]^{\alpha_X\quad} & {\mathcal O(X) \subset \g}\\
{\mathbf{K}/{\mathbf H}}\ar[ur]_{\widehat{\alpha}_X} &}
\end{equation}
\item[(c)]
Let $\h := \h_X$ denote the Lie algebra of $\bH$ and $\p$ be any
complementary space to $\h$ in $\mathfrak{k}$, then
$\D\alpha(\unity) = - \ad{X}$
is a surjective homomorphism with $\mathrm{ker}\,\ad{X} = \h$
and commutative diagram
\begin{equation}
\label{isomorphismlawII-h}
\xymatrix{
\g \ar[d]_{\D\Pi(\unity)} \ar[r]^{\D\alpha_X(\unity)\qquad} &
\rT_X\mathcal{O}(X) \subset \g\\
{\p \cong \mathfrak{k}/\h \;;}\ar[ur]_{\D\widehat{\alpha}_X([\unity])} &}
\end{equation}
Moreover, the tangent space of $\mathcal{O}(X)$ at $Y = \Ad{K}X$
is given by
\begin{equation*}
\rT_Y \mathcal{O}(X)  =  \ad{Y}\mathfrak{k} = 
\ad{Y}(\Ad{K^{-1}}\mathfrak{p}).
\end{equation*}
\item[(d)]
${\mathcal O(X)} \cong \bK/\bH$ is naturally reductive. More precisely,
$\mathfrak{p} := \h^{\perp}$ yields a  naturally reductive
decomposition of $\mathfrak{k}$ with $\Adr_\bH$-invariant scalar
product on $\p$ is given by the restriction of
$\braket{\cdot}{\cdot}$.
\item[(e)]
There is a well-defined $\alpha$-invariant metric on ${\mathcal O(X)}$
given by
\begin{equation}
\label{eq:admetric}
\braket{\xi}{\eta}_{\Ad{K}X} :=
\braket{p_{\xi}}{p_{\eta}}
\end{equation}
with $\xi = \ad{Y}\big(\Ad{K}p_{\xi}\big)$,
$\eta = \ad{Y}\big(\Ad{K}p_{\eta}\big)$ and
$p_{\xi},p_{\eta} \in \p$.
\item[(f)]
All geodesics through $Y = \Ad{K}X \in \mathcal{O}(X)$ with respect
to the metric given in part (e) are of the form
\begin{equation}
\label{eq:adgeodesic}
t \mapsto \Ad{\exp(t \Ad{K}p)}Y
\end{equation}
with $p \in \p$.
\end{enumerate}
\end{corollary}

{\bf Proof.}\hspace{.5em}
Part (a) and (b) follow immediately from Theorem \ref{thmorbit}
and Corollary \ref{cor:orbit}. 

\medskip

(c) For $k \in \mathfrak{k}$ we have
$$
\frac{\rm d}{{\rm d} t} \Ad{\exp(tk)}X \,\Big|_{t=0}
= -\ad{X}k
$$
and thus $\D\alpha(\unity) = - \ad{X}$. All other statemants
are again consequences of Theorem \ref{thmorbit}.

\medskip

(d) First, observe that the bi-invariance of $\braket{\cdot}{\cdot}$
implies that $\mathfrak{k} = \h \oplus \p$ with $\p := \h^{\perp}$
is reductive. Now, let $P$ denote the orthogonal projection
onto $\p$. In turn, the bi-invariance of $\braket{\cdot}{\cdot}$
yields
$$ 
\braket{P\ad{g}h}{k} = \braket{\ad{g}h}{k} = - \braket{h}{\ad{g}k}
= - \braket{h}{P\ad{g}k}
$$
for all $g,h,k \in \p$, cf. Proposition \ref{propalgebravsgroup}.
Therefore, ${\mathcal O(X)} \cong \bK/\bH$ is naturally reductive.

\medskip

(e) Let $Y \in \mathcal{O}(X)$ and $\widetilde{K} \in \bK$.
A straightforward calculation using the identities
$\D\alpha_{\widetilde{K}}(Y) \xi = \Ad{\widetilde{K}} \xi$
for $\xi \in \rT_Y \mathcal{O}(X)$
and
$\Ad{\widetilde{K}}(\ad{Y} k) =
\ad{\Ad{\widetilde{K}}Y}(\Ad{\widetilde{K}}k)$
for all $k \in \mathfrak{k}$ yields the required invariance.

\medskip

Part (f) follows immediately from Theorem \ref{thmgeodesicsquotientspaceB} 
and the identity $\h_Y = \Ad{K}\h_X$ for $Y = \Ad{K}X$ which implies
$\h_Y^{\perp} = \Ad{K}\p$. \hfill $\square$

\subsubsection*{Gradient Flows on Riemannian Homogeneous Spaces}

Applying the previous results on gradient flows to
quality functions $\widehat{f}$ 
on Riemannian homogeneous spaces $\bG/\bH$, we obtain by the
$\bG$-invariance of the Riemannian metric---similar to
(\ref{Liegrad})---the gradient equality
\begin{equation}
\label{quotientgrad}
\grad\, \widehat{f}\,([G]) = \mathrm{D}\widehat{r}_G([\unity])
\grad (\widehat{f}\circ\widehat{r}_G)\,([\unity])
\end{equation}
for all $G \in \bG$, where $\widehat{r}_G$ denotes the mapping
$[G'] \mapsto [G'G]$. 
Therefore, the gradient of $\widehat{f}$ is completely determined
by
\begin{equation}
\label{quotientgradII}
G \mapsto \grad (\widehat{f}\circ\widehat{r}_G)([\unity])
\in \mathfrak{p}.
\end{equation}
However,  Eqn.~\eqref{quotientgradII} does not induce a
mapping from $\bG/\bH$ to $\mathfrak{p}$, as in general
\begin{equation*} 
\grad (\widehat{f}\circ\widehat{r}_G)([\unity]) \neq
\grad (\widehat{f}\circ\widehat{r}_{HG})([\unity])
\end{equation*}
for $H \in \bH \setminus \{\unity\}$.
The corresponding gradient system reads 
\begin{equation}
\label{quotientgradflowII}
\dot{[G]}  = \mathrm{D}\widehat{r}_G([\unity])
\grad (\widehat{f}\circ\widehat{r}_G)([\unity]).
\end{equation}
For analysing the asymptotic behaviour of
Eqn.~(\ref{quotientgradflowII}), Subsection \ref{sec:mfds} again
provides the appropriate tools. For instance, if $\bG/\bH$ is compact
we have.

\begin{corollary}
\label{corIIpointwiseconvergence}
Let $\bG/\bH$ be a compact Riemannian homogeneous space and let
$\widehat{f}: \bG/\bH \to \R$ be real analytic. Then any solution
of Eqn.~(\ref{quotientgradflowII}) converges to a critical point
of $\widehat{f}$ for $t \to +\infty$.
\end{corollary}

{\bf Proof.}\hspace{.5em}
This follows immediately from Proposition \ref{propgradomegalimit}
and Theorem \ref{thmIIpointconvergence} as a Riemannian homogeneous
space constitutes always a real analytic Riemannian manifold,
cf. \cite{Helgason78, DuiKolk00}.
\hfill$\square$

\medskip
 
Finally, we return to our starting point 
and ask for the relation between (\ref{Liegradflow}) and
(\ref{quotientgradflowII}) in the case of an $\bH$-equivariant
quality function $f$. Here, $f$ induces a quality function
$\widehat{f}$ on $\bG/\bH$ via 
\begin{equation}
\label{inducedqualityfunction}
\widehat{f}([G]) := f(G) 
\end{equation}
for all $G \in \bG$.
Moreover, assume $\bG$ carries a bi-invariant metric
$\braket{\cdot}{\cdot}$ and $\bG/\bH$ is a homogeneous space
with reductive decomposition $\mathfrak{g}= \mathfrak{h} \oplus \mathfrak{p}$
and $\mathfrak{p} := \mathfrak{h}^{\perp}$. This implies that the restriction
of $\braket{\cdot}{\cdot}$ to ${\mathfrak{p}}\times{\mathfrak{p}}$ is
$\Ad{\bH}$-invariant. Now, the identity $\widehat{f} \circ \Pi = f$
yields
\begin{equation}
\label{eq:D-equivariant}
\mathrm{D}\,\widehat{f}([G]) \cdot \mathrm{D}\,\Pi(G) = \mathrm{D}\,f(G) 
\quad\mbox{for all $G \in \bG$.}
\end{equation}
and hence
\begin{equation}
\label{gradequivariantI}
\big(\mathrm{D}\,\Pi(G)\big)^* \grad\widehat{f}([G])  = 
\grad f(G) 
\end{equation}
for all $G \in \bG$, where $\Pi$ denotes the canonical projection and 
$(\cdot)^*$ the adjoint mapping. By identifying $\mathfrak{p}$ with the
tangent space of $\bG/\bH$ at $[\unity]$, the map
$\mathrm{D}\,\Pi(\unity)$ represents the orthogonal projector
$h+p \mapsto p$ for $h \in \mathfrak{h}$ and $p \in \mathfrak{p}$.
Hence, we obtain
\begin{equation}
\label{eq:D-proj-p}
\mathrm{D}\,\Pi(\unity)  \big(\mathrm{D}\,\Pi(\unity)\big)^* =
\mathrm{id}_{\mathfrak{p}}.
\end{equation}
In the same way, using the identity
$\Pi \circ r_G = \widehat{r}_G \circ \Pi$, one shows
\begin{equation}
\label{eq:tan-proj-p}
\mathrm{D}\,\Pi(G)  \big(\mathrm{D}\,\Pi(G)\big)^* =
\mathrm{id}_{\rT_{[G]}\bG/\bH}
\end{equation}
for all $G \in \bG$. Thus (\ref{gradequivariantI}) yields
\begin{equation}
\label{gradequivariantII}
\grad\widehat{f}([G])  = \mathrm{D}\,\Pi(G) \grad f(G) 
\end{equation}
for all $G \in \bG$. Therefore, we have proven the following result:

\begin{theorem}
\label{thmgradientflowequivariant}
Suppose $\bG/\bH$ satisfies the above assumptions and $f: \bG \to \R$
is a $\bH$-equivariant quality function with induced quality function
$\widehat{f}: \bG/\bH \to \R$. Then the canonical
projection of the gradient flow of Eqn.~(\ref{Liegradflow}) onto
$\bG/\bH$ yields the gradient flow of Eqn.~(\ref{quotientgradflowII}),
i.e., if $G(t)$ is a solution of Eqn.~(\ref{Liegradflow})
then $\Pi(G(t))$ is one of Eqn.~(\ref{quotientgradflowII}) . 
\end{theorem}




\subsection{Examples}\label{sec:ex}

Often practically relevant quality functions take the form of a linear
functional restricted to an adjoint orbit $\mathcal O(X)$. For instance,
in quantum dynamics the unitary orbit
$\mathcal O(A):= \{UAU^\dagger\;|\; U \in SU(N)\}$ 
of an initial state $A$ plays a central role, because it defines the
largest reachability set under
closed Hamiltonian dynamics. Then the set of feasible expectation values
is such a linear map,
since it is the projection onto an observable $C$ in the sense
of a Hilbert-Schmidt scalar product. These expectation values can be
generalised to
arbitrary complex square matrices $A,C\in \C^{N\times N}$ such as to
coincide with the $C$"~{\em numerical range}
\begin{equation}\label{eqn:RCA}
W(C,A):= \{ \tr (C^\dagger UAU^\dagger)\;|\; U\in SU(N)\}\quad.
\end{equation}
As $C$"~numerical ranges are well established in the mathematical
literature \cite{GS-77,Li94},
in the sequel we will adopt the notation.

Note that finding the maximum absolute value, i.e., the
$C$"~{\em numerical radius}
\begin{equation}\label{eqn:WCA}
r(C,A):=\maxover{U\in SU(N)} | \tr \{C^\dagger UAU^\dagger\}|
\end{equation}
is straightforward for Hermitian $A,C$ (it amounts to sorting the
respective eigenvalues, cf.~Corollary \ref{cor:doublebracket}), 
while for arbitrary complex $A,C$ there is no general analytical
solution. Moreover, when restricting to local unitary
operations $K \in SU_{\rm  loc}(2^n):= SU(2)^{\otimes n}$,
the maximisation task becomes non-trivial even for Hermitian
$A,C$ \cite{WONRA_tosh,WONRA_dirr}.

Therefore, we now illustrate the previous theory by gradient flows
on the entire unitary group $SU(2^n)$, on the local unitary group
$SU(2)^{\otimes n}$ as well as their adjoint orbits.



\subsubsection*{Worked Example: $SU(N)$}

Recall that $SU(N)$ is a compact connected Lie group of real
dimension $N^2-1$. Its Lie algebra, i.e. its tangent space at
the identity is given by set $\mathfrak{su}(N)$ of all skew-Hermitian
matrices $\Omega$ with ${\rm tr}\, \Omega = 0$, i.e.,
\begin{equation}
\mathfrak{su}(N) :=
\{\Omega \in {\C}^{N \times N} \;|\;
\Omega^{\dagger} = - \Omega,\, {\rm tr}\, \Omega = 0\}.
\end{equation}
So elements $\Omega \in\mathfrak{su}(N)$ relate to Hamiltonians $H$
via $\Omega = iH$.
The tangent space at an arbitrary element $U \in SU(N)$ is
\begin{equation}
\rT_{U}SU(N) = \su(N) U = \{\Omega U\,|\,\Omega \in \mathfrak{su}(N)\},
\end{equation}
cf. Eqn.~\eqref{tangenspacerelation}.
Moreover, let $SU(N)$ be endowed with the bi-invariant 
Riemannian metric 
\begin{equation}
\label{SUmetric}
\braket{\Omega U}{\Xi U}_U :=
\tr(\Omega^\dagger\Xi),
\end{equation}  
defined on the tangent spaces $\rT_U SU(N) $, cf. Eqn~\eqref{invariance}.

Now set
\begin{eqnarray*}
g:SU(N) &\to& \C^{N \times N},\quad g(U) := \Cd UAU^\dagger \\
f:SU(N) &\to& \R,\quad
f(U) := \Re \tr \{\Cd UAU^\dagger \} 
\end{eqnarray*}
For computing the tangent map of $g$, we exploit the fact that
$SU(N)$ is an embedded submanifold of $\C^{N \times N}$. Therefore,
the tangent map is obtained by restricting the ordinary Fr{\'e}chet
derivative $\D\!g(U)$ to the tangent space $\rT_U SU(N)$,
cf. Appendix A. 
Thus, by applying the product rule, one easily finds
\begin{equation}
\begin{split}
\D\! g(U)(\Omega U)
& = \Cd \Omega UAU^\dagger + \Cd UA(\Omega U)^\dagger\\
&= \Cd \Omega UAU^\dagger  - \Cd UAU^\dagger\Omega \;.
\end{split}
\end{equation}
%
Now, the chain rule 
as well as the short-hand notations $\tilde A:= UAU^\dagger$ and 
$[\cdot,\cdot]_S$ to denote the skew-Hermitian part of the commutator
$[\cdot,\cdot]$ give
\begin{equation}\label{eqn:gradfU-worked}
\begin{split}
\D f(U)(\Omega U)
&= \D (\Re\,\tr)(g(U)) \circ \D(g(U))(\Omega U)\\
&= \Re \tr \{\Cd \Omega \tilde A - \Cd \tilde A \Omega \}\\
&= \Re \tr \{[\tilde A, \Cd] \Omega \}\\
&= \tr \{[\tilde A, \Cd]_S \Omega \}\\
&= \braket{[\tilde A, C]_S^\dagger}{\Omega}\\
&= \braket{[\tilde A, \Cd]_S^\dagger U}{\Omega U} \,,
\end{split}
\end{equation}
where the last identity explicitly invokes the right-invariance
of the Riemannian metric on $SU(N)$, cf. Eqn.~\eqref{SUmetric}.
Now, identifying the above expression with
\begin{equation}
Df(U)(\Omega U) = \braket{\grad f(U)}{\Omega U}
\end{equation}
one gets the gradient vector field
\begin{equation}\label{eqn:gradfU}
\grad f(U) = [\tilde A, \Cd]_S^\dagger\; U
\end{equation}
and thus the gradient system
\begin{equation}
\dot U = \grad f(U) = - [\tilde A, \Cd]_S \; U \quad.
\end{equation}
By making use of the Riemannian exponential, see Eqns.~\eqref{eqn:RiemExp2}
and \eqref{eq:gradII}, we finally arrive at the discretisation 
\begin{equation}
U_{k+1} = e^{-\alpha_k [U_k A U_k^\dagger , C]_S}\, U_k,
\end{equation}
where $\alpha_k \geq 0$ denotes an appropriate step size.


\subsubsection*{Gradient Flows on $SU(N)$}

Consider a fully controllable system on $SU(N)$ such as
\begin{equation}
\label{eq:su-system}
(\Sigma)\quad\quad
\dot{U}(t) = -\ri\Big(
H_{d}+\sum_{k=1 \atop \alpha \in \{x,y\}}^{n}u_{k}(t)H_{k,\alpha}\Big) U(t)
\end{equation}
with $H_{d} := \sum_{k < l} J_{kl} \sigma_{k,z}\sigma_{l,z}$,
$H_{k,\alpha} := \sigma_{k,\alpha}$, $\alpha \in \{x,y\}$
and connected spin-spin coupling topology, cf. \cite{TOSH-Diss}.
Here and in the sequel, 
\begin{equation}
\label{eqn:Pauli_sigma}
\sigma_x :=
\begin{bmatrix}
0 & {\rm i}\\{\rm i} & 0
\end{bmatrix},
\quad
\sigma_y :=
\begin{bmatrix}
0 & -1\\ 1 & 0
\end{bmatrix},
\quad
\sigma_z :=
\begin{bmatrix}
{\rm i}  & 0\\ 0 &{\rm -i}
\end{bmatrix}
\end{equation}
denote the Pauli matrices, which form an orthogonal basis of
${\mathfrak{su}(2)}$. Moreover, $\sigma_{k,\alpha}$,
$\alpha \in \{x,y,z\}$ is defined by
\begin{equation}
\label{sigma_k}
\sigma_{k,\alpha} :=
\unity_2 \otimes \cdots \otimes \unity_2 \otimes
\sigma_{\alpha} \otimes \unity_2 \otimes \cdots \otimes \unity_2,
\end{equation}
where the term $\sigma_{\alpha}$ appears in the $k$-th position of
the Kronecker product and $\unity_2$ denotes the $2\!\times\!2$-identity
matrix.

As $(\Sigma)$ is fully controllable by assumption,  
the entire group $SU(N)$ can be generated by evolutions
under the Hamiltonian of the system plus the available controls.
If $A$ is an initial density operator or a matrix collecting its
signal-relevant terms, the orbit of the canonical semigroup action
of $(\Sigma)$ on $A$ yields in this case the entire unitary orbit
$\mathcal{O}(A) = UAU^{\dagger}$.
Recall its \/`projection\/' on some observable $C$ (or its
signal-relevant terms) forms
the {\em $C$-numerical range of $A$} (see Eqn.\ref{eqn:WCA}).

In this setting, there are two geometric optimisation tasks
of particular practical relevance as they determine maximal signal
intensity in coherent spectroscopy \cite{Science98}.
\begin{enumerate}
\item[(a)]
{\em Find all points on the unitary orbit of $A$ that 
minimise the Euclidean distance to $C$.}
\item[(b)]
{\em Find all points on the unitary orbit of $A$ that
minimise the angle to the $1$-dimensional, complex subspace
spanned by $C$.}
\end{enumerate}

Clearly, the distance
\begin{equation}
\begin{split}
\Vert UAU^{\dagger} &- \;C \Vert_2^2 = \\[1mm]
   & \Vert A \Vert_2^2 + \Vert C \Vert_2^2 - 2 {\Re}\{ {\rm tr}
\left(\Cd UAU^{\dagger}\right)\}
\end{split}
\end{equation}
is minimal if the overlap ${\Re}\{ {\rm tr} \left(\Cd UAU^{\dagger}\right)\}$
is maximal.
Moreover, making use of the definition of the angle between $1$-dimensional
complex subspaces 
\begin{equation}
\phantom{X}\quad
\cos^2 ( \measuredangle\{ UAU^{\dagger},C\} ) \,
        :=\,\frac{|\,{\rm tr}\{\Cd UAU^{\dagger}\}|^2}{{\|A\|}_2^2\cdot{\|C\|}_2^2}\; .
\end{equation}
problem (b) is equivalent to maximising the function
$|{\rm tr} (\Cd UAU^{\dagger})|$. Its maximal value is the
$C$-numerical radius of $A$ (see Eqn.~\ref{eqn:RCA}). 
Obviously, $r_C(A) \leq {\|A\|}_2 \cdot {\|C\|}_2$ with equality if and
only if $UAU^{\dagger}$ and $C$ are complex collinear for
some $U \in  SU(N)$.
Note that the two tasks (a) and (b) are equivalent whenever
the $C$-numerical range forms a circular disk in the complex plane
(centred at the origin); conditions for circular symmetry have
been characterised in \cite{Li91}.

Extending concepts of Brockett \cite{Bro88+91} from the orthogonal
to the special unitary group \cite{Science98, TOSH-Diss, NMRJOGO},
the above optimisation problems (a) and (b) can be treated by the
previously presented gradient-flow methods, cf. also \cite{HM94,Bloch94}.

For fixed matrices $A,C \in \mathbb{C}^{N \times N}$ define
\begin{eqnarray}\label{function1}
f_1 &:& SU(N) \to {\R},\quad f_1(U):= {\rm Re}\, {\rm tr} (\Cd UAU^{\dagger})
\end{eqnarray}
and 
\begin{eqnarray}
\label{function2}
f_2 &:& SU(N) \to {\R}, \quad f_2(U):= |{\rm tr} (\Cd UAU^{\dagger})|^2.
\end{eqnarray}

Observe that the distance problem (a) is solved by maximising
$f_1$, while the angle problem is solved for maximal $f_2$.


Now, the differential and the gradient of $f_1$  with
respect to the bi-invariant Riemannian metric Eqn.~\eqref{SUmetric}
is precisely given by the previous example as
\begin{eqnarray*}
Df_1(U)(\Omega U)
& = &
{\rm Re}\, {\rm tr}([UAU^{\dagger},\Cd]\Omega),\\
\grad f_1(U) & = & [UAU^{\dagger}, \Cd]_S^\dagger\; U.
\end{eqnarray*}
The differential and the gradient of $f_2$ can be obtained in the same
manner as 
\begin{eqnarray*}
Df_2(U)(\Omega U)
& = &
{{\rm tr}(\Cd UAU^{\dagger})}^*
\cdot{\rm tr}([UAU^{\dagger},\Cd]\Omega)\\
&& -\,
{\rm tr}(\Cd UAU^{\dagger})\cdot
{\rm tr}([UAU^{\dagger},\Cd]^{\dagger}\Omega),\\
\grad f_2(U) & = & 
2\big( {f_2(U)}^* \cdot [UAU^{\dagger}, \Cd]\big)_S^\dagger U.
\end{eqnarray*}
This yields the following result. 

\begin{theorem}
\label{thm:gradSU}
The gradient systems of $f_\nu$, $\nu=1,2$ with respect to the
bi-invariant Riemannian metric (\ref{SUmetric}) are given
by
\begin{eqnarray}
\label{gradflow12}
\dot U & = & \Omega_\nu(U)U
\end{eqnarray}
with 
\begin{equation}
\label{grad12}
\Omega_1(U) := [UAU^{\dagger}, \Cd]_S^\dagger
\quad\text{and}\quad
\Omega_2(U) :=
2\big( {f_2(U)}^* \cdot [UAU^{\dagger}, \Cd]\big)_S^\dagger.
\end{equation}
respectively.
Each solution of (\ref{gradflow12}) 
converges to a respective critical point for $t\to +\infty$. Thereby, the critical
points of $f_{\nu}$ are characterised by $\Omega_{\nu}(U) = 0$, $\nu=1,2$.
\end{theorem}

{\bf Proof:} \hspace{.5em}
The above computations immediately yield Eqn.~\eqref{gradflow12}. 
As $f_{\nu}$, $\nu = 1,2$ are real analytic, the
convergence of each solution to a critical point is guaranteed by
Proposition \ref{propgradomegalimit} and Theorem
\ref{thmIIpointconvergence}, cf. \cite{NMRJOGO}.
\hfill$\square$

\medskip

An implementable numerical integration scheme for the
above gradient systems making use of the Riemannian exponential,
see Eqns.~\eqref{eqn:RiemExp2} and \eqref{eq:gradII}, is given by
\begin{equation}
\label{iterate}
U_{k+1}^{(\nu)}\,=\,
\exp\big(\alpha_k^{(\nu)} \Omega_{\nu}(U_k^{(\nu)})\big)\,U_k^{(\nu)},
\quad U_0 = \unity_N
\end{equation}
A suitable choice of step sizes $\alpha^{(\nu)}_k>0$ ensuring convergence
can be found in \cite{Science98, TOSH-Diss, NMRJOGO}.
Generically, it drives $U_{k}^{(\nu)}$ into final states attaining the
maxima of the quality functions $f_\nu$, $\nu=1,2$.
However, there is no guarantee that the gradient flows always reach
the {\em global} maxima.
Standard numerical integration procedures such as {\em e.g.} the Euler
method are not applicable here as they would not preserve unitarity.


\subsubsection*{Gradient Flows on the Local Subgroup $\SUloc(2^n)$}
\label{gradflowonSUloc}

The quality functions 
introduced in the previous subsection may be restricted to the
subgroup of local action, i.e. to
\begin{equation}
\SUloc(2^n) :=
\underbrace{SU(2)\otimes\cdots\otimes SU(2)}_{\mbox{$n$-times}}
\subset SU(2^n).
\end{equation} 
The Lie subalgebra to $\SUloc(2^n) \subset \su(2^n)$ can be specified by
\begin{equation*}
\label{su2tensor}
\begin{split}
\suloc(2^n) & :=\\
\bigg\{\sum_{j=1}^{n}\unity_2 & \otimes \cdots \otimes \unity_2
\otimes \Omega_j \otimes \unity_2\otimes \cdots \otimes \unity_2
\,\Big|\, \Omega_j \in \mathfrak{su}(2)\bigg\},
\end{split}
\end{equation*}
where the term $\Omega_j \in \mathfrak{su}(2)$ appears at the $j$-th
position, cf.~Eqn.~\eqref{sigma_k}.
Therefore, the tangent space of $SU_{\mathrm{loc}}(2^n)$ at an
arbitrary element $U$ is given by
\begin{equation}
\label{TUnC}%
{\rm T}_U SU_{\mathrm{loc}}(2^n) =
\{\Omega U \;|\; \Omega \in \suloc(2^n)\}.
\end{equation}%
Finally, $\SUloc(2^n)$ is endowed with the bi-invariant
Riemannian metric induced by $SU(2^n)$, i.e.
\begin{equation}\label{RiemannMetric}
\langle\Omega U,\Xi U\rangle_U := \tr(\Omega^\dagger\Xi)
\end{equation}%
for $\Omega U,\Xi U \in {\rm T}_U SU_{\mathrm{loc}}(2^n)$.

\begin{lemma}
\label{lem:project}
Let $\bH \subset GL(N,\C)$ be any closed subgroup with Lie algebra
$\h \subset \mathfrak{gl}(N,\C) := \C^{N \times N}$. Moreover let
$h_1, \dots, h_m$ be a real orthonormal basis of $\h$ with respect
to the real scalar product
\begin{equation}
\label{glrealproduct}
(g_1|g_2) := \Re \tr\,(g_1^{\dagger}g_2),
\quad g_1,g_1 \in \C^{N \times N},
\end{equation}
i.e. $\mathrm{span}_{\R}\{h_1, \dots, h_m\} = \h$ and
$(h_i|h_j) = \delta_{ij}$.
\begin{enumerate}
\item[(a)]
Then the orthogonal projection
$P : C^{N \times N} \to C^{N \times N}$ onto $\h$ is given by 
\begin{equation}
\label{projector2}
g \mapsto P g :=
\sum\limits_{j=1}^m \Re \tr\{h_j^\dagger g\} h_j\; .
\end{equation}
\item[(b)]
The orthogonal projection
$P^{\perp}:C^{N \times N} \to C^{N \times N}$
onto the orthogonal complement $\h^{\perp}$ is given by
\begin{equation}
g \mapsto P^{\perp}g = g - P g.
\end{equation}
\end{enumerate}
\end{lemma}

{\bf Proof:}\hspace{.5em}
Both, (a) and (b) are basic and well-known facts from linear algebra.
\hfill$\square$

\begin{remark}
For the unitary case, i.e. for $\h \subset \su(N)$, the real part
in Eqn.~\eqref{projector2} can be neglected and the projector $P$
can be rewritten in the more convenient matrix form $\widehat{P}$ as
\begin{equation}
\label{proj:dirac1}
\widehat{P} := \sum\limits_{j=1}^m \vec (h_j) \vec (h_j) ^{\dagger},
\end{equation}
where the terms $\vec (h_j) \vec (h_j) ^{\dagger}$ represent the 
rank-$1$ projectors $P_j = \ketbra{h_j}{h_j}$ in vec-notation.
\end{remark}

\begin{corollary}
\label{cor:projection}
The orthogonal projection $P: C^{N \times N} \to C^{N \times N}$
onto $\suloc(2^n)$ with respect to (\ref{glrealproduct}) is given by
\begin{equation}
\begin{split}
P g := \frac{1}{2^n}\sum_{k=1}^{n}\Big(
&
\Re\big({\rm tr}(g^{\dagger}X_k)\big)X_k +
\Re \big({\rm tr}(g^{\dagger}Y_k)\big)Y_k
\\ \label{locprojection}
+ &
\Re \big({\rm tr}(g^{\dagger}Z_k)\big)Z_k
\Big),
\end{split}
\end{equation}
where $X_k, Y_k$ and $Z_k$ are defined by, cf. Eqn.~\eqref{sigma_k}
\begin{equation}
\label{XYZ}
X_k := \sigma_{k,x},
\quad
Y := \sigma_{k,y},
\quad
Z := \sigma_{k,z}.
\end{equation}
\end{corollary}

{\bf Proof:}\hspace{.5em}
This follows straightforwardly from the orthogonality of the set
$\{X_k, Y_k, Z_k \;|\; k = 1, \dots, n$\}
and Lemma \ref{lem:project}. \hfill$\square$

\medskip

\begin{theorem}%
\label{thm:gradSUloc}
Let $f_{\rm loc}$ be the restriction of (\ref{function1})
to $SU_{\mathrm{loc}}(2^n)$.
\begin{enumerate}
\item[(a)]
The gradient of $f_{\rm loc}$ with respect (\ref{RiemannMetric}) and
the corresponding gradient system are given by
\begin{equation}%
\grad f_{\rm loc}(U)=P([\Cd,UAU^\dagger])U
\end{equation}%
and
\begin{equation}
\label{SUlocgradflow}
\dot{U}=P([\Cd,UAU^\dagger])U,
\end{equation}%
respectively, where $P$ denotes the orthogonal projection
$P: \gl(2^n,\C) \to \gl(2^n,\C)$ onto $\suloc(2^n)$.
More explicitly, (\ref{SUlocgradflow}) is equivalent to a
system of $n$ coupled equations
\begin{equation}
\label{coupledgradientflow}
\dot{U}_k = \Omega_kU_k, \quad k=1,\cdots,n
\end{equation}
on $SU(2)$, where
\begin{equation*}
\begin{split}
\Omega_k = \frac{1}{2^n} \Big(
&
\Re({\rm tr}([\Cd,UAU^\dagger]^{\dagger} X_k))X \\
+ &
\Re({\rm tr}([\Cd,UAU^\dagger]^{\dagger} Y_k))Y \\
+ &
\Re({\rm tr}([\Cd,UAU^\dagger]^{\dagger} Z_k))Z \Big).
\end{split}
\end{equation*}
Each solution of (\ref{SUlocgradflow}) 
converges for $t\to\pm\infty$ to a critical point of $f_{\rm loc}$.
Thereby, the critical points are characterised by
\begin{equation}
\label{CriticalCond1}
P([\Cd,UAU^\dagger])=0.
\end{equation}
\item[(b)]
The Hessian form $\Hess f_{\rm loc}(U)$ and the Hessian operator
$\Hessop f_{\rm loc}(U)$ of $f_{\rm loc}$ at $U$ are given by
\begin{eqnarray}
\lefteqn{\Hess f_{\rm loc}(U)(\Omega U,\Xi U) \; = \;}
\nonumber\\
& = &
\frac{1}{2}
\Big(
\Re \big(\tr(\Omega^{\dagger}[\Cd,[\Xi,UAU^{\dagger}]])\big)\\
&&
+ \Re \big(\tr(\Omega^{\dagger}[UAU^{\dagger},[\Xi,\Cd]])\big)
\Big).
\nonumber
\end{eqnarray}
and
\begin{eqnarray}
\Hessop f_{\rm loc}(U)\Omega U & = & \big(\mathbf{S}(U)\Omega\big)U,
\end{eqnarray}
respectively, with
$\Omega \in \mathfrak{su}_{\rm loc}(2^n)$ and
\begin{eqnarray}
\lefteqn{S(U)\Omega \; := \;}\\
&&
\frac{1}{2}P
\Big(
[\Cd,[\Omega,UAU^{\dagger}]] +[UAU^\dagger,[\Omega,\Cd]]
\Big).
\nonumber
\end{eqnarray}
\item[(c)]
For all initial points $U_0 \in \SUloc(2^n)$ the discretization
scheme
\begin{equation}
\label{DiscretizationScheme1}%
U_{k+1}:=\exp\left( \alpha_k P\big([\Cd,U_kAU_k^\dagger]\big)
\right)U_k
\end{equation}%
with step size $\alpha_k := $
\begin{eqnarray}
\label{alpha}
&&
{\textstyle\frac{\big\|P \left([\Cd,U_kAU_k^\dagger]\right) \big\|^2_{\phantom{|}} }
{\big\|
\left[\Cd,P \left([\Cd,U_kAU_k^\dagger]\right)\right]
\big\|^{\phantom{|}}
\cdot\;
\big\|
\left[[P\left([\Cd,U_kAU_k^\dagger]\right),U_kAU_k^\dagger\right]
\big\|} }\quad
\end{eqnarray}
converges to the set of critical points of $f_{\rm loc}$.
\end{enumerate}
\end{theorem}

{\bf Proof:} \hspace{.5em} The subsequent arguments follow
our conference report \cite{loc_WCA}, which also contains a complete proof
for the flow on the entire groups such as $SU(2^n)$.
\begin{enumerate}
\item[(a)]
Since $\SUloc(2^n)$ is a closed subgroup of $SU(2^n)$, it is also an
embedded Lie subgroup and thus a submanifold of $SU(2^n)$, cf.
Remark \ref{remmatrixLiegroup}. Therefore, the gradient of
$f_{\rm loc}$ is well-defined by (\ref{restrictedgrad}).
Furthermore, by (\ref{projector}) and (\ref{grad12}) we obtain  
\begin{eqnarray*}
\lefteqn{\grad f_{\rm loc}(U) \; = \; }\\
& = &
P(\grad f_1(U)) \; = \; P\big([UAU^{\dagger},\Cd]^{\dagger}\big) U\\
& = &
P\big([\Cd,UAU^{\dagger}]\big) U,
\end{eqnarray*}
where the last equality follows from
$P([UAU^{\dagger},\Cd]^{\dagger}) = - P([UAU^{\dagger},\Cd])$
and the skew-symmetry of the commutator. Moreover,
Eqn.~(\ref{coupledgradientflow}) is derived by Corollary
\ref{cor:projection} and the identity
\begin{eqnarray*}
\lefteqn{\quad\frac{\mathrm{d}}{\mathrm{d}t}
\Big(U_1(t) \otimes \dots \otimes U_n(t)\Big)
\; = \;}\\\\
&&
\Bigg(\sum_{k=1}^{n}
\unity_2 \otimes \dots \otimes
\dot{U}_k(t)U^{-1}_k(t)
\otimes \dots \otimes \unity_2
\Bigg)\times\\\\
&&
\times\Big(U_1(t) \otimes \dots \otimes U_n(t)\Big).
\end{eqnarray*}
Now, compactness of $\SUloc(2^n)$ and real analyticity of
$f_{\rm loc}$ imply that each solution converges to critical
points for $t\to +\infty$, cf. Proposition \ref{propgradomegalimit}
and Theorem \ref{thmIIpointconvergence}.
\item[(b)]
By (\ref{hessianIII}),  the Hessian of $f_{\rm loc}$ at $U$
is determined by evaluating the second derivative of
$\varphi := f \circ \gamma$ at $t=0$, where $\gamma$ is any geodesics.
This yields
\begin{eqnarray}
\lefteqn{\Hess f_{\rm loc}(U)(\Omega U,\Omega U)
\; := \; \varphi''(0)}
\nonumber\\
\label{quadraticform}
& = &
\Re \big(\tr(\Cd[\Omega,[\Omega,UAU^{\dagger}]])\big),
\end{eqnarray}
for $\Omega\in \suloc(2^n)$. The Hessian then is obtained from
quadratic form (\ref{quadraticform}) by a standard polarisation
argument, i.e.
\begin{eqnarray*}
\lefteqn{\Hess f_{\rm loc}(U)(\Omega U,\Xi U) \; = \;}\\
& = &
\frac{1}{2}
\Big(\Re \big(\tr(\Cd[\Omega,[\Xi,UAU^{\dagger}]])\big)\\
&&
+\Re \big(\tr(\Cd[\Xi,[\Omega,UAU^{\dagger}]])\big)\Big).
\end{eqnarray*}
Finally, by the identity $\tr [X,Y]Z = - \tr Y[X,Z]$ we conclude
\begin{eqnarray*}
\lefteqn{\Hess f_{\rm loc}(U)(\Omega U,\Xi U) \; = \;}\\
& = &
\frac{1}{2}
\Big(
\Re \big(\tr(\Omega^{\dagger}[\Cd,[\Xi,UAU^{\dagger}]])\big)\\
&&
+ \Re \big(\tr(\Omega^{\dagger}[UAU^{\dagger},[\Xi,\Cd]])\big)
\Big).
\end{eqnarray*}
Therefore, the Hessian operator of $f_{\rm loc}$ at $U$ is
given by 
\begin{eqnarray}
\Hessop f_{\rm loc}(U)\Omega U & = & \big(\mathbf{S}(U)\Omega\big)U
\end{eqnarray}
with
$\Omega\in \mathfrak{su}_{\rm loc}(2^n)$ and 
\begin{eqnarray}
\lefteqn{{\mathbf{S}}(U)\Omega \; := \;}\\\nonumber
&&
\frac{1}{2}P
\Big(
[\Cd,[\Omega,UAU^{\dagger}]] +[UAU^\dagger,[\Omega,\Cd]]
\Big).
\end{eqnarray}
\item[(c)]
Estimating the second derivative 
\begin{equation}%
\varphi''(t)=
\Re \big(\tr([\Cd,\Omega][\Omega,\e^{t\Omega}UAU^\dagger\e^{-t\Omega}])\big)
\end{equation}
for 
$\Omega:=\grad f_{\rm loc}(U) = P([\Cd,UAU^\dagger])$
and $U \in \SUloc(2^n)$ yields
\begin{eqnarray*}
\label{varphidotdot}%
|\varphi''(t)|
& \le &
\Big\|[\Cd,\Omega]\Big\|\cdot
\Big\|[\Omega,\e^{t\Omega}UAU^\dagger\e^{-t\Omega}]\Big\|\\
& = &
\Big\|[\Cd,\Omega]\Big\|\cdot
\Big\|[\Omega,UAU^\dagger]\Big\|.
\end{eqnarray*}%
Therefore, we get the estimate
\begin{eqnarray*}%
\lefteqn{\max_{t \geq 0} 
\Big|
\frac{\mathrm{d^2}}{\mathrm{d}t^2}
f_{\rm loc} (\exp_U(\Omega t))
\Big|}\\
& \leq &
\Big\|[\Cd,\Omega]\Big\|\cdot \Big\|[\Omega,UAU^\dagger]\Big\|
\end{eqnarray*}%
for $\Omega:=\grad f_{\rm loc}(U)$. Now, a standard Lyapunov-type
argument, similar to the proof of Theorem 3.3 in cf. \cite{HM94},
yields the desired result.
\end{enumerate}
\hfill$\square$

\medskip

\noindent
For similar discretisation schemes in different contexts or other
intrinsic Riemannian methods see also 
\cite{Bro93,HM94,Science98,DiHeKlSchuGAMM06}.


\subsubsection*{Double-Bracket Flows on Naturally Reductive Homogeneous Spaces}

The well-known \emph{double-bracket flows} have established themselves
as useful tools for diagonalising matrices (usually real symmetric
ones) as well as for sorting lists \cite{Bro88+91,Bro93,Bloch94,HM94}.
Moreover, they relate to Hamiltonian integrateble systems \cite{BBR90,BBR92}.
(Note again that in many-particle physics gradient flows were later 
introduced independently for diagonalising Hamiltonians \cite{Weg94,Kehr06}.)
In summarising the most important results we show that double-bracket
flows can be viewed as special cases of gradient flows on {\em naturally reductive}
homogeneous spaces $\bG/\bH$ in terms of Sec.~\ref{sec:flows_homsp}, where $\bH$
is a stabiliser group, which is typically not normal. Then
the homogeneous space $\bG/\bH$ does not constitute a group itself.

Let 
\begin{equation}
\mathcal O(A) := \{ UAU^\dagger | U \in SU(N) \}
\end{equation}
denote the unitary orbit of some $A\in\mathbb{C}^{N\times N}$.
Note that the \emph{adjoint action}
$(U,A) \mapsto \Adr_U A  := UAU^\dagger$ of $SU(N)$ constitutes a
\emph{left} action on the Lie algebra $\g := \mathbb{C}^{N \times N}$.
However, this should not cause any confusion for the reader since
the key result we refer to---Corollary \ref{adjointorbit}---
was presented for left actions.

 
Let $C \in\mathbb{C}^{N\times N}$ be another complex matrix.
For minimising the (squared) Euclidean distance $\Vert X - C \Vert_2^2$ 
between $C$ and the unitary orbit of $A$ we derive a gradient flow
maximising the target function
\begin{equation}
\widehat{f}(X) := \Re \tr \{\Cd X\}
\end{equation}
over $X \in \mathcal O(A)$. Clearly, this is but an alternative
to tackling the problem by  a gradient flow on the unitary group,
since as in Sec.~\ref{sec:flows_homsp}, we have the equivalence
\begin{equation}\label{eqn:maxequiv}
\maxover{X\in\mathcal O(A)} \widehat{f}(X)
\;=\;
\maxover{U\in SU(N)} f(U) 
\end{equation}
for $f(U):= \Re \tr \{C^\dagger UAU^\dagger\}$.

Building upon Corollary \ref{adjointorbit}, we have the following
facts:
$\mathcal O(A)$ constitutes a compact and
connected naturally reductive homogeneous space isomorphic to
$SU(N) / \bH$. Here,
\begin{equation}
\bH := \{U \in SU(N) | \Adr_U A  = A \}
\end{equation}
denotes the stabiliser group of $A$.
Recalling that the Lie algebra of $SU(N)$ is given by 
\begin{equation}
\su(N) :=
\{ \Omega \in \C^{N\times N}\;|\; \Omega^\dagger = - \Omega\},
\end{equation}
we further obtain for the tangent space of $\mathcal O(A)$ at
$X = \Adr_U A$ the from
\begin{equation}
T_X \mathcal O(A) =
\left\{ \adr_X\Omega \;\big|\; \Omega \in \su(N)\right\}
\end{equation}
with $\adr_X \Omega  := [X,\Omega]$.
%
%
Moreover, the kernel of $\adr_A: \su(N) \to \g$ reads
\begin{equation}
\mathfrak h = 
\{\Omega \in \su(N) \;|\;  [A,\Omega] = 0 \}.
\end{equation}
and forms the Lie subalgebra to $\bH$.  Now, by the standard
Hilbert-Schmidt scalar product
$(\Omega_1,\Omega_2) \mapsto \tr\{\Omega_1^\dagger \Omega_2\}$
on $\su(N)$ one can define the ortho-complement to the above
kernel as
\begin{equation}
\p := \mathfrak h^\perp\quad.
\end{equation}
This induces a unique decomposition of any skew-Hermitian matrix 
$\Omega = \Omega^\h + \Omega^\p$ with $\Omega^\h \in \mathfrak h$
and $\Omega^\p \in \mathfrak p$.
Finally, we obtain an $\Adr_{SU(N)}$-invariant Riemannian metric
on $\mathcal O(A)$ via
\begin{equation}
\label{eqn:RiemMetrOrbitMfd}
\braket{\adr_X(\Ad{U}\Omega_1)}{\adr_X(\Ad{U}\Omega_2)}_X
        := \tr \{{\Omega_1^\p}^\dagger \Omega_2^\p\}
\end{equation}
for $X := \Ad{U}A$, which is equivalent to saying
\begin{equation}
\label{eqn:RiemMetrOrbitMfdII}
\braket{\adr_X(\Omega_1)}{\adr_X(\Omega_2)}_X
        := \tr \{{\Omega_1^{\p_X}}^\dagger \Omega_2^{\p_X}\}
\end{equation}
with $\p_X := \Ad{U}\p$. Now, the main results on double-bracket
flows read as follows:
\begin{theorem}
\label{thm:doublebracket}
Set $\widehat f: \mathcal O(A) \to \R$,
$\widehat f(X) := \Re \tr \{\Cd X\}$. Then one finds
\begin{enumerate}
\item[(a)] 
The gradient of $\widehat f$ with respect to the Riemannian metric
defined by Eqn.~\eqref{eqn:RiemMetrOrbitMfd} is given by 
\begin{equation}
\label{eqn:dbgrad}
\grad \widehat f(X) = [X,[X,C^\dagger]_S],
\end{equation}
where $[X,C^\dagger]_S$ denotes the skew-Hermitian part of $[X,C^\dagger]$.
\item[(b)]
The gradient flow
\begin{equation}
\label{eqn:dbflow}
\dot X = \grad \widehat f(X) = [X,[X,C^\dagger]_S] 
\end{equation}
defines an isospectral flow on $\mathcal O(A) \subset \g$.
The solutions exist for all $t \geq 0$ and converge to
a critical point $X_\infty$ of $\widehat f(X)$ characterised by
$[X_\infty,C^\dagger]_S=0$.
\end{enumerate}
\end{theorem}

{\bf Proof.}\hspace{.5em}
(A detailed proof for the real case can be found in \cite{HM94};
for an abstract Lie algebraic version see also \cite{Bro93}.)

(a) For $X = \Ad{U}A$ and $\xi = \ad{X}\Omega \in T_X \mathcal O(A)$
we obtain
\begin{eqnarray*}
\D \widehat{f}(X)\ad{X}\Omega
& = & 
\frac{\rm d}{{\rm d}t}
\Re \tr \{\Cd \e^{-t\Omega}X\e^{t\Omega}\}\,\Big|_{t=0}\\
& = & 
\Re \tr \{\Cd \ad{X}\Omega\}.
\end{eqnarray*}
Therefore, the gradient of $\widehat{f}$ has to satisfy
\begin{equation*}
\Re \tr \{\Cd \ad{X}\Omega^{\p_X}\} =  
\braket{\grad \widehat{f}(X)}{\ad{X}\Omega^{\p_X}}_X
\end{equation*}
for all $\Omega^{\p_X} \in \p_X$. Applying
Eqn.~\eqref{eqn:RiemMetrOrbitMfd} to $X=A$ gives
\begin{equation*}
\Re \tr \{\Cd \ad{A}\Omega^{\p}\} = 
\tr \{{\Gamma^{\p}}^\dagger \Omega^{\p}\} 
\end{equation*}
for all $\Omega^{\p} \in \p$, where $\Gamma^{\p}$ is defined by
$\grad \widehat{f}(A) = \ad{A}\Gamma^{\p}$ with $\Gamma^{\p} \in \p$.
Thus, we finally arrive at
\begin{equation*}
\tr \{(\ad{A^\dagger}C)_S^\dagger \Omega^{\p}\} = 
\tr \{{\Gamma^{\p}}^\dagger \Omega^{\p}\} 
\end{equation*}
for all $\Omega^{\p} \in \p$, where $(\ad{A^\dagger}C)_S$ denotes
the skew-Hermitian part of $\ad{A^\dagger}C$. Hence,
$\Gamma^{\p} = (\ad{A^\dagger}C)_S^{\p}$. Moreover, for
$\Omega^{\h} \in \h$, we have
\begin{eqnarray*}
\tr \{(\ad{A^\dagger}C)^\dagger \Omega^{\h}\}
& = & 
- \tr \{\ad{A}C^\dagger \Omega^{\h}\}\\
& = & 
\tr \{C^\dagger \ad{A}\Omega^{\h}\} \; = \; 0.
\end{eqnarray*}
Hence, $(\ad{A^\dagger}C)_S \in \p$ and therefore
\begin{equation*}
\grad \widehat{f}(A) = \ad{A} (\ad{A^\dagger}C)_S 
=[A,[A,C^\dagger]_S].
\end{equation*}
The same arguments apply to $X = \Ad{U}A$ and thus
\begin{equation*}
\grad \widehat{f}(X) = [X,[X,C^\dagger]_S].
\end{equation*}

%
\medskip

(b) Since Eqn.~\eqref{eqn:dbgrad} evolves on the unitary orbit of
$A$, the associated flow is isospectral by construction. The
compactness of $\mathcal O(A)$ then implies that each solution
$X(t)$ of Eqn.~\eqref{eqn:dbgrad} exists for all $t \geq 0$
and converges to the set of critical points
cf.~Proposition \ref{propgradomegalimit}. 
Moreover, from Theorem \ref{thmIIpointconvergence} we derive that
$X(t)$ converges actually to a single
critical point $X_{\infty}$ of $\widehat{f}$, i.e. to a point
$X_{\infty}$ which satisfies
\begin{equation}
\label{bdcritical}
[X_{\infty},[X_{\infty},C^\dagger]_S] = 0.
\end{equation}
Since $[X_{\infty},C^{\dagger}]_S \in \p_{X_{\infty}}$,
Eqn. \eqref{bdcritical} is equivalent to 
\begin{equation*}
[X_{\infty},C^{\dagger}]_S = 0.
\end{equation*}
\hfill$\square$

In order to obtain a numerical algorithm for maximising $\widehat{f}$
one can discretise the continuous-time gradient flow
\eqref{eqn:dbgrad} as in the previous examples via
\begin{equation}
\label{eqn:db-recursionI}
X_{k+1} =
e^{-\alpha_k [X_k,C^\dagger]_S}\; X_k\; e^{\alpha_k [X_k,C^\dagger]_S}
\end{equation}
with appropriate step sizes $\alpha_ k > 0$. Note that
Eqn.~\eqref{eqn:db-recursionI} heavily exploits the fact that the
adjoint orbit $\mathcal O(A)$ constitutes a
{\em naturally reductive homogeneous space} and thus the
knowledge on its geodesics, cf.~Corollary \ref{adjointorbit}.

\begin{remark}
As an alternative to Eqn.~\ref{eqn:db-recursionI},
taking the standard Euler-type iteration
\begin{equation}
\label{eqn:db-recursionII}
X_{k+1} = X_k + \alpha_k [X_k,[X_k,C^\dagger]_S]
\end{equation}
does {\em not retain} the isospectral nature of the flow. Therefore,
it should only be used as a computationally inexpensive, rough
scheme in the neighbourhood of equilibrium points, if at all.
\end{remark}

For $A,C$ complex Hermitian (real symmetric) and the full unitary
(or orthogonal) group or its respective orbit the gradient flow 
\eqref{eqn:dbgrad} is well understood, cf.~Corollary
\ref{cor:doublebracket}.
However, for non-Hermitian $A$ and $C$, 
the nature of the flow and in particular the critical points have
not been analysed in depth, because the Hessian at critical points
is difficult to come by. 
Even for $A,C$ Hermitian, a full critical point analysis
becomes non-trivial as soon as the flow is restricted to a closed
and connected  {\em subgroup} $\mathbf K \subset SU(N)$.
Nevertheless, the techniques from Theorem \ref{thm:doublebracket}
can be taken over to establish a gradient flow and a respective
gradient algorithm on the orbit $\mathcal O_\bK$ in a straightforward
manner.

\begin{corollary}
\label{cor:loc-doublebracket}
The gradient flow \eqref{eqn:dbgrad} restricts to the subgroup orbit 
$\mathcal O_{\bK}(A):=\{KAK^\dagger\;|\; K \in \bK\subset SU(N)\}$
by taking the respective orthogonal projection $P_{\mathfrak k}$
onto the subalgebra $\mathfrak k \subset \mathfrak su(N)$ of $\bK$
instead of projecting onto the skew-Hermitian part, i.e.
$\dot{X} =
[X, P_{\mathfrak k}[X,C^\dagger]].$
\end{corollary}

\noindent
The corresponding discrete integration scheme takes the form
\begin{equation}
\label{eqn:loc-db-recursionI}
{X}_{k+1} =
e^{-\alpha_k P_{\mathfrak k}[{X}_k,C^\dagger]}\;
{X}_k\; e^{\alpha_k P_{\mathfrak k}[{X}_k,C^\dagger]}
\end{equation}
with appropriate step sizes $\alpha_ k > 0$. 

In view of unifying the interpretation of unitary networks, e.g., for the task of computing
ground states of quantum mechanical Hamiltonians $H\equiv A$, the double-bracket flows
for complex Hermitian $A,C$ on the full unitary orbit $\mathcal O_u(A)$
as well as on the subgroup orbits $\mathcal O_{\bK}(A)$ for different
partitionings $\bK := \{K \in SU(N_1)\otimes SU(N_2)\otimes\cdots\otimes SU(N_r)| \prod_{j=1}^r N_j = 2^n\}$
have shifted into focus \cite{Eisert07}. Therefore, we have given the foundations
for the the recursive schemes of Eqns.~(\ref{eqn:db-recursionI}) and (\ref{eqn:loc-db-recursionI}), 
which are listed in Table~\ref{tab:flows} as U1P and U1KP.

\medskip

Finally, we summarise what is known about the nature of critical points
for the real symmetric or
complex Hermitian case. For a detailed discussion of the real symmetric
case and the orthogonal group see e.g.~\cite{HM94}.


\begin{corollary}
\label{cor:doublebracket}
Let $C$ and $A$ be real symmetric or complex Hermitian
and assume for simplicity that they show distinct eigenvalues in either case. 
Then one finds:
\begin{enumerate}
\item[(a)] For $A,C$ real symmetric, define with respect to the special
orthogonal group $SO(N)$
and $Y\in\mathcal O_o(A):=\{OAO^\top|O\in SO(N)\}$ a pair of target
functions on the group and on the respective orbit by
\begin{eqnarray}
         g(O) &:=& \tr\{C^\top O A O^\top \}\\[2mm]
         \widehat g(Y) &:=& \tr\{C^\top Y \}\quad.
\end{eqnarray}
Then the gradient flow 
\begin{equation}\label{eqn:ggradR}
        \dot O := \grad\; g(O) = [O A O^\top, C]\; O
\end{equation}
shows $2^{(N-1)} N!$ critical points, while the double-bracket flow 
\begin{equation}\label{eqn:ogradR}
        \dot Y := \grad\; \widehat g(Y) = [Y,[Y,C]]
\end{equation}
only shows $N!$ equilibrium points. 

\item[(b)]
For $A,C$ complex Hermitian, and
$X\in\mathcal O_u(A):=\{UAU^\dagger |U\in SU(N)\}$
\begin{eqnarray}
         f(U) &:=& \tr\{C^\dagger U A U^\dagger \}\\[2mm]
         \widehat f(X) &:=& \tr\{C^\dagger X \}
\end{eqnarray}
the gradient flow on the special unitary group $SU(N)$
\begin{equation}\label{eqn:ggradC}
        \dot U := \grad\; f(U) = [U A U^\dagger, C] U
\end{equation}
shows a continuum of critical points, while the double-bracket flow
on the unitary orbit 
\begin{equation}\label{eqn:ogradC}
        \dot X := \grad\; \widehat f(X) = [X,[X,C]]
\end{equation}
again shows only $N!$ equilibrium points.
\item[(c)]
On the orbit, the respective target function has a unique global maximum
which is given by the diagonalisation $\diag(\lambda_1, \dots, \lambda_N)$,
$\lambda_1 > \dots > \lambda_N$ of $A$, if $C$
is assumed to be diagonal of the form 
$C = \diag(\mu_1, \dots, \mu_N)$, $\mu_1 > \dots > \mu_N$.
Moreover, the respective gradient flow converges to the unique global
maximum for almost all initial values with an exponential bound on the
rate.
\end{enumerate} 
\end{corollary}

{\bf Proof.}\hspace{.5em}

(a) and (b) The counting arguments follow immediately from the fact
that in either case for $C$ diagonal with distinct eigenvalues, the
set of critical points
$\mathcal C_{\infty} :=
\{X_{\infty}\in \mathcal O(A) | [X_{\infty}, C] =0\}$ 
on the orthogonal or unitary {\em orbit} is given by 
$N!$ different diagonalisations of $A$ and 
remains therefore invariant under conjugation by any permutation
matrix. 

Moreover, on the orthogonal group $O(N)$, the stabiliser group
of $A$ is given by 
$$  \{\diag(\pm 1, \pm 1, \dots, \pm 1)\}\quad,$$
which adds $2^N$ independent further degrees of freedom.
Finally, restricting to $SO(N)$ we obtain $2^{N-1}N!$ critical points
on the group level. 

In contrast, for the unitary case $SU(N)$, the stabiliser group
of $A$ reads
$$ 
\Big\{ \diag(e^{\ri\phi_1}, \dots, e^{\ri\phi_\nu}, \dots, e^{\ri\phi_N})
\;\Big|\; \sum_{\nu=1}^{N}\phi_\nu \in 2\pi\Z,\; \phi_\nu \in \R \Big\}\,,
$$
which is always continuous.

(c) Since $C$ is symmetric or Hermitian, we can assume without
loss of generality that $C$ is diagonal. Then, the critical point
condition $[X_{\infty},C]$ yields that the critical points of
$\widehat g$ and, respectively, $\widehat{f}$ are given by the
diagonalisations of $A$. Moreover, analysing the Hessian
at critical points shows that {\em there is only one global maximum}
in both cases and no local ones \cite{HM94}.

The exponential convergence of the gradient flows Eqns.~\eqref{eqn:ogradR} 
and \eqref{eqn:ogradC} to the respective unique
global maximum for almost all initial values is also established 
via the Hessian, i.e.~by linearising the respective gradient flows
at critical points \cite{HM94}.
\hfill$\square$


\subsubsection*{Gradient Flows on Naturally Reductive Coset Spaces}

More generally,
let $\bG$ be a Lie group with bi-invariant metric and let $f$
be an {\em equivariant} quality function with respect to the closed
subgroup $\bH$, i.e. for all $H \in \bH$ one has $f(G) = f(HG)$, so
$$
f|_{\bH G} =
\mathrm{constant}
$$
for every $G \in \bG$. 
Moreover, assume that $\bG/\bH$ is a naturally reductive coset space.
Establishing a gradient method for the induced quality function
$\widehat{f}$ on $\bG/\bH$ finally  yields a recursion scheme
which looks like the corresponding one on the group level,
cf.~Theorem \ref{thmgeodesicsquotientspaceA} and
\ref{thmgradientflowequivariant}. 
This, however, is not surprising, as the equivariance of
$f$ guarantees that the gradient (taken on the group level)
at $G \in \bG$ is orthogonal to the coset $\bH G$.
Thus, its \/`pullback\/' to the Lie algebra automatically
belongs to $\p$ inducing a gradient flow on
$\bG/\bH$.---This can be illustrated as follows.

With $\bG/\bH$ being naturally reductive, there is the reductive
decomposition $\fg = \fh \oplus \fp$ with $\p := \h^{\perp}$,
so any $\Omega\in \fg$ decomposes uniquely into
$\Omega=\Omega^\fh + \Omega^\fp$. Then, by the equivariance
of $f$ one finds
\begin{equation*}
\braket{\grad f(G) }{\Omega^\fh G}
 = \rD f(G)\, \Omega^\fh G = 0
\end{equation*}
for all $\Omega^\fh \in \h$. Therefore, the \/`pullback\/' of
the gradient of $f$ to $\g$ satisfies $\grad f(G)G^{-1} \in \p$.
Furthermore, combining Eqns.~\eqref{eq:tangentspace-GH},
\eqref{gradequivariantII} and the identity
\begin{equation*}
\rD (\Pi \circ l_G)(\unity)\,\Omega = \rD \Pi (G) \,G\Omega 
\end{equation*}
for all $\Omega \in \g$ (cf. Remark \ref{multiplicationmap})
yields
\begin{equation}
\label{eq:grad-coset}
\grad \widehat f([G]) =
\rD (\Pi \circ l_G)(\unity) \big(G^{-1} \grad f(G)\big) .
\end{equation}
Thus, from Eqn.~\eqref{quotientgeodesicsA} we finally obtain
\begin{equation*}
\exp_{[G]}\big(t \grad \widehat f([G])\big)
= \big[\exp\big(t \grad f(G)\,G^{-1}\big)\, G\,\big]
\end{equation*}
for all $t \in \R$, where $\exp_{[G]}$ denotes the Riemannian
exponential map at $[G]$, cf. Eqns.~\eqref{eq:gradI} and
\eqref{eq:gradII}.---This precisely explains what we meant by
the above statement \/`looking like the one on the group level\/'.

\medskip

\noindent
{\bf Example} 
Let $f: SU(2^n) \to \R{}$ be an arbitrary smooth function 
that is equivariant under local
unitary operations of the $n$"~fold tensor product
$\SUloc(2^n):= SU(2)\otimes \cdots\otimes SU(2)$.
This includes, e.g., any measure of entanglement
$\mu_E(U)$ that varies smoothly with $U$.
Since by equivariant construction $\grad f|_{\SUloc(2^n)} = 0$,
we may consider the flow to $[\dot U] = \grad \widehat f ([U])$
on the homogeneous space
$$
\bG/\bK = SU(2^n)/\SUloc(2^n)\quad,
$$
which is naturally reductive for all $n$ and Cartan-like only
for $n=2$. 
This can be seen, because (i) $SU(2^n)$ carries a bi-invariant
metric induced by the Killing form allowing to define $\fp:=\fk^\perp$,
which gives the reductive decomposition $\fg = \fk \oplus \fp$, yet only
for $n=2$ one recovers the commutator inclusions 
$[\fk,\fk]\subseteq\fk,\, [\fp,\fp]\subseteq
\fk,\, \text{and}\,[\fk,\fp]\subseteq\fp$;
(ii) in any case, by Prop.~\ref{propinvariantquotientmetric}
there is an $\Ad{\bK}$"~invariant scalar product on $\fp$; 
and (iii) Eqn.~\eqref{naturallyreductive} is fulfilled
for  all $\{a,b,c\}\subseteq \fp$, as 
$
\tr\{[a,b]^\dagger c\} =
- \tr\{b^\dagger [a,c]\}\;,
$
cf. Remark \ref{rem:naturallyreductive}. ---
Clearly, this example generalises analogously to functions that
are equivariant under actions of other partitionings of the full unitary
group giving flows on the  corresponding reductive homogeneous spaces
$$
\bG/\bK =
SU(N)/\big(SU(N_1)\otimes SU(N_2)\otimes\cdots\otimes SU(N_r)\big)
$$
with $\prod_{j=1}^r N_j = N$.

\medskip

Moreover, it is important to note that
quality functions $\widehat{f}$ directly defined
on $\bG/\bH$ (without resorting to equivariance) can be handled with the very same
techniques as above, once the relation 
$$
\grad \widehat f([G]) =
\rD (\Pi \circ l_G)(\unity) \Ad{G^{-1}}\Omega^{\p}
$$
for some $\Omega^{\p} \in \p$ is established,
cf.~Eqn.~\eqref{eq:grad-coset}.

\section{Applications to Quantum Control and Quantum Information}
\label{Sec:applications}


\subsection{A Geometric Measure of Pure-State Entanglement}
\label{entanglement}


The Euclidean distance of a pure state to the set $\mathcal{S}_{pp}$
of all pure product states may be seen as a geometric measure of
\emph{entanglement} \cite{Thirring, Guehne04, NC00}. Since
$\mathcal{S}_{pp}$ coincides with the local unitary orbit
\begin{equation}
\mathcal{O}_{\rm loc}(yy^{\dagger})
:= \{Uyy^{\dagger}U^{\dagger} \;|\; U \in \SUloc(2^n)\}
\end{equation}
of any pure product state $y \in \mathcal{S}_{pp}$, it relates to the
following optimisation task
\begin{eqnarray}
\label{measureI}
\Delta(x)
& := &
\min_{U \in \SUloc(2^n)}
\| xx^{\dagger} - Uyy^{\dagger} U^{\dagger}\|^2,
\end{eqnarray} 
where $x \in \C^{2^n}$ denotes a normalised pure state and
$y \in \C^{2^n}$ a pure product state, e.g.
$y = (1, 0, \dots, 0)^{\top} = (e_1 \otimes \dots \otimes e_1)$.
This notation replaces $\ket x$ by $x$ and $\ketbra x x$ by
$xx^\dagger$ for the sake of convenient generalisation to
higher order tensor products.
Obviously, 
minimising (\ref{measureI}) is equivalent to maximising the
so-called \emph{local transfer} 
\begin{equation}
\label{measureII}
\max_{U \in \SUloc(2^n)}
\Re \big(\tr (xx^{\dagger}Uyy^{\dagger} U^{\dagger})\big),
\end{equation} 
between $xx^{\dagger}$ and $yy^{\dagger}$.
Further, since
\begin{equation*}
\tr (xx^{\dagger}Uyy^{\dagger} U^{\dagger})
                = \big|\tr (x^{\dagger}Uy)\big|^2
\end{equation*}
taking the real part in (\ref{measureII}) is redundant.

Now, the techniques developed in Section \ref{gradflowonSUloc} match
perfectly to tackle problem (\ref{measureII}). Let $C := xx^{\dagger}$,
$A := \diag (1, 0, \dots, 0)$ and define the so-called local unitary
transfer between $C$ and $A$ by the real-valued function
\begin{equation}
\label{restfunction}
f_{\rm loc}(U) :=  \tr\,(CUAU^\dagger)\quad.
\end{equation} 
Then the gradient flow (\ref{SUlocgradflow}) or more precisely its
discretisation (\ref{DiscretizationScheme1}) will generically solve
(\ref{measureII}). For explicit numerical results see Subsection
\ref{numresultsI} and \cite{loc_WCA, CuDiHeICIAM07}.

In general, neither an algebraic characterisation of the maximal
value of $f_{\rm loc}$ nor the structure of its critical points
is known,
the major difficulty arising from the fact that $U$ is restricted
to $\SUloc(2^n)$. As soon as $U$ may
be taken from the entire special unitary group,
the solution is well-known: it is simply obtained by arranging
the (real) eigenvalues of both $A$ and $C$ magnitude-wise 
in the same order \cite{NEUM-37,Bro88+91,OLE-89,HM94,OLE-95}.


\subsection{Generalised Local Subgroup Optimisation}


\subsubsection*{Bipartite Systems and Relations to Singular-Value
Decompositions}

An exceptional case, where the restricted problem (\ref{measureII})
can be solved are bipartite pure systems.
These systems are particularly simple in as much as
the maxima of $f_{\rm loc}$ can be linked
to the singular-value decomposition ({\sc svd})
of the matrices $X$ and $Y$ associated to $x$ and $y$
by $x:=\vec X$ and $y:=\vec Y$.
Since these ideas readily extend to arbitrary finite dimensional
bipartite systems, we generalise the formulation of Problem
\eqref{measureII} thus leading to Eqn.~\eqref{bipartite}, before
going into multi-partite systems.
\begin{proposition}
\label{propbipartite}
Let $X = V_X \Sigma_X W_X^{\dagger}$,
$Y = V_Y \Sigma_Y W_Y^{\dagger}$ be singular value decompositions
with $V_X,V_Y \in \U(N_1)$, $W_X,W_Y \in \U(N_2)$ and
$\Sigma_X,\Sigma_Y$ sorted by magnitude.
Moreover, let $x := \vec X$ and $y := \vec Y$. Then the maximum value
of the local transfer between $xx^{\dagger}$ and $yy^{\dagger}$
is bounded by
\begin{equation}
\label{bipartite}
\begin{split}
\max_{U \in \SU(N_2) \otimes \SU(N_1)} 
& \Re \big(\tr (xx^{\dagger}Uyy^{\dagger} U^{\dagger})\big)\\
& \leq (\tr \Sigma_X^{\dagger}\Sigma_Y )^2.
\end{split}
\end{equation}
Equality is actually achieved for
$V_X, V_Y \in \SU(N_1)$, $W_X, W_Y \in \SU(N_2)$ and
$U_* := ({W}^*_X \otimes V_X)
\cdot (W_Y^{\top} \otimes V_Y^{\dagger})$.
\end{proposition}

{\bf Proof:}\hspace{.5em}
For $U := W \otimes V \in SU(N_2) \otimes SU(N_1)$ we obtain
\begin{eqnarray}
\label{equibipar}
\lefteqn{\tr (xx^{\dagger}U yy^{\dagger}U^{\dagger}) \;=\;}
\nonumber\\
&=&
\tr \big(xx^{\dagger} (W \otimes V) yy^{\dagger}
(W^{\dagger}\otimes V^{\dagger})\big)
\nonumber\\
&=&
\tr \big(xx^{\dagger} \vec(VYW^{\top})\vec(VYW^{\top})^{\dagger})\big)
\nonumber\\
&=&
\big|x^{\dagger} \vec(VYW^{\top})\big|^2
\\
&=&
\big|\tr (X^{\dagger} V Y W^{\top})\big|^2.
\nonumber
\end{eqnarray}
Here, we have used the identities
\begin{eqnarray*}
& \vec (VYW)  = (W^{\top} \otimes V) \vec Y, &\\
& (\vec X)^{\dagger} \vec Y = \tr X^{\dagger}Y &
\end{eqnarray*}
for all $X,Y \in \C^{N_1 \times N_2}$. Now, (\ref{equibipar}) implies
\begin{eqnarray}
\label{equibiparII}
&&
\max_{U \in \SU(N_2) \otimes \SU(N_1)}
\Re \tr (xx^{\dagger}U yy^{\dagger}U^{\dagger}) =\\
&&
= \max_{V \in \SU(N_1) \atop W \in \SU(N_2)}
\big|\tr (X^{\dagger} V Y W^{\top})\big|^2
\leq (\tr \Sigma_X^{\dagger}\Sigma_Y)^2,
\nonumber
\end{eqnarray}
where the last inequality is due to von Neumann,
cf.~\cite{GS-77, NEUM-37}.
If $V_X, V_Y \in \SU(N_1)$ and $W_X, W_Y \in \SU(N_2)$,
equality is assumed in Eqn.~(\ref{equibiparII}) for
\begin{equation*}
U_* := (W_YW_X^{\dagger})^{\top} \otimes V_XV_Y^{\dagger} =
({W}^*_X \otimes V_X) \cdot (W_Y^{\top} \otimes V_Y^{\dagger}).
\end{equation*}
\hfill$\square$

\medskip

\begin{corollary}
Set $x:= \vec A$ and $y:=\vec C$. Then the maximum local transfer
between $xx^{\dagger}$ and $yy^{\dagger}$ in the sense of
Proposition \ref{propbipartite} is bounded by 
\begin{equation*}
||A||_C^2 :=
\max_{V \in \U(N_1) \atop W \in \U(N_2)}
\big|\tr (C^{\dagger} V A W^{\dagger})\big|^2,
\end{equation*}
which is known as the $C$-spectral norm of $A$, cf. \cite{Li94}.
\end{corollary}

Note that in the context of finding maximal distances between {\em global unitary}
orbits for the purpose of geometric discrimination of generic non-pure quantum states \cite{Zycz08},
results similar to \cite{OLE-89,OLE-95} show up, while
here we treat {\em local unitary} orbits of pure bipartite states 
as explicit in Eqn.~(\ref{bipartite}).

\subsubsection*{Multipartite Systems and Relations to Best Rank-$1$
Approximations of Higher Order Tensors}

Proposition \ref{propbipartite} has a straightforward generalisation
to multipartite systems, which relates to best rank-$1$ approximations
of higher order tensors. To outline this relation, we define the concept
of a \emph{generalised local subgroup}  
\begin{equation}
\label{genlocsub}
\SUloc( N_1, \dots, N_r) := \SU(N_1) \otimes \cdots \otimes \SU(N_r).
\end{equation}
of type $( N_1, \dots, N_r)$ with $N_k \in \N$, $k = 1, \dots, r$.
Thus, the associated general local subgroup optimisation Problem
can be stated as follows.

\begin{center}
\textbf{Generalised Local Subgroup Optimisation\\ Problem {\sc (glsop)}}
\end{center}
For $C,A \in \C^{N\times N}$ with $N := N_1 \!\cdot\!  N_2 \cdots N_r $
find
\begin{equation}
\label{measureIII}
\max_{U \in \SUloc(N_1, \dots, N_r)}
\Re \big(\tr (CUA U^{\dagger})\big).
\end{equation} 

\noindent
To our knowledge, the {\sc glsop} seems to be unsolved so far. 
To introduce higher order tensors, we have to fix some further notation.
For simplicity, we regard a
\emph{tensor} of \emph{order} $r \in \N$ as an array
\begin{equation*}
X = (X_{i_1 \dots i_r})_{1 \leq i_1 \leq N_1, \dots, 1 \leq i_r \leq N_r}
\end{equation*}
of size $N_1 \times \dots \times N_r$. 
The space of all
$N_1 \times \dots \times N_r$-tensors is denoted by
$\C^{N_1 \times \dots \times N_r}$. A natural scalar product
for tensors of the same size is given by
\begin{equation}
\label{tensorscalarproduct}
\braket{Y}{X} :=
\sum_{i_1 \dots i_r} {Y}^*_{i_1 \dots i_r} X_{i_1 \dots i_r}\quad.
\end{equation}
Moreover, a tensor $X$ is called a \emph{rank-$1$ tensor} if there
exist $x^k \in \C^{N_k}$, $k = 1, \dots, r$ such that
\begin{equation}
\label{outerproduct}
X = x^1 \otimes_{\!a} x^2 \otimes_{\!a} \cdots \otimes_{\!a} x^r,
\end{equation}
where the $(i_1 \dots i_r)$-entry of the outer product $\otimes_{\!a}$
is defined by
\begin{equation*}
(x^1 \otimes_{\!a} x^2 \otimes_{\!a} \cdots \otimes_{\!a} x^r)_{i_1 \dots i_r}
:=
x^1_{i_1} \!\cdot\! x^2_{i_2} \cdots x^r_{i_r}.
\end{equation*}

Thus, the question of decomposing a given tensor by tensors of lower
rank leads to the following fundamental approximation problem:

\begin{center}
\textbf{Best Rank-$1$ Approximation Problem ({\sc brap})}
\end{center}
Let $\|\cdot\|$ denote the norm induced by scalar product
(\ref{tensorscalarproduct}). For $X \in \C^{N_1 \times \dots \times N_r}$
solve
\begin{eqnarray}
\label{bestapp}
\min_{C \in \C, \|x^k\|=1 \atop k = 1, \dots, r}
\|X - C \cdot x^1 \otimes_{\!a} \dots \otimes_{\!a} x^r  \|^2.
\end{eqnarray} 

\noindent
Note that the above notation $\otimes_{\!a}$ is necessary to
distinguish between two different types of outer products:
the Kronecker product $\otimes$ (of column-vectors), which maps
$r$-tuples of column-vectors to a column-vector of larger size,
and the \/`abstract\/' outer product $\otimes_{\!a}$, which maps $r$-tuples
of column-vectors to arrays ($=$ tensors) of order $r$. The relation
between both is given by the canonical isomorphism 
$\mathfrak{vec}: \C^{N_1 \times \dots \times N_r} \to \C^{N}$ with
$N := N_1 \!\cdot\! N_2 \dots N_r$, which is uniquely determined by
\begin{equation}
\label{caniso}
x^1 \otimes_{\!a} x^2 \otimes_{\!a} \cdots \otimes_{\!a} x^r
\mapsto x^1 \otimes x^2 \otimes \cdots \otimes x^r,
\end{equation}
i.e. $\mathfrak{vec}$ assigns to each array
$X \in \C^{N_1 \times \dots \times N_r}$ a column-vector in $\C^{N}$
by arranging the entries of $X$ in a lexicographical order. With
these notations at hand, the relation between {\sc glsop} and
{\sc brap} can be stated as follows.

\begin{theorem}
\label{thm:tensor}
Let $X \in \C^{N_1 \times \dots \times N_r}$ be a tensor of
order $r$ and let $x := \mathfrak{vec}(X) \in \C^{N}$ with
$N := N_1 \cdot N_2 \cdots  N_r$. Then the {\sc brap} is
equivalent to the {\sc glsop}
\begin{equation}
\label{bestappII}
\max_{U \in \SUloc(N_1, \dots, N_r)} 
\Re \big(\tr (xx^{\dagger}Uyy^{\dagger} U^{\dagger})\big),
\end{equation}
where $y \in \C^{N}$ can be any pure product state, e.g.
$y = (1, 0, \dots, 0)^{\top} = e_1 \otimes \dots \otimes e_1$.
More precisely,
\begin{enumerate}
\item[(a)]
If $U_1 \otimes \dots \otimes U_r$ is a solution of (\ref{bestappII})
then $x^k := U_k e_1$, $k = 1, \dots, r$ and
$C := \braket{X}{x^1 \otimes_{\!a} \dots \otimes_{\!a} x^r}$ solve
(\ref{bestapp}).
\item[(b)]
If $C \in \C$ and $x^k$, $k = 1, \dots, r$ solve
(\ref{bestapp}) then any $U_1 \otimes \dots \otimes U_r$ with
$x^k = U_k e_1$, $k = 1, \dots, r$ yields a solution of
(\ref{bestappII}).
\end{enumerate}
\end{theorem}

\noindent
For proving Theorem \ref{thm:tensor} we need the following
technical lemma. 

\begin{lemma}
\label{lembestapp}
The pair $(x^1 \otimes_{\!a} \dots \otimes_{\!a} x^r, C)$ solves
(\ref{bestapp}) if and only if $x^1 \otimes_{\!a} \dots \otimes_{\!a} x^r$
is a maximum of
\begin{equation}
\label{bestappIII}
\max_{\|z^k\|=1, k = 1, \dots, r} 
\big|\braket{X}{z^1 \otimes_{\!a} \dots \otimes_{\!a} z^r}\big|
\end{equation}
and $C = \braket{X}{x^1 \otimes_{\!a} \dots \otimes_{\!a} x^r}$.
\end{lemma}

{\bf Proof:}\hspace{.5em}
Consider the following identity
\begin{eqnarray*}
\lefteqn{
\|X - C \cdot z^1 \otimes_{\!a} \dots \otimes_{\!a} z^r  \|^2}\\
& = &
\|X\|^2 + |C|^2
- 2 \Re \big({C}^*
\braket{X}{z^1 \otimes_{\!a} \dots \otimes_{\!a} z^r}\big)\\
& = &
\|X\|^2
+ \big|C - \braket{X}{z^1 \otimes_{\!a} \dots \otimes_{\!a} z^r}\big|^2\\
&&
- \big|\braket{X}{z^1 \otimes_{\!a} \dots \otimes_{\!a} z^r}\big|^2.
\end{eqnarray*}
Thus, we obtain
\begin{equation*}
\begin{split}
\min_{C \in \C, \|z^k\|=1 \atop k = 1, \dots, r} 
& \|X - C \cdot z^1 \otimes_{\!a} \dots \otimes_{\!a} z^r  \|^2 =\\
= \|X\|^2 - 
& \max_{\|z^k\|=1 \atop k = 1, \dots, r} 
\big|\braket{X}{z^1 \otimes_{\!a} \dots \otimes_{\!a} z^r}\big|^2.
\end{split}
\end{equation*}
This yields the desired result.
\hfill$\square$

\medskip

{\bf Proof of Theorem \ref{thm:tensor}:}\hspace{.5em}
Let $y = e_1 \otimes \dots \otimes e_1$. Then
\begin{equation*}
\big(U_1 \otimes \cdots \otimes U_r\big) y
 = 
(U_1e_1) \otimes \cdots \otimes (U_re_1)
\end{equation*}
and thus
\begin{eqnarray*}
\tr (xx^{\dagger}Uyy^{\dagger} U^{\dagger})
& = &
\tr (x^{\dagger}Uyy^{\dagger} U^{\dagger}x)\\
& = &
\big|x^{\dagger}Uy \big|^2\\
& = &
\big|\braket{X}{(U_1e_1) \otimes_{\!a} \cdots \otimes_{\!a} (U_re_1)}\big|^2.
\end{eqnarray*}
Therefore, we obtain
\begin{eqnarray*}
\lefteqn{
\max_{U \in \SUloc(N_1, \dots, N_r)}
\Re \big(\tr (xx^{\dagger}Uyy^{\dagger} U^{\dagger})\big) = }\\
& = &
\max_{U \in \SUloc(N_1, \dots, N_r)} 
\big|\braket{X}{(U_1e_1) \otimes_{\!a} \cdots \otimes_{\!a}(U_re_1)}\big|^2\\
& = &
\max_{\|z^k\|=1 \atop k = 1, \dots, r} 
\big|\braket{X}{z^1 \otimes_{\!a} \dots \otimes_{\!a} z^r}\big|^2.
\end{eqnarray*}
and hence Lemma \ref{lembestapp} implies (a) and (b).
\hfill$\square$

\begin{remark}
\begin{enumerate}
\item 
The isomorphism $\mathfrak{vec}$ coincides \/`almost\/' with the
standard $\vec$-operation on matrices for $r = 2$, more precisely
$\mathfrak{vec}(X) = \vec (X^{\top})$.
\item 
Since any phase factor can readily be absorbed into
$x^1 \otimes_{\!a} \dots \otimes_{\!a} x^r$, it is easy to show that 
\begin{equation*}
\begin{split}
\max_{\|x^k\|=1, k = 1, \dots, r} 
\big|\braket{X}{x^1 &\otimes_{\!a} \dots \otimes_{\!a} x^r}\big|\\
= \quad \max_{\|x^k\|=1, k = 1, \dots, r} 
& \Re \big(\braket{X}{x^1 \otimes_{\!a} \dots \otimes_{\!a} x^r}\big).
\end{split}
\end{equation*}
Therefore, maxima of the \/`real-part-expression\/' on the right-hand side
are always maxima of the \/`absolute-value-term\/' on the left.
\item 
By replacing $yy^{\dagger}$ in (\ref{bestappII}) with an appropriate
sum $\sum_{i=1}^{l} y_iy_i^{\dagger}$, the above ideas can be
extended to best approximations of higher rank, i.e. to best
approximations of the form
\begin{eqnarray*}
\label{highbestapp}
\min_{C_i \in \C, \|x^{i,k}\|=1}
\|X - \sum_{i=1}^{l}
C_i \cdot x^{i,1} \otimes_{\!a} \dots \otimes_{\!a} x^{i,r} \|^2,
\end{eqnarray*} 
with $l \leq \min\{N_1, \dots, N_r\}$ and all
$x^{i,1} \otimes_{\!a} \dots \otimes_{\!a} x^{i,r}$ mutually
orthogonal, cf. \cite{Kol01}. 
\item 
Unfortunately, an analogue of Proposition \ref{propbipartite}
involving the tensor {\sc svd} as defined in \cite{LMV-I}
does not hold for higher order tensors. Even the classical
\emph{Eckart-Young} Theorem, which asserts that the best rank-$k$
approximation of a matrix is given by its truncated {\sc svd},
is false for higher order tensors, cf.~\cite{LMV-II}.
\end{enumerate}
\end{remark}


\subsubsection*{Numerical Results}
\label{numresultsI}

For comparing our gradient-flow approach to tensor-{\sc svd} techniques,
here we focus on two examples that are well-established
in the literature, since analytical solutions \cite{Goldbart03} as well
as numerical results from semidefinite programming
are known \cite{Guehne04}. First, consider a pure 3"~qubit state
depending on a real parameter $s \in [0,1]$
\begin{equation}\label{psi1}
\ket{X(s)} := 
\sqrt{s}\ket W + \sqrt{1-s}\ket{V}\quad,
\end{equation}
where one defines
\begin{eqnarray*}
\ket W
&:=&
\tfrac{1}{\sqrt{3}}\Big(\ket{001}+\ket{010}+\ket{100}\Big)\quad{\rm and}\\
\ket{V} &:=& \tfrac{1}{\sqrt{3}}\Big(\ket{110}+\ket{101}+\ket{011}\Big)
\end{eqnarray*}
with the usual short-hand notation of quantum information
$\ket 0 := \left(\begin{smallmatrix}1\\0\end{smallmatrix}\right)$,
$\ket 1 := \left(\begin{smallmatrix}0\\1\end{smallmatrix}\right)$
and $\ket{001}:= \left(\begin{smallmatrix}1\\0\end{smallmatrix}\right)\otimes%
        \left(\begin{smallmatrix}1\\0\end{smallmatrix}\right)\otimes%
        \left(\begin{smallmatrix}0\\1\end{smallmatrix}\right)$ etc.
With these stipulations one finds the corresponding $2\times 2\times 2$
tensor representations for $\ket W$ and $\ket{V}$ to take the form
\begin{equation}
W_{(1,:,:)} = \tfrac{1}{\sqrt 3} 
\begin{bmatrix}0&1\\1&0\end{bmatrix}
\quad
W_{(2,:,:)} = \tfrac{1}{\sqrt 3}
\begin{bmatrix}1&0\\0&0\end{bmatrix}
\end{equation}
and
\begin{equation}
{V}_{(1,:,:)} = \tfrac{1}{\sqrt 3} 
\begin{bmatrix}0&0\\0&1\end{bmatrix}
\quad
{V}_{(2,:,:)} = \tfrac{1}{\sqrt 3}
\begin{bmatrix}0&1\\1&0\end{bmatrix}\;.
\end{equation}
Likewise, observe the pure 4-qubit-state
\begin{equation}\label{psi2}
\ket{\widehat{X}(s)} 
:=\sqrt{s}\, \ket{GHZ'} - \sqrt{1-s}\, \ket{X^+}\otimes\ket{X^+}\quad,
\end{equation}
with the definitions
\begin{eqnarray*}
\ket{\rm GHZ'}
&:=&
\tfrac{1}{\sqrt{2}}\Big(\ket{0011}+\ket{1100}\Big)\\
\ket{X^+} 
&:=&
\tfrac{1}{\sqrt{2}}\Big(\ket{10}+\ket{01}\Big)\quad .
\end{eqnarray*}

\begin{figure}[Ht!]
\mbox{\sf (a)\hspace{35mm}(b)\hspace{23mm}}
\includegraphics[scale=0.22]{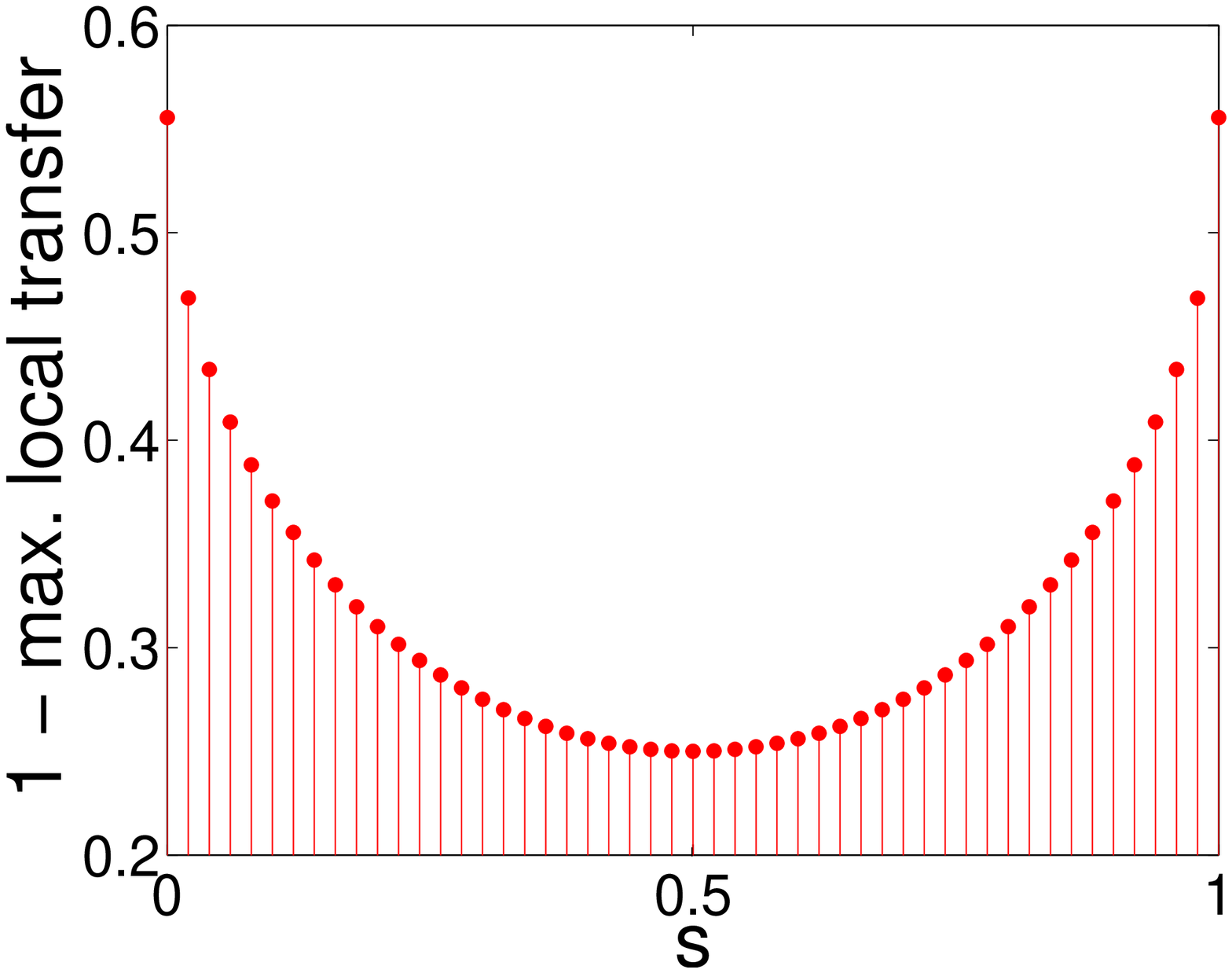}
\includegraphics[scale=0.22]{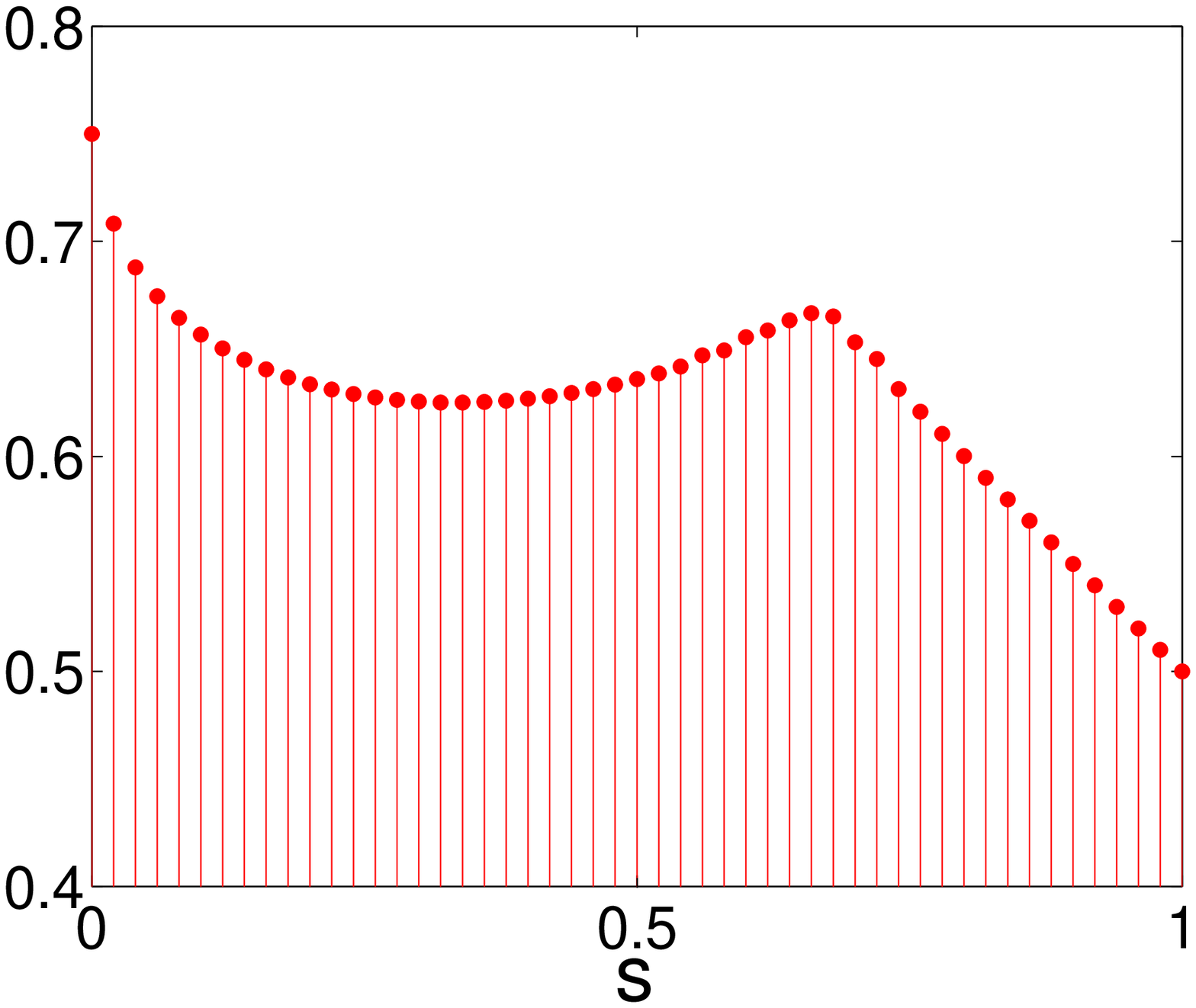}
\caption{\label{fig:EuclDist} (Colour online)
Numerical results by gradient flow on the local unitary
group $\mathbf K=\SUloc(2^n)$ determining
(a) the Euclidean distance of the 3-qubit state 
$\ket{X(s)} =\sqrt{s}\ket W + \sqrt{1-s}\ket{V}$
(see Eqn.~\ref{psi1})
to the nearest product state as a function of $s$; 
(b) the distance of 4-qubit state 
$\ket{\widehat{X}(s)}=\sqrt{s}\ket{\rm GHZ'} - 
\sqrt{1-s}\ket{X^+}\otimes\ket{X^+}$(see Eqn.~\ref{psi2})
to the nearest product state.}
\end{figure}

\begin{table*}[Ht!]
\caption{\label{tab:locals_table} {\sc cpu} Times for Determining the
Euclidean Distance to the Orbit of Separable Pure States as in
Fig.~\ref{fig:EuclDist} }
\begin{ruledtabular}
\begin{tabular}{cccccccc}
no. of
&semidefinite
&\multicolumn{2}{c}{\em gradient flow}
&\multicolumn{2}{c}{tensor-{\sc svd} I ({\sc hopm}) \cite{LMV-II,KoldaBaderMATLAB}}
&\multicolumn{2}{c}{tensor-{\sc svd} II ({\sc hooi}) \cite{LMV-II,KoldaBaderMATLAB}}\\ 

qubits 
&programming 
&\multicolumn{2}{c}{\em on local unitaries} 
&\multicolumn{2}{c}{higher-order power method} 
&\multicolumn{2}{c}{higher-order orthogonal iteration} \\[2mm]

& {\sc cpu} time \;[sec]\footnote{Eisert {\em et al.}
(processor with 2.2 GHz, $1$ GB RAM) \cite{Guehne04}}
& {\sc cpu} time \;[sec]\footnote{average of 50 runs, Athlon XP1800+
(1.1 GHz, $512$ MB RAM)}
& speed-up
& {\sc cpu} time \;[sec]$^b$ 
& speed-up
& {\sc cpu} time \;[sec]$^b$ 
& speed-up
\\[1mm]
\hline\\[-2.5mm]
3 &10.92   &{\em 0.30}  &{\em 36.4} &2.39 &4.6 &5.37 &2.0\\
4 &103.97   &{\em 0.71}  &{\em 147.0} &3.93 &26.5 &7.03 &14.8\\
\end{tabular}
\end{ruledtabular}
\end{table*}

Consider the target function $f(K)=\tr\{C^\dagger KAK^\dagger\}$ with $C=\diag(1,0,0,\dots,0)$
and $A:= \ketbra{\widehat{X}(s)}{\widehat{X}(s)}$.
As shown in Fig.~\ref{fig:EuclDist} with the gradient flow restricted
to the local unitaries $K\in \SUloc(2^n)$ one obtains results perfectly
matching the analytical solutions of \cite{Goldbart03} as well as
the numerical ones from semidefinite  programming ensuring global
optimality---yet in drastically less {\sc cpu} time as compared to
\cite{Guehne04}, see Table~\ref{tab:locals_table}. 
Gradient flows are some 30 to 150 times faster in {\sc cpu} time than
semidefinite programming methods for the 3"~qubit and 4"~qubit example,
respectively. 

\begin{figure}[Ht!]
\includegraphics[scale=0.4]{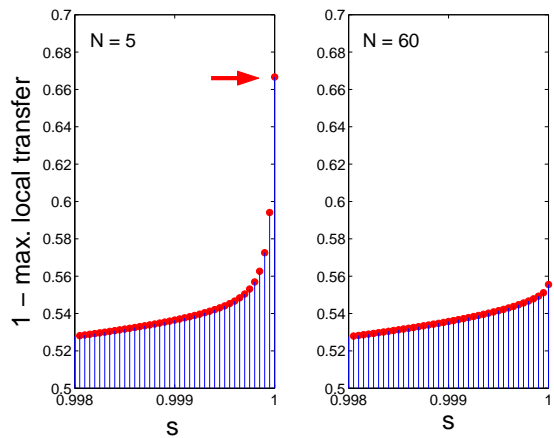}
\caption{\label{fig:EuclDist2} (Colour online)
Tensor-{\sc svd} results for Euclidean distance of the
3-qubit state $\ket{X(s)}$ to the nearest product state as
in Fig.~\ref{fig:EuclDist}(a). With the standard of $N=5$ iterations,
both methods (here shown for {\sc hopm}) give systematic errors as indicated by the arrow.
$N=60$ iterations are needed for quantitatively matching the well-established
distance values. The high number of iterations required slows down the method as indicated
in Tab.~\ref{tab:locals_table}.
}
\end{figure}

In the tensor-{\sc svd} algorithms \cite{LMV-II} such as the
higher-order power method ({\sc hopm}) or the higher-order orthogonal
iteration ({\sc hooi}) as implemented in the {\sc matlab} package
\cite{KoldaBaderMATLAB},
$N = 50$ to $N = 60$ iterations are required for quantitative agreement
with the algebraically established results. In the 3"~qubit example,
all minimal distances are also reproduced correctly with $N=5$
iterations--except for the limiting values $s$ near $0$ and near $1$, 
for which the minimal distances of
$\Delta(\ket{X(0)})=\Delta(\ket{X(1)})=2/3$ 
are obtained by either tensor method instead of the correct analytical
value of $5/9$,  which requires $N=60$ iterations as shown in
Fig.~\ref{fig:EuclDist2}. 
In the 4"~qubit example, however, for $N=5$ iterations, both tensor
methods suffer from apparently random numerical instabilities, which only
vanish when allowing for $N=50$ iterations in either method. 
It is the considerably high number of iterations that makes the tensor
methods substantially slower than our gradient-flow algorithm as shown
in Tab.~\ref{tab:locals_table}.

Therefore, at least for lower order tensors, gradient flows
provide an appealing alternative to tensor-{\sc svd} methods for best
rank-1 approximations of higher order tensors. Moreover, one should
take into account that the above gradient methods are developed
to solve the {\sc glsop} and thus a considerable speed-up can
be expected by adjusting them to the local orbit
$\mathcal{O}_{\rm loc}(yy^\dagger)$ of a pure product state. 
For a similar result obtained by an intrinsic Newton and conjugated
gradient method see also \cite{CuDiHeICIAM07}. --- We anticipate that
these numerical approaches will prove useful tools in tensor and rank
aspects of entanglement and kinematics of qubit pairs as addressed, e.g., in 
Refs.\cite{Bry02,EnMe02}.


\subsection{Locally Reversible Interaction Hamiltonians}
\label{locinversion}


\subsubsection*{Joint Local Reversibility}

\begin{figure}[Hb!]
\includegraphics[scale=0.4]{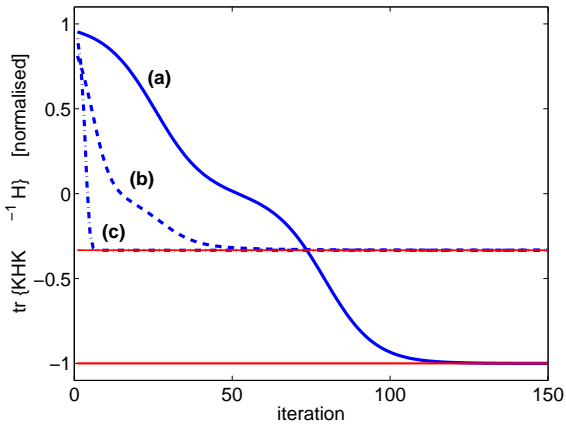}
\caption{\label{fig:inversionI} (Colour online)
Gradient-flow driven local reversion of different Heisenberg 
interaction Hamiltonians:
(a) the $ZZ$ interaction on a cyclic four-qubit topology $C_4$
can be locally reversed,
(c) the $ZZ$ interaction on a cyclic three-qubit topology $C_3$
cannot be reversed locally,
(c) nor the $XXX$ isotropic interaction between two qubits.
}
\end{figure}

In a recent study \cite{PRA_inv}, we have addressed the decision problem
whether a time-independent (self-adjoint) Hamiltonian $H$ normalised to ${||H||}_2=1$
generates a one-parameter
unitary group $U(t)=\{e^{-itH}| t\in\R\}$ that is jointly invertible
for all $t$ by {\em local} unitary operations $K\in \SUloc(2^n)=SU(2)^{\otimes n}$ in the sense 
\begin{equation}
K H K^\dagger = - H\quad.
\end{equation}
Apart from complete algebraic classification, in \cite{PRA_inv} we used
that the question obviously finds an affirmative answer,
if there is an element $K\in \SUloc(2^n)$ such that
\begin{equation}
||KHK^\dagger + H||_2 = 0\quad,
\end{equation}
which amounts to {\em minimising} the transfer function
\begin{eqnarray}
f(K) &=& \Re \,\tr\{H KHK^\dagger\}\quad.
\end{eqnarray}
With $P$ denoting the projector onto $\mathfrak k$, i.e. the Lie algebra of $\mathbf K = \SUloc(2^n)$,
we therefore used the gradient flow 
\begin{eqnarray}
\dot K &=& -\grad f(K) = -P\big([KHK^\dagger,H]\big)\; K
\end{eqnarray}
as an other application of Theorem \ref{thm:gradSUloc}.
If (due to normalisation) $\Re \,\tr \{H KHK^\dagger\} = -1$ can be
reached, the interaction Hamiltonian is locally reversible. 

\begin{remark}
There is an interesting relation to {\em local} $C$"~numerical ranges as described in detail in
Refs.~\cite{WONRA_tosh, WONRA_dirr}: if the local $C$"~numerical range 
$$W_{\rm loc}(H,H) := \{ \tr(HKHK^{-1}) | K\in\mathbf K\} = [-1;+1]$$
then the interaction Hamiltonian $H$ is locally reversible. 
The references also establish the interconnection to local $C$"~numerical ranges of circular
symmetry and multi-quantum interaction components transforming like irreducible
spherical spin tensors.
\end{remark}

In Fig.~\ref{fig:inversionI}, we give some examples: 
e.g., the Heisenberg $ZZ$ interaction in a cyclic four-qubit coupling topology
is locally reversible, while in the cyclic three-qubit
topology or for the isotropic $XXX$ interaction it is not.
Thus numerical tests provide convenient answers in problems where an
algebraic assessment becomes more tedious than in these
examples, which are fully understood on algebraic grounds \cite{PRA_inv}.

\begin{figure}[Hb!]
\includegraphics[scale=0.4]{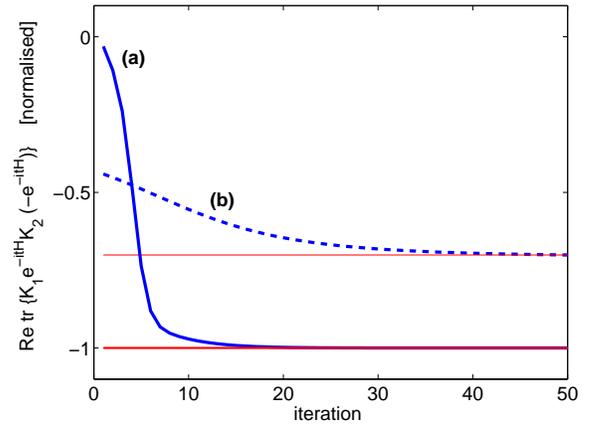}
\caption{\label{fig:inversionII} (Colour online)
Gradient-flow driven local inversion of exponential of the Hamiltonian 
$H = \tfrac{1}{2}\;(\sigma_z\otimes\unity + 
\unity\otimes\sigma_z + \sigma_z\otimes\sigma_z)$
and $U(\tau):=\e^{-i\tfrac{\pi}{4}\;H}$
(a) by a gradient flow with $K_1$ and $K_2$
(b) by a gradient flow with $K_1 = K_2^\dagger =:K$.}
\end{figure}

\subsubsection*{Pointwise Local Reversibility}

In \cite{PRA_inv} we also generalised the above problem to
the question, whether for a fixed $\tau\in\R$ there is a pair
$K_1,K_2\in {\mathbf K}=\SUloc(2^n)$ so that
\begin{eqnarray}
K_1 e^{-i\tau H} K_2 &=& e^{+i\tau H}
\end{eqnarray}
which upon setting $A:=e^{-i\tau H}$ and $C:=e^{+i\tau H}$ is equivalent to 
\begin{eqnarray}
{||K_1 A K_2 - C||}_2 &=& 0\quad.
\end{eqnarray}
Thus one may choose a gradient flow to {\em minimise}
\begin{equation}
f(K_1,K_2) := -\tfrac{1}{2^n} \Re \,\tr\{\Cd K_1 A K_2\}
\end{equation}
by the coupled system
\begin{eqnarray}
\dot K_1 &=& \grad f(K_1) = P(K_1 A K_2 \Cd)\;K_1 \\
\dot K_2 &=& \grad f(K_2) = P(K_2 \Cd K_1 A)\;K_2\quad.
\end{eqnarray}
So if $f(K_1,K_2) = -1$
can be reached, then 
$U(t)=e^{-i\tau H}$ is locally reversible at time $t=\tau$.
See Fig.~\ref{fig:inversionII} for examples comparing pointwise and universal
local reversibility.

\bigskip
\noindent

\begin{figure*}[Ht!]
\fbox{\hspace{-3mm}\parbox[c][8.0cm][s]{1.00\textwidth}
{ \vspace{2mm} {\bf Summary: General Gradient Algorithm for
Steepest Ascent on Riemannian Manifolds}\\[3mm]
\begin{enumerate}
\item[] {\em Requirements:} Riemannian manifold $M$, e.g.~Lie group
$\bG$ with (bi-invariant) metric $\braket{\cdot}{\cdot}$ or its group orbits;
smooth target function $f: M \to \R{}$;
associated gradient system $\dot X = \grad f(X)$.
\item[] {\em Input:} initial state $X(0)\in M$, parameters for target
function.
\item[] {\em Output:} sequence of iterative pairs
$\{\big(X_k, f(X_k)\big)\}$ approximating critical points $X_*$
and their critical values $f(X_*)$.
\item[] {\em Initialisation:}
If possible, generate generic initial state $X_0$, e.g.~for compact
Lie groups pick random $G_0\in \bG$ according to Haar measure
(for $SU(N)$ see \cite{Mez07}) and set $X_0:= G_0\cdot X(0)$,
otherwise identify $X_0:= X(0)$; calculate $f(X_0)$, $\grad f(X_0)$,
and step size $\alpha_0$ according to Section~\ref{Sec:theory}.
\item[] {\em Recursion:}\\ {\bf while} 
                $k = 0,1,2, \dots, k_{\rm limit}$ \;
                {\bf and}\; 
                $\alpha_k > \alpha_{\rm threshold}>0$\quad 
                {\bf do}
\begin{itemize}
\item[] {\bf 1:} iterate $X_{k+1} = \exp_{X_k}\big(\alpha_k \grad f(X_k)\big)$ 
        (see collection of examples in Tab.~\ref{tab:flows}).
\item[] {\bf 2:} calculate $f(X_{k+1})$ .
\item[] {\bf 3:} update step size $\alpha_{k+1}$ according to Section~\ref{Sec:theory}.
\item[] {\bf 4:} {\bf go to} step 1.
\end{itemize}
\item[] {\bf end}
\end{enumerate}
}
}
\caption{\label{fig:summary} Summerising scheme for steepest-ascent gradient flows on
Riemannian manifolds. For related methods, like conjugate gradients,
Jacobi- or Newton-type schemes, step (1) has to be
modified in a straight-forward way according to Sec.~\ref{Sec:overview},
for details see Refs.~\cite{Gabay, Diss-Smith, Diss-Kleinsteuber}.
If the dynamic stepsize selection of Sec.~\ref{Sec:theory} is too costly
{\sc cpu}-timewise, one may start out with constant stepsizes, and
halve them whenever $\big(f(X_{k+1})-f(X_k)\big) \leq 0$, cf.~Armijo's
rule. In cases, where local extrema exist (see Sec.~\ref{Sec:theory}),
make sure to run with a sufficient number of generic initial conditions.}
\end{figure*}

\begin{table*}[Ht!]
\caption{\label{tab:flows}
Examples of Optimisation Tasks and Related Gradient Flows}
\begin{ruledtabular}
\begin{tabular}{clll}
\\[-3mm]
no. & target function & discretised gradient flow & Ref.\\[1mm]
\hline\hline\\[-2.5mm]
\multicolumn{3}{l}{I. Unconstrained Optimisation}\\[1mm]
\hline\\[-2.5mm]
\multicolumn{3}{l}{maximisation over the orthogonal group:
$O\in SO(N,\R)$ and $A,\Delta \in \mathbb{R}^{N \times N}$
with $\Delta$ diagonal, $\alpha_k>0$ stepsize}\\[2mm]
O1 &{$f(O) = \tr \{\Delta^\top OAO^{\top} \}$}
&{$O_{k+1} = \exp\{-\alpha_k [O_k A O_k^{\top}, \Delta^\top]\} O_k$}
&{ \cite{Bro88+91,HM94,Bloch94} }\\[2mm]
\hline\\[-2.5mm]
\multicolumn{3}{l}{maximisation over the unitary group:
$U,V\in SU(N)$ and $A,C\in \mathbb{C}^{N \times N}$; %
	$[\cdot,\cdot]_S$ and $(\cdot)_S$ denote skew-Hermitian parts }\\[2mm]
U1 &{$f(U) = \Re \tr \{C^\dagger UAU^\dagger\}$}
&{$U_{k+1} = \exp\{-\alpha_k [U_k A U_k^\dagger, C^\dagger]_S\} U_k$}
&{ \cite{Science98,TOSH-Diss} }\\[2mm]
U2 &{$f(U) = | \tr \{C^\dagger UAU^\dagger\}|^2$}   
&{$U_{k+1} = \exp\{-\alpha_k \big([A_k, C^\dagger]
f^*(U_k)-[A_k, C^\dagger]^\dagger f(U_k)\big)\} U_k$}
&{ \cite{Science98,TOSH-Diss} }\\
&   &{\hspace{10mm} where \; $A_k := U_k A U_k^\dagger$} 
&{  }\\[2mm]
U3 &{$f(U,V) = \Re \tr \{C^\dagger UAV \}$}
&{$U_{k+1} = \exp\{-\alpha_k (U_k A V_k C^\dagger)_S\} U_k$}
&{ \cite{Bloch94,PRA_inv} }\\
&   &{$V_{k+1} = \exp\{-\beta_k (V_k C^\dagger U_k A)_S\} V_k$}
&{  }\\[2mm]
\hline\\[-2.5mm]
\multicolumn{3}{l}{maximisation restricted to subgroups
$\mathbf K \subset U(N)$ of the unitary group}&\\
&\multicolumn{3}{l}{with $K\in\mathbf K$ and $P_{\mathfrak k}$ as projection from
$\mathfrak{gl}(N,\C)$ onto $\mathfrak k$, i.e. the Lie algebra to
$\mathbf K$}\\[2mm]
U1K &{$f(K) = \Re \tr \{C^\dagger KAK^\dagger\}$}
&{$K_{k+1} = \exp\{-\alpha_k P_{\mathfrak k}[K_k A K_k^\dagger, C^\dagger]\} K_k$}
&{ [here\footnote{work presented in part at the MTNS 2006 \cite{loc_WCA}}] }\\[2mm]
U2K &{$f(K) = | \tr \{C^\dagger KAK^\dagger\}|^2$}   
&{$K_{k+1} = \exp\{-\alpha_k \big(P_{\mathfrak k}[A_k, C^\dagger]
f^*(K_k)-P_{\mathfrak k}[A_k, C^\dagger]^\dagger f(K_k)\big)\} K_k$}
&{ [here$^a$] }\\
&   &{\hspace{10mm} where \; $A_k := K_k A K_k^\dagger$}
&{  }\\[2mm]
U3K &{$f(K_1,K_2) = \Re \tr \{C^\dagger K_1AK_2\}$}
&{$K_{k+1}^{(1)} = \exp\{-\alpha_k P_{\mathfrak k}(K_k^{(1)}
A K_k^{(2)} C^\dagger)\} K_k^{(1)}$}
&{ \cite{PRA_inv} }\\[1mm]
&   &{$K_{k+1}^{(2)} = \exp\{-\beta_k P_{\mathfrak k}(K_k^{(2)} C^\dagger K_k^{(1)} A)\}
K_k^{(2)}$}
&{  }\\[2mm]
\hline\\[-2.5mm]
\multicolumn{3}{l}{maximisation restricted to homogeneous spaces
$\bG/\bH$ of the orthogonal group}&\\
&\multicolumn{3}{l}{with $X\in\mathbf \bG/\bH$ and $A,C$ real symmetric}
\\[2mm]
O1P &{$f(X) = \tr \{C X\}$ \; with $X_k:=\Ad {O_k}(A)$}
&$X_{k+1} =
e^{-\alpha_k[X_k,C]}\;X_k\; e^{+\alpha_k[X_k,C]} $
&{ \cite{BBR90,HM94} }\\[2mm]
\hline\\[-2.5mm]
\multicolumn{3}{l}{maximisation restricted to homogeneous spaces
$\bG/\bH$ of the unitary group}&\\
&\multicolumn{3}{l}{with $X\in\mathbf \bG/\bH$ and $A,C$ {\em arbitrary complex square }
and $P_{\mathfrak k}$ as projection from $\mathfrak{gl}(N,\C)$ onto $\mathfrak k$}
\\[2mm]
U1P &{$f(X) = \Re \tr \{\Cd X\}$ \; with $X_k:=\Ad{U_k}(A)$}
&$X_{k+1} =
e^{-\alpha_k[X_k,\Cd]_S}\;X_k\; e^{+\alpha_k[X_k,\Cd]_S} $
&{ [here] }\\[2mm]
U1KP &{$f(X) = \Re \tr \{\Cd X\}$ \; with $X_k:=\Ad{K_k}(A)$}
&$X_{k+1} =
e^{-\alpha_kP_{\mathfrak k}[X_k,\Cd]}\;X_k\; e^{+\alpha_kP_{\mathfrak k}[X_k,\Cd]} $
&{ [here] }\\[2mm]
\hline\\[-2.5mm]
\multicolumn{3}{l}{II. Constrained Optimisation}\\[1mm]
\hline\\[-2.5mm]
\multicolumn{3}{l}{maximising $L(U)$ with penalty parameter
$\lambda\in\R$ over the unitary group: $U\in SU(N)$;
$A,C,D,E\in \mathbb{C}^{N \times N}$ }\\[2mm]
U1C &{$L(U) = \Re f_C(U) -\lambda \Im^2 f_C(U) $}   
&{$U_{k+1} = \exp\{-\alpha_k \big([A_k, C^\dagger]_S + 
2i \lambda \Im f_C(U_k) [A_k, C^\dagger]_H \big)\} U_k$}
&{ \cite{TOSH-Diss} }\\[0.5mm]
&{\hspace{2mm} with $f_C(U) := \tr \{C^\dagger UAU^\dagger\}$ } 
&{\hspace{10mm} where \; $A_k := U_k A U_k^\dagger$ and
$X_{H,S}:=\tfrac{1}{2}(X\pm X^\dagger)$}\\[2mm]
U2C &{$L(U) = |f_C(U)|^2 -\lambda (f_E(U) - ||E||_2^2) $}   
&{$U_{k+1} = \exp\{-\alpha_k \big((2 f_C^*(U_k) [A_k, C^\dagger])_S -
\lambda [E_k, E^\dagger] \big)\} U_k$}
&{ \cite{TOSH-Diss} }\\[0.5mm]
&{\hspace{2mm} $f_C(U)$ (s.a.) and $f_E(U) := \tr \{E^\dagger UEU^\dagger\}$} 
&{\hspace{10mm} where \; $A_k := U_k A U_k^\dagger$ and
$E_k := U_k E U_k^\dagger$}\\[2mm]
U3C &{$L(U) = |f_C(U)|^2 -\lambda |f_D(U)|^2$}   
&{$U_{k+1} = \exp\{-2\alpha_k \big((f_C^*(U_k) [A_k, C^\dagger])_S -
\lambda (f_D^*(U_k) [A_k, D^\dagger])_S\big)\} U_k$}
&{ \cite{TOSH-Diss} }\\[0.5mm]
&{\hspace{2mm} $f_C(U)$ (s.a.) and $f_D(U) := \tr \{D^\dagger UAU^\dagger\}$} 
&{\hspace{10mm} where \; $A_k := U_k A U_k^\dagger$}\\[2mm]
\end{tabular}
\end{ruledtabular}
\end{table*}



\subsection{Constrained Optimisation in Quantum Control:\\
Intrinsic vs Penalty Approach}


In practical quantum control, one may face the problem to maximise
a quality function $f$ on the reachable set of a quantum system under
\emph{additional} state space contraints. For instance, find the maximal
unitary transfer from matrix (state) $A$ to $C$ subject to leaving
another state $E$ invariant (provided $A$ and $E$ do not share the same
stabiliser group). Another variant amounts to optimising the contrast
between the transfer from $A$ to $C$ and the  transfer from $A$ to $D$; so
the task is to maximise the transfer from $A$ to $C$ subject to suppressing
the transfer from $A$ to $D$.

For tackling those types of problems, we address two basically different
approaches---a purely intrinsic one and a combined method joining
intrinsic and penalty-type techniques.
Both methods will be briefly illustrated for the problem of maximising
the transfer from $A$ to $C$ while leaving $E$ invariant, i.e.
\begin{equation}
\label{eqn:inv_copt}
\maxover {U \in U(N)} |\tr\{UA\Ud \Cd\}|
\quad\mbox{ subject to $UE\Ud= E$}.
\end{equation}
It is straightforward to see that the stabiliser group 
\begin{equation}\label{eq:stab_E}
\mathbf K_E := \{ K\in U(N) \,|\, KEK^\dagger = E \}
\end{equation}
of $E$ forms a compact connected Lie subgroup of $U(N)$.
Differentiating the identity $\e^{tk} E \e^{-tk} = E$ for $t=0$
yields its Lie algebra
\begin{equation}\label{eq:stab_E2}
\mathfrak k_E :=
\{ k\in \mathfrak{u}(N) \,|\, \adr_{k}(E)\equiv \comm{k}{E} = 0 \}.
\end{equation}
By the Jacobi identity 
\begin{equation}\label{jacobi}
[[A,B],E] + [[B,E],A] + [[E,A],B] = 0
\end{equation}
for all  $A,B,E \in \C^{N \times N}$, one can easiliy verify that
$\mathfrak k_E$ is a Lie subalgebra of $\mathfrak{u}(N)$.
Moreover, from the compactness of $\mathbf K_E$ we conclude that
not only does $\mathfrak k_E$ generate the stabiliser group $\mathbf K_E$
in the sense $\langle\exp(\mathfrak k_E)\rangle = \mathbf K_E$,
but also does so in the stronger sense $\exp(\mathfrak k_E) = \mathbf K_E$.

A set of generators of $\mathfrak k_E$ may constructively be found
by solving a system of homogeneous linear equations, i.e.
\begin{equation}\label{eqn:centraliser}
\begin{split}
\mathfrak k_E
&= \ker\adr_E \cap \,\mathfrak{u}(N) \\
&= \{ k \in \mathfrak{u}(N) \;|\;
(\unity\otimes E - E^{\top}\otimes\unity) \vec(k) = 0\}\;.
\end{split}
\end{equation}
In particular, if $E$ is of the form $E=\mu\unity + \Omega$ with
$\mu\in\C{}$ and $\Omega\in\mathfrak{u}(N)$, then $\mathfrak k_E$ is
identical to the centraliser of $\Omega$ in $\mathfrak{u}(N)$.

By ortho-normalising the elements $k_j\in \mathfrak k_E$ of the generating
set $\mathfrak k_E$ with $j=1,2,\dots,n_E$, one obtains the projectors
$ P_j := \ketbra{k_j}{k_j}$ according to Eqn.~\eqref{projector2} to give
the total projection operator $ P := \sum_j P_j$. With this definition,
the gradient flow U2K of the summarising Table~\ref{tab:flows} applies and solves
Eqn.~\eqref{eqn:inv_copt}.
Therefore, the constraint of leaving a neutral state $E$ invariant
during the transfer from $A$ to $C$ can be approached {\em intrinsically}
by restricting the flow from the full unitary group to a compact connected
Lie subgroup, the stabiliser group $\mathbf K_E$ of $E$.

However, since it may be tedious to check for the stabiliser group
$\mathbf K_E$ in each and every practical instance and then
project the gradients onto the corresponding subalgebra $\mathfrak k_E$,
a more versatile programming tool would be welcome.
In \cite{TOSH-Diss}, we therefore presented a {combined approach}
based on the \LAG-type function
\begin{equation}
L(U) = f_2(U) -\lambda\left(\tr\{\Ed UE\Ud\}-||E||_2^2\right)
\end{equation}
with $f_2(U):=|\tr\{C^\dagger UAU^\dagger\}|^2$. and {\em penalty term}
$\lambda\left(\tr\{\Ed UE\Ud\}-||E||_2^2\right)$.
Here, the constraint $UE\Ud -E = 0$ was rewritten in the more convenient
form $ \tr\{\Ed UE\Ud\}-\fnormsq{E} = 0$.
The algorithm given in Table~\ref{tab:flows} as U2C implements a
discretised gradient flow of $L$ obtained from the identity
\begin{equation*}
\begin{split}
D\,&L(U)\,(\Omega U) \\
& = \tr \Big\{\Big(2\,\big({f_2(U)}^* \cdot
[UAU^\dagger,C^\dagger] \big)_S - \lambda\GEE\Big)\, \Omega \Big\}\,.
\end{split}
\end{equation*}
Note that the \emph{penalty parameter} $\lambda$ is increased within
the recursion to guarantee that the constraint is (at least approximately)
satisfied in the limit. 

Thus, for the constrained optimisation task of maximising the transfer
from $A$ to $C$ subject to leaving the state $E$ invariant, 
one has the choice of taking either the intrinsic approach U2K or
the combined approach of U2C. Note, however, that the intrinsic
approach restricts the flow to the stabiliser group $\mathbf K_E$
{\em at any time}, whereas the combined method is designed
such as to start  {\em arbitrarily} on $U(N)$ but \emph{finally} to
give an equilibrium point on $\mathbf K_E$.
Therefore, the intrinsic approach has the advantage that the constraint
is (at least in principal) properly satisfied for the entire iteration.
However, there are situations where an intrinsic method is impractical
as the computational costs 
are too expensive. The combined method, in contrast, does not suffer from
this shortcoming and thus has a wider range of applications.
Note that the intrinsic approach paves the way to perform (or approximate)
a transfer from $A$ to $C$
robustly by taking $\mathbf K_E$ as the stabiliser group 
resistent against a certain error class in the sense familiar from stabiliser 
codes \cite{Grassl07,CRSS98,CS96}. 
The extrinsic approach, on the other hand,
could be taken to transfer one protected state $A$ to another one $C$ via intermediate states
that are no longer necessarily protected against errors as in the intrinsic case. 

Finally, in \cite{TOSH-Diss, WONRA_tosh}, 
we devised a penalty-type
gradient flow algorithm for solving the constrained optimisation
problem 
\begin{equation*}
\label{eq:const_opt}
\maxover U |\tr\{\Cd UA\Ud\}|\quad\text{\rm subject to}\quad
\tr\{\Dd UA\Ud\} = \min.
\end{equation*}
To this end, we introduced the \LAG-type function
\begin{equation*}
L(U):= |\tr\{\Cd UA\Ud\}|^2-\lambda\,|\tr\{\Dd UA\Ud\}|^2\quad,
\end{equation*}
to maximise the transfer from $A$ to $C$ while suppressing the
transfer from $A$ to $D$. This leads to the recursive scheme U3C in 
Table~\ref{tab:flows}.


\section{CONCLUSIONS}

The ability to calculate optima of quality functions for quantum
dynamical processes and to determine steerings in concrete experimental
settings that actually achieve these optima is tantamount to
exploiting and manipulating quantum effects in future technology.

To this end, we have presented a comprehensive account of gradient
flows on Riemannian manifolds (see general scheme of Fig.~\ref{fig:summary})
allowing for generically convergent
quantum optimisation algorithms---an ample array of explicit examples
being given in Tab.~\ref{tab:flows}.
Since the state space 
of quantum dynamical systems can often be represented by smooth
manifolds, the unified foundations given are illustrated by many
applications for numerically addressing two categories of problems:
(a) we have focussed on abstract optimisation tasks over the dynamic
group, and (b) we have sketched the relation to optimal control tasks
in specified experimental settings.

In the present work on closed systems, a variety of applications
are addressed by relating the dynamics to Lie group actions of
the unitary group and its closed subgroups.
Since symmetries give rise to stabiliser groups, particular emphasis
has been on gradient flows on homogeneous spaces. 
Theory and algorithms have been structured and tailored for the
following scenarios:
\begin{enumerate}
\item[(i)] for Lie groups with bi-invariant metric,
\item[(ii)] for closed subgroups 
\item[(iii)] for compact Riemannian symmetric spaces,
\end{enumerate}
or, more generally,
\begin{enumerate}
\item[(iv)]  for naturally reductive homogeneous spaces.
\end{enumerate}
As soon as the homogeneous spaces are no longer naturally reductive,
the \/`usual\/' way of obtaining geodesics on quotient spaces
(by projecting geodesics from the group level to the quotient) fails. 
Alternatives of local
approximations have been sketched in these cases in order to
structure future developments.

Techniques based on the Riemannian exponential are easy to implement
on Lie groups (with bi-invariant metric) and their closed subgroups.
In particular, gradient flows on subgroups of the unitary group allow to
address different partitionings of $m$-party quantum systems, the finest one 
being the group of purely local operations 
$SU(2)\otimes SU(2)\otimes\cdots\otimes SU(2)$.
The corresponding gradient flows have several applications in quantum dynamics:
for instance they prove useful tools to decide whether effective
multi-qubit interaction Hamiltonians generate time evolutions that
can be reversed in the sense of Hahn's spin echo solely by local
operations.  

As a new application, gradient flows on 
$SU(N_1)\otimes SU(N_2)\otimes\dots\otimes SU(N_m)$
turned out to be a valuable and reliable alternative to
conventional tensor-{\sc svd} methods for determining best
rank-1 tensor approximations to higher-order tensors.
In the case of $m$"~party multipartite pure quantum states,  
they can readily be applied to optimising entanglement witnesses.

Double-bracket flows have been characterised as a special case
of a broader class
of gradient flows on naturally reductive homogeneous orbit spaces.
Here, in view of using gradient techniques for ground-state
calculations \cite{Eisert07}, it is important to note that
double-bracket flows can also be established for any closed subgroup
of $SU(N)$: by allowing for different partitionings
$SU(N_1)\otimes SU(N_2)\otimes\dots\otimes SU(N_m)$, 
one may set up a common frame to compare different types of 
unitary networks \cite{Eisert07,Eisert08} for calculating and simulating 
large-scale quantum systems.

Moreover, we have shown how techniques of restricting a gradient
flow to subgroups also prove a useful tool for addressing
constrained optimisation tasks by ensuring the constraints
are fulfilled intrinsically. As an alternative, we have devised 
gradient flows that respect the constraints extrinsically, i.e., by way of
penalty-type Lagrange parameters. These methods await application, e.g.,
in error-correction and robust state transfer.

Finally, in a follow-up study, we discuss the
dynamics of open quantum systems in terms of Lie semigroups
\cite{DHKS08}.
We sketch relations between the theory of Lie semigroups
and completely positive semigroups. In particular in open systems,
an easy characterisation of reachable sets
arises only in very simple cases. It thus poses a current limit to an
abstract optimisation approach on reachable sets.
However, in these cases, gradient-assisted optimal control methods again prove valuable.

Therefore, not only does the current work give the justification
to some recent developments, it also provides new techniques
to the field of quantum control. 
It shows how to exploit the differential geometry in Lie theoretical terms
for optimisation of dynamics on quantum-state manifolds.
Thus the comprehensive theoretical treatment
illustrated by known examples and new practical applications
has been given to fill a gap. ---
We anticipate the ample array of methods and their exemplifications
will find broad application. 
The account of theoretical foundations is also meant to 
structure and trigger further basic research thus widening the
set of useful tools.


\section{APPENDIX SECTION}
\section*{Appendix A: Fr{\'e}chet Differentials and Tangent Maps}
\label{app:frechet-diff}

For basic differential geometric terms and definitions we refer
to \cite{AMR88}. Here, we recall only the fundamental relation
between Fr{\'e}chet derivatives and tangent maps for
\emph{finite dimensional} smooth manifolds. 

\begin{definition}[Fr{\'e}chet derivative]
Let $\mathcal H$ and $\mathcal H'$ be real or comlpex Hilbert
spaces. Let $F: \mathcal U \to \mathcal H'$ be a map defined
in some open neighbourhood $\mathcal U \subset \mathcal H$ of $X$. Then
$F$ is \emph{(Fr{\'e}chet) differentiable} in $X$, if there exists a
linear operator $DF(X): \mathcal H \to \mathcal H'$ and a map
$r_{X}: \mathcal H \to \mathcal H'$ such that
\begin{equation}
\label{dervative}
F(X + \Delta) - F(X) = DF(X)\Delta + r_{X}(\Delta) \norm{\Delta}
\end{equation}
for $\Delta \in \mathcal H$ with $X + \Delta \in \mathcal U$ and
$\norm{r_{X}(\Delta)} \to 0$ for $\norm{\Delta} \to 0$.
The linear operator $DF(X)$ is called the \emph{(Fr{\'e}chet) derivative}
of $F$ in $X$ and $F$ is said to be \emph{smooth}, if it is
Fr{\'e}chet differentiable of any order for all $X \in \mathcal U$.
\end{definition}

\begin{remark}
\begin{enumerate}
\item[(a)]
It is more common to stress the term \/`Fr{\'e}chet\/' derivative
in an infinite dimensional setting  to distinguish from other
non-equivalent differentiability notions, cf. \cite{HillePhillips}.
In finite dimensions, however, $DF(X)$ is often simply called
\/`the\/' derivative of $F$.
\item[(b)]
Note, that if $DF(X)$ is complex linear for all $X \in \mathcal U$,
then $F$ is immediately holomorphic, i.e. $F$ can be locally expanded
in a power series, cf.~\cite{HillePhillips}.
\end{enumerate}
\end{remark}

Now, let $M$ be a smooth manifold. If $M$ is embedded in a
Hilbert space $\mathcal H$, then the \emph{tangent space}
$T_X M$ to $M$ at the point $X$ is the set of all tangent vectors
$\xi$ at $X$, i.e. the set of all \/`velocity\/' vectors
\begin{equation*}
\xi = \frac{\mathrm{d}}{\mathrm{d}t}\gamma(t)
\,\Big|_{t=0}
\end{equation*}
of smooth curves in $M$ through $X = \gamma(0)$.
If $M$ is not embedded in a Hilbert space $\mathcal H$, the situation is
slightly more complicated. Then, $T_X M$ is defined as the set of all
equivalence classes $\xi = [\gamma]$ of smooth curves $\gamma$ in $M$
through $X = \gamma(0)$ having the same \/`velocity\/' at $t=0$, i.e.
$\gamma$ and  $\widetilde{\gamma}$ represent the same equivalence class
$\xi$ if and only if
\begin{equation*}
\frac{\mathrm{d}}{\mathrm{d}t}(\phi \circ\gamma)(t)
\,\Big|_{t=0} =
\frac{\mathrm{d}}{\mathrm{d}t}(\phi \circ\widetilde{\gamma})(t)
\,\Big|_{t=0}
\end{equation*}
for one and thus for all charts $\phi$ at $X$.  These equivalence
classes $\xi$ are called \emph{tangent vectors}. Note that
either way the tangent space $T_XM$ is a vector space isomorphic
to the chart space.

\begin{definition}[Tangent map]
Let $M$ and $N$ be smooth manifolds. Take $f:M \to N$ as a
continuous map.
\begin{enumerate}
\item[(a)]
Then $f$ is called \emph{smooth}, if $\psi \circ f \circ \phi^{-1}$
is smooth for all admissible charts $\phi$ and $\psi$.
\item[(b)]
If $f$ is smooth, the linear map $Df(x): T_X M \to T_{f(X)}N$ given by
\begin{equation*}
\xi = [\gamma] \mapsto Df(X)(\xi) := [f\circ\gamma]
\end{equation*}
is called the \emph{tangent map} to $f: M \to N$ at $X \in M$.
\end{enumerate}
\end{definition}


\noindent
Therefore, differentiability on manifolds is locally
expressed in terms of charts. The associated tangent maps 
obey the standard rules of differential calculus,
like the chain rule, etc.

Finally, we quote two useful results for computing tangent
maps in the case of embedded manifolds. 

\begin{app:fact}
If $N$ is embedded in a Hilbert space $\mathcal H$,
the above definition of the tangent map reads
\begin{equation*}
\xi \mapsto Df(X)(\xi) =
\frac{\mathrm{d}}{\mathrm{d}t}(f\circ\gamma)(t)\,\Big|_{t=0}\;,
\end{equation*}
where $\gamma$ denotes a representative of the  tangent vector $\xi$.
\end{app:fact}

\begin{app:fact}
\label{embedded}
Let $\mathcal H$, $\mathcal H'$ be Hilbert spaces and
let $M \subset \mathcal H$, $N \subset \mathcal H'$ be
embedded submanifolds. Furthermore, let
$F:\mathcal H \to \mathcal H'$
be a smooth map with $F(M) \subset N$. Then the tangent map of
$f:= F|_M:M \to N$ at $X$ is given by the restriction of the
derivative $DF(X)$ to the  tangent space of $M$ at $X$,
i.e.
\begin{equation*}
Df(X) = DF(X)|_{T_XM} : T_XM \to T_{f(X)}N.
\end{equation*}
\end{app:fact}

\section*{Appendix B: Polarisation Procedure}
\label{app:polartrick}

What is meant by the term \emph{polarisation procedure}?
Below, we give a short description of the concept and
a sketch of its proof.

Let $\mathcal{H}$ be a real Hilbert space and $\beta: \mathbb{H} \to \R$
a bounded \emph{quadratic form}, i.e., there exists a bounded symmetric
bi-linear form $B:\mathcal{H} \times \mathcal{H} \to \R$ such that 
\begin{equation}
\label{bilinearform}
\beta(v) = B(v,v),
\end{equation}
for all $v \in \mathcal{H}$.

\begin{claimA}
Let $\beta$ be a bounded quadratic form. The bounded symmetric
bi-linear form $B$ which satisfies (\ref{bilinearform}) is
uniquely determined by $\beta$.
\end{claimA}

{\bf Proof.}\hspace{.5em}
By the symmetry and bi-linearity of $B$ we have
\begin{equation*}
B(v+w,v+w) = B(v,v) + B(w,w) + 2B(v,w)
\end{equation*}
and hence
\begin{eqnarray}
B(v,w)
& = &
\frac{1}{2}\Big(B(v+w,v+w) - B(v,v) - B(w,w)\Big)
\nonumber\\ 
\label{polarization}
& = &
\frac{1}{2}\Big(\beta(v+w) - \beta(v) - \beta(w)\Big)
\end{eqnarray}
for all $v,w \in \mathcal{H}$. Therefore, $B$ is uniquely
determined by the quadratic form $\beta$. \hfill$\square$

\medskip

\noindent
The above identity (\ref{polarization}) is frequently called the
\emph{law of polarisation}. Next, we show that $\beta$ can always
be represented by a selfadjoint linear operator.

\begin{claimA}
For any bounded quadratic form $\beta$ there exists a unique
bounded selfadjoint linear operator
$\mathbf{B}: \mathcal{H} \to \mathcal{H}$ such that 
\begin{equation}
\label{claimF3}
\braket{v}{\mathbf{B}v} = \beta(v),
\quad\mbox{for all $v\in \mathcal{H}$}.
\end{equation}
\end{claimA}

{\bf Proof.}\hspace{.5em}
A straightforward application of the Riesz Representation Theorem
yields that any bounded symmetric bi-linear form $B$ on $\mathcal{H}$
can be represented by a bounded selfadjoint linear operator,
cf.~\cite{ReedSimonI}. From this the result follows immediately.
\hfill$\square$

\begin{remarkA}
If $\beta$ is any form such that $\beta(t v) = t^2 \beta(v)$ and
$\beta(v) \leq K \|v\|^2$ is satisfied for all $t \in \R$ and
$v \in \mathcal{H}$, then $\beta$ is in general not induced by a
bounded symmetric bi-linear form, or in other words,
(\ref{polarization}) does in general not define a bounded symmetric
bi-linear form. However, if $\beta$ meets the parallelogram
identity, i.e.
\begin{equation}
\label{parallelogram}
\beta(v+w) + \beta(v-w) = 2 \beta(v) + 2 \beta(w)
\end{equation}
for all $v,w \in \mathcal{H}$, then (\ref{polarization}) does
define a bounded symmetric bi-linear form.
\end{remarkA}

\medskip
\begin{acknowledgments}
Fruitful discussion with Jens Eisert on \cite{Eisert07} is gratefully
acknowledged.
We wish to thank Otfried G{\"u}hne for drawing our attention to witness
optimisation and
Ref.~\cite{Guehne04}.
This work was supported in part by the integrated EU programme QAP and
by {\em Deutsche Forschungsgemeinschaft}, DFG, in the incentives SPP 1078
and SFB 631.
Support and exchange enabled by the two Bavarian PhD programmes of excellence
{\em Quantum Computing, Control, and Communication} (QCCC) as well as
{\em Identification, Optimization and Control with Applications in Modern
Technologies}  is gratefully acknowledged.
\end{acknowledgments}

\bibliography{control21}
\end{document}